\documentclass{emulateapj} 

\usepackage {epsf}
\usepackage{epstopdf}
\usepackage {lscape, graphicx}
\usepackage {amssymb,amsmath}

\gdef\1054{MS\,1054--03}

\def\farcs{\hbox{$.\!\!^{\prime\prime}$}}
\def\simgeq{{\raise.0ex\hbox{$\mathchar"013E$}\mkern-14mu\lower1.2ex\hbox{$\mathchar"0218$}}} 

\begin {document}

\title {Smooth(er) stellar mass maps in CANDELS: constraints on the longevity of clumps in high-redshift star-forming galaxies}

\author{Stijn Wuyts\altaffilmark{1}, Natascha M. F\"{o}rster Schreiber\altaffilmark{1}, Reinhard Genzel\altaffilmark{1},
Yicheng Guo\altaffilmark{2}, 
Guillermo Barro\altaffilmark{3}, 
Eric F. Bell\altaffilmark{4}, 
Avishai Dekel\altaffilmark{5},
Sandra M. Faber\altaffilmark{3}, 
Henry C. Ferguson\altaffilmark{6}, 
Mauro Giavalisco\altaffilmark{2}, 
Norman A. Grogin\altaffilmark{6}, 
Nimish P. Hathi\altaffilmark{7}, 
Kuang-Han Huang\altaffilmark{8}, 
Dale D. Kocevski\altaffilmark{3}, 
Anton M. Koekemoer\altaffilmark{6}, 
David C. Koo\altaffilmark{3}, 
Jennifer Lotz\altaffilmark{6}
Dieter Lutz\altaffilmark{1}, 
Elizabeth McGrath\altaffilmark{3}, 
Jeffrey A. Newman\altaffilmark{9}, 
David Rosario\altaffilmark{1}, 
Amelie Saintonge\altaffilmark{1}, 
Linda J. Tacconi\altaffilmark{1}, 
Benjamin J. Weiner\altaffilmark{10}, 
Arjen van der Wel\altaffilmark{11}}

\altaffiltext{1}{Max-Planck-Institut f\"{u}r extraterrestrische Physik, Postfach 1312, Giessenbachstr., D-85741 Garching, Germany}
\altaffiltext{2}{Astronomy Department, University of Massachusetts, 710 N. Pleasant Street, Amherst, MA 01003, USA}
\altaffiltext{3}{UCO/Lick Observatory, Department of Astronomy and Astrophysics, University of California, Santa Cruz, CA 95064, USA}
\altaffiltext{4}{Department of Astronomy, University of Michigan, 500 Church Street, Ann Arbor, MI 48109, USA}
\altaffiltext{5}{Racah Institute of Physics, The Hebrew University, Jerusalem 91904, Israel}
\altaffiltext{6}{Space Telescope Science Institute, 3700 San Martin Drive, Baltimore, MD 21218, USA}
\altaffiltext{7}{Observatories of the Carnegie Institution of Washington, Pasadena, CA 91101, USA}
\altaffiltext{8}{Johns Hopkins University, 3400 North Charles Street, Baltimore, MD 21218, USA}
\altaffiltext{9}{Department of Physics and Astronomy, University of Pittsburgh, 3941 O'Hara Street, Pittsburgh, PA 15260, USA}
\altaffiltext{10}{Steward Observatory, 933 N. Cherry Street, University of Arizona, Tucson, AZ 85721, USA}
\altaffiltext{11}{Max-Planck-Institut f\"{u}r Astronomie, K\"{o}nigstuhl 17, D-69117 Heidelberg, Germany}

\begin{abstract}
We perform a detailed analysis of the resolved colors and stellar populations of a complete sample of 323 star-forming galaxies at $0.5 < z < 1.5$, and 326 star-forming galaxies at $1.5 < z < 2.5$ in the ERS and CANDELS-Deep region of GOODS-South.  Galaxies were selected to be more massive than $10^{10}\ M_{\sun}$ and have specific star formation rates above $1 / t_H$.  We model the 7-band optical ACS + near-IR WFC3 spectral energy distributions of individual bins of pixels, accounting simultaneously for the galaxy-integrated photometric constraints available over a longer wavelength range.  We analyze variations in rest-frame color, stellar surface mass density, age, and extinction as a function of galactocentric radius and local surface brightness/density, and measure structural parameters on luminosity and stellar mass maps.  We find evidence for redder colors, older stellar ages, and increased dust extinction in the nuclei of galaxies.  Big star-forming clumps seen in star formation tracers are less prominent or even invisible on the inferred stellar mass distributions.  Off-center clumps contribute up to $\sim 20\%$ to the integrated SFR, but only 7\% or less to the integrated mass of all massive star-forming galaxies at $z \sim 1$ and $z \sim 2$, with the fractional contributions being a decreasing function of wavelength used to select the clumps.  The stellar mass profiles tend to have smaller sizes and M20 coefficients, and higher concentration and Gini coefficients than the light distribution.  Our results are consistent with an inside-out disk growth scenario with brief (100 - 200 Myr) episodic local enhancements in star formation superposed on the underlying disk.  Alternatively, the young ages of off-center clumps may signal inward clump migration, provided this happens efficiently on the order of an orbital timescale.
\end{abstract}


\section {Introduction}
\label{intro.sec}

Understanding the nature of typical $\sim M^*$ star-forming galaxies (SFGs) during the peak epoch of cosmic star formation remains one of the key goals in galaxy evolution.  Starting already from the first lookback observations with HST/WFPC2, an increased fraction of irregular rest-frame UV morphologies was noted among galaxies at higher redshift (Griffiths et al. 1994, Windhorst et al. 1995, Abraham et al. 1996).  With the exquisite survey capabilities of the Advanced Camera for Surveys (ACS), the statistics on irregular galaxy number counts have since increased significantly (e.g., Elmegreen et al. 2007).  Since in the nearby universe, such irregular morphologies tend to be caused by interactions and mergers, they were initially often associated with the same physical processes at larger lookback times.

Two reasons for concern about the above line of reasoning are the following.  First, the rest-frame UV part of the spectral energy distribution (SED) is heavily dominated by the emission from short-lived O and B stars.  If the star formation history (SFH) is not uniform across the face of the galaxy, the rest-frame UV light may therefore differ from the underlying stellar mass distribution.  In a scenario where the SFH averaged over longer timescales is uniform (or uniform at a given galactocentric radius), but varies spatially on the order of the lifetime of OB associations, irregular (clumpy) rest-frame UV morphologies would originate, while the underlying stellar mass distribution would be smooth.

Second, evidence has been accumulating over the past years that not all (or in fact only a minority of) high-z galaxies are in the process of merging.  Counts of galaxy pairs (Bundy et al. 2009; Lopez-Sanjuan et al. 2009; de Ravel et al. 2009; Williams et al. 2011) as well as disturbed morphologies (Lotz et al. 2008, 2011a; Conselice et al. 2009) of mass-selected samples of galaxies out to $z=2$ report pair/merger fractions ranging from a few to 25\% .  Moreover, the gas-phase kinematics of a substantial fraction of SFGs at $z \sim 1-2$, even including those with irregular morphologies, is dominated by ordered disk rotation (see, e.g., Genzel et al. 2006; F\"{o}rster Schreiber et al. 2006, 2009; Shapiro et al. 2008; Epinat et al. 2012), unlike what would be expected in the case of frequent (major) merging (although see also Robertson et al. 2006).  The relative paucity of major mergers is also reflected in the inferred gradual evolution of galaxy-averaged star formation histories (Robaina et al. 2009; Bouch\'{e} et al. 2010; Renzini 2010; Genzel et al. 2010; Wuyts et al. 2011a; Rodighiero et al. 2011), and leads to a well-defined relation between the rate of ongoing star formation and the amount of assembled stellar mass of normal SFGs, dubbed the 'main sequence of star formation' (MS; Noeske et al. 2007; Elbaz et al. 2007; Daddi et al. 2007).  Fitting single-component Sersic profiles, Wuyts et al. (2011b) found the surface brightness distributions of normal SFGs, located on the MS, to be best described by exponential disks, unlike the cuspier systems that lie below or above the MS.  

The star formation rates (SFRs) and molecular gas mass fractions (Tacconi et al. 2010) of typical SFGs rise steeply from $z \sim 0$ to $z \sim 2$.   If not mergers, other mechanisms must play a dominant role in fueling the high level of star formation activity.  In fact, a continuous fueling of the gas reservoir is required to sustain the observed levels at $z \sim 2$.  Cosmological simulations indicate that at that epoch, cold, filamentary streams of gas can efficiently penetrate through halos of galaxies and deposit new fuel for star formation (Keres et al. 2005, 2009; Dekel \& Birnboim 2006; Dekel et al. 2009a).  A picture has hence emerged in which the star-forming clumps that are characterstic features of the H$\alpha$ and rest-frame UV maps of high-z galaxies originate from gravitational instabilities within continuously replenished gas-rich disks.

Recently, the fate of these clumps has become a hot topic of debate.  If they correspond to local mass overdensities, dynamical friction would lead them to migrate inwards, where they may coalesce to form a bulge (Noguchi 1999; Immeli et al. 2004a,b; Bournaud et al. 2011; Genzel et al. 2008; Dekel et al. 2009b, Ceverino et al. 2010).  Whether such an avenue for bulge formation, alternative to the classical merger paradigm, is viable, depends sensitively on the longevity of the clumps.  The clump survival timescale may be severely limited by internal feedback processes.  First evidence for outflowing material emerging from individual clumps, with mass loading factors ($\dot{M}_{out}/SFR$) as high as $\sim 8$ has been reported by Genzel et al. (2011) and Newman et al. (2012).  Their observations suggest that at least some of the clumps could be disrupted on timescales of a few 100 Myr.

A crucial step forward in addressing the first concern, and shedding light on the nature of clumps and their potential impact on galaxy evolution, is an expansion of the analysis into the observed near-infrared, tracing the rest-frame optical emission of $z \sim 2$ SFGs on the relevant $\sim 1$ kpc scales.  First efforts in this direction date back to Thompson et al. (1999) and Dickinson et al. (2000), but until recently the statistical relevance of such studies, mostly carried out from space, was limited by the small field of view of the NICMOS camera onboard HST.  Moreover, while the rest-frame optical light is a better proxy for stellar mass than rest-frame UV emission, significant variations in the rest-frame $V$-band mass-to-light ($M/L$) ratio are known to exist between galaxies near and far, and likewise may be expected spatially within individual high-z galaxies.  Resolved color information is therefore a key ingredient in tracing spatial variations in the stellar populations, and assessing the above evolutionary scenario.  Pioneering work on resolved stellar populations in distant galaxies using NICMOS was carried out by Elmegreen et al. (2009) and F\"{o}rster Schreiber et al. (2011b), but again mostly limited to a small number of objects.

With the advent of the Wide Field Camera 3 (WFC3) onboard HST, this situation has changed.  With twice the sensitivity and 45 times the field of view of NIC2\footnote{The gain in FOV with respect to NIC3 is lower (a factor 6.4), but NIC3 severely undersamples the PSF, which makes it less optimal for structural analyses.}, WFC3 is efficient in mapping large areas, opening the long-awaited high-resolution window on the rest-frame optical appearance of large samples of distant galaxies in the act of formation.  Particularly the Early Release Science program (ERS; Windhorst et al. 2011) and the Cosmic Assembly Near-infrared Deep Extragalactic Legacy Survey (CANDELS; Grogin et al. 2011; Koekemoer et al. 2011) offer exciting opportunities.  The first exploitation of these rich data sets have focussed either on monochromatic structural measurements for large, highly complete samples (e.g., Wuyts et al. 2011b; Bell et al. 2011; Kartaltepe et al. 2011; Lotz et al. 2011b; Papovich et al. 2012), or on detailed analyses of the resolved colors within a smaller subset of galaxies (Guo et al. 2011a,b).

Here, we aim to combine the two approaches: studying the resolved colors and stellar populations for a complete sample of 326 (323) star-forming galaxies at $1.5 < z < 2.5$ ($0.5 < z < 1.5$) above $10^{10}\ M_{\sun}$ ($\log (M) > 10$ \& $\log (SFR/M) > -\log(t_H)$).  Exploiting the deep 7-band ACS+WFC3 imaging in GOODS-South, we construct rest-frame light, color and stellar mass maps, and investigate variations in stellar population properties with radius and surface brightness or surface density.  In addition, we present insights on the galaxy-averaged stellar population properties emerging from our resolved analysis, and contrast structural properties measured on light and stellar mass maps.  While our emphasis is on $z \sim 2$ galaxies, we treat galaxies at $z \sim 1$ with identical procedures, allowing us to identify similarities and evolutionary aspects where appropriate.  The paper is structured as follows.  In Section\ \ref{obs_sample.sec}, we present the observations and sample.  The methodology of the resolved SED modeling (similar in spirit to the techniques applied by Zibetti et al. 2009 and Welikala et al. 2008, 2011 at $z \sim 0$ and $z \sim 1$) is detailed in Section\ \ref{methodology.sec}, followed by an inspection of a few case examples in Section\ \ref{examples.sec}.  We present our results on global stellar population properties, resolved color and stellar population profiles, and structural measurements on light and mass maps of the complete sample in Section\ \ref{results.sec}.  Section\ \ref{discussion.sec} discusses the implications for the nature of high-z SFGs, aided by a toy model for their evolution in structure and star formation activity.  Finally, we synthesize our results in Section\ \ref{summary.sec}.

Throughout this paper, we quote magnitudes in the AB system, and
adopt the following cosmological parameters: $(\Omega _M, \Omega
_{\Lambda}, h) = (0.3, 0.7, 0.7)$.

\section{Observations and Sample}
\label{obs_sample.sec}

\subsection{HST imaging in GOODS-South}
\label{HST.sec}
We make use of the 7-band high-resolution data in the GOODS-South field.  The GOODS survey (Giavalisco et al. 2004) collected ACS imaging in the $B_{435}$, $V_{606}$, $i_{775}$, and $z_{850}$ passbands, with typical exposure times per pointing of 7 ks, 5.5 ks, 7 ks, and 18ks respectively (release version 2.0).  Additional $Y_{098}$, $J_{125}$, and $H_{160}$ imaging with WFC3 was carried out as part of the ERS (Windhorst et al. 2011) program, and $Y_{105}$, $J_{125}$ and $H_{160}$ as part of CANDELS-Deep.  ERS exposure times amounted to 2 orbits per pointing per filter, and the first 4 epochs of CANDELS-deep reach a similar depth ($H_{160} < 27$, 5 $\sigma$).  The WFC3 images were drizzled to a $0\farcs 06$ per pixel scale, and we block-averaged the ACS images by a factor 2 to the same platescale.  For details on the layout and observations of the CANDELS multi-cycle treasury program, we refer the reader to Grogin et al. (2011).  Koekemoer et al. (2011) details the CANDELS data reduction.  Processing of the ERS data followed similar steps.  The combined ERS + CANDELS-Deep data set spans an area of 116 arcmin$^2$ with deep multi-band imaging.

\subsection{Sample definition} 
\label{sample.sec}

\begin {figure}[htbp]
\plotone{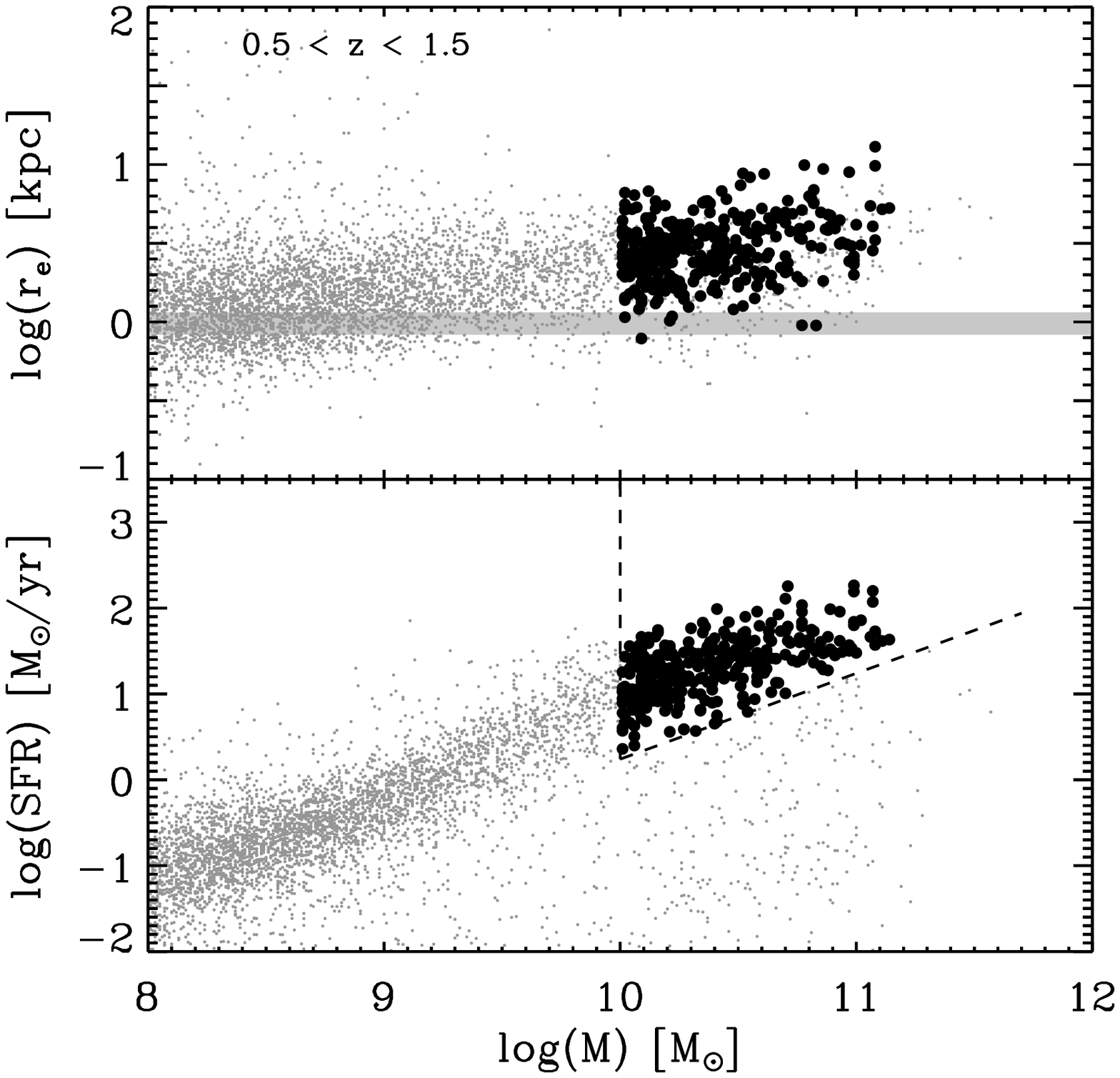}
\plotone{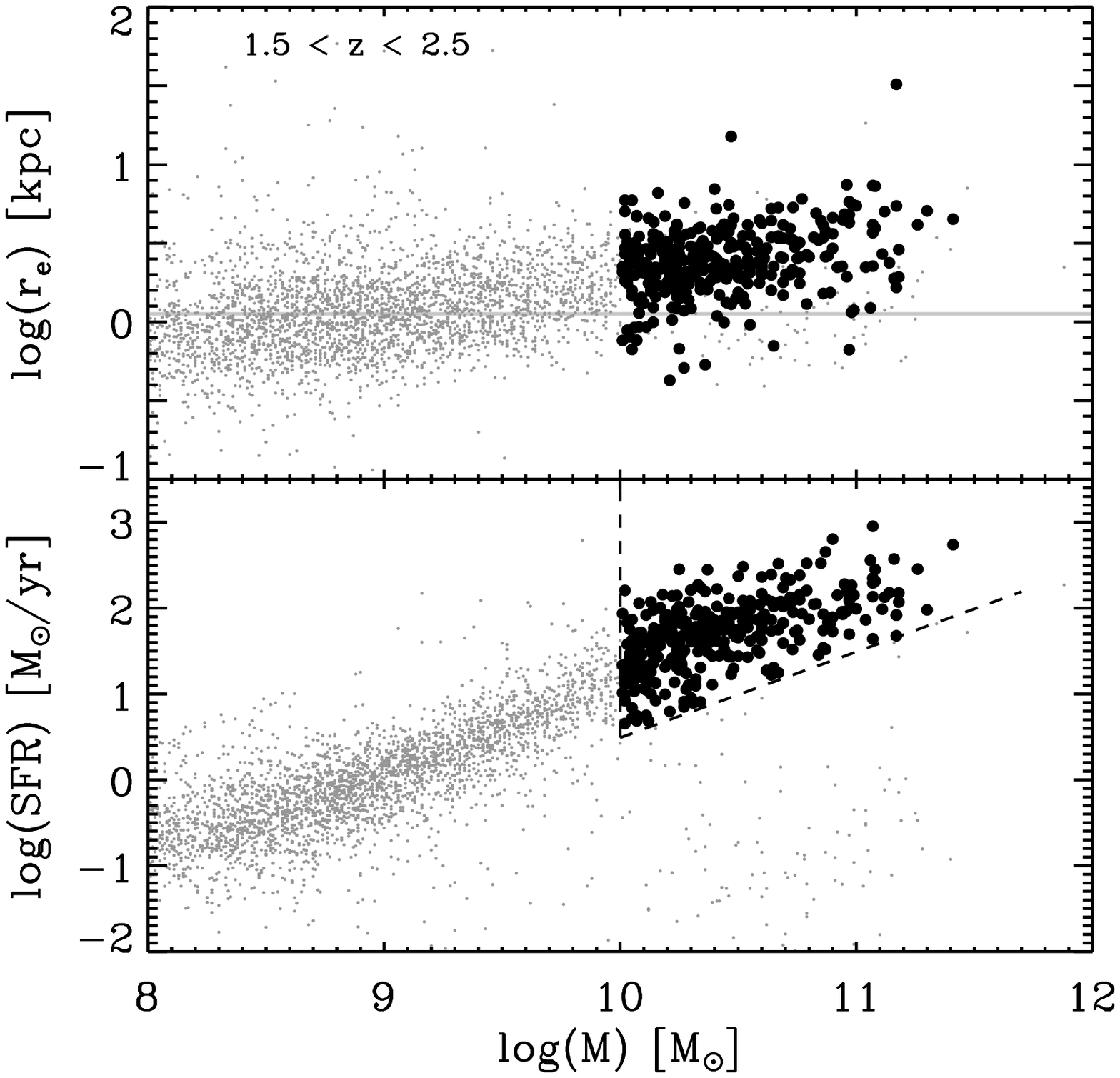}
\caption{
Size-mass and SFR-mass relation of galaxies in our $0.5 < z < 1.5$ and $1.5 < z < 2.5$ samples ({\it large black circles}), contrasted with the underlying galaxy population ({\it small gray dots}) extracted from the same $H_{160}$-selected catalog ($H_{160} < 27$).  Dashed lines indicate the sample selection limits on mass and $SFR/M$.  The horizontal gray bar marks the range of physical sizes to which the half-light radius of the WFC3 $H_{160}$-band PSF corresponds at the respective redshift interval.  A tight MS is observed in SFR-mass space.  By selecting only massive star-forming systems, we naturally limit ourselves to predominantly well-resolved galaxies.
\label{sample.fig}}
\end {figure} 

The parent catalog from which our sample was extracted, is selected in the $H_{160}$ band, and contains consistent photometry in 14 passbands from $U$ to 8 $\mu$m.  For further details on the source extraction and integrated multi-wavelength photometry, we refer the reader to Y. Guo et al. (2012, in preparation).  Using the redshift, stellar mass, and SFR information from Wuyts et al. (2011b), we construct two samples, consisting of 323 SFGs at $0.5<z<1.5$, and 326 SFGs at $1.5<z<2.5$.  In order to have a sufficiently high signal to noise ($S/N$) ratio for resolved SED modeling, and to guarantee a 100\% completeness, we limit our sample to galaxies with masses above $10^{10}\ M_{\sun}$.  Since we aim to study the spatially resolved stellar populations of galaxies during their lifetime prior to quenching, we furthermore apply a cut in specific SFR.  Given that the zeropoint of the MS increases with redshift, a slightly different selection (in absolute terms) is applied for the $z \sim 1$ and $z \sim 2$ sample, corresponding to $\frac{1}{t_H}$ for the middle of each redshift interval (i.e., $\log (SFR/M) > -9.76$ at $z \sim 1$, and $\log (SFR/M) > -9.51$ at $z \sim 2$).  In other words, if all galaxies were forming stars at a constant rate, we exclude those objects that would need more than the Hubble time at the observed epoch to double their mass.  Figure\ \ref{sample.fig} illustrates that this selection effectively removes systems that already formed the bulk of their stars and lie below the MS, although we note that our results do not sensitively depend on the precise value of the $SFR/M$ limit.

We excluded 17 objects that are located in the immediate vicinity of a bright star or foreground galaxy to avoid contamination of the photometry by blended light from the neighbor.  An additional 24 X-ray detected sources were excluded from our analysis because the HST photometry was heavily affected by a non-stellar, blue point source from an active galactic nucleus (AGN).  The typical X-ray luminosity in the rest-frame 2-10 keV band of this subset exceeds $10^{43}\ erg/s$.  We verified that the remaining 62 X-ray detected sources (typically with $L[2-10keV] < 10^{43}\ erg/s$) in our total sample of 649 SFGs at $0.5 < z < 2.5$ follow the same general behavior as the SFGs that lack an X-ray detection in the 4 Ms Chandra catalog (Rangel et al. in prep), and our results would remain unchanged if they were excluded.  We did not impose a selection on inclination, but checked that our general conclusions remain valid for subsets of our sample that were visually classified to be face-on, or edge-on separately.

Also illustrated in Figure\ \ref{sample.fig} is the location of our SFG sample with respect to the overall galaxy population (down to $H_{160} = 27$) in the size versus stellar mass diagram.  The relation between size, mass and SFR has been discussed in depth by Wuyts et al. (2011b, see also references therein), and the sizes plotted here stem from the same database.  Of relevance to the present analysis, is the observation that, by applying a mass and $SFR/M$ selection, we naturally obtain a sample that is dominated by well-resolved objects.  For the sake of completeness, and in order to avoid biases in our analysis due to an additional selection criterion based on the half-light radius measured in a particular observed band, we refrain from removing the few galaxies that are only marginally (or not) resolved in the $H_{160}$ band, even though the amount of resolved information that can be extracted from them is necessarily limited.

\section {Methodology}
\label{methodology.sec}

\subsection{PSF matching}
\label{PSFmatch.sec}

When extracting the physical information from the resolved color maps, it is of utmost importance that the colors are measured over the same physical region.  We therefore smeared the ACS images to WFC3 resolution, gaining consistency at the expense of the superior optical resolution.  Kernels to match the ACS F435W, F606W, F775W, F850LP, and WFC3 F098m, F105W, F125W PSFs to WFC3's broader F160W PSF were derived by deconvolving the latter with the former using the PSFMATCH algorithm in IRAF.  We applied a cosine bell tapering of the highest frequencies to avoid noise in the input PSFs to induce low-level artifacts in the resulting kernel.  This procedure was carried out separately for ERS and CANDELS-Deep, since these regions were observed with different roll angles and number of visits.  As a sanity check, we compared the profiles and growth curves of the convolved $B_{435},V_{606},i_{775},z_{850},Y_{098},Y_{105},J_{125}$ PSFs with those of the $H_{160}$ PSF.  The ratios of the growth curves deviate by less than 1\% from unity.  The resolution of the PSF-matched images is $0\farcs18$ full-width half-maximum (FWHM).

\subsection{Pixel binning}
\label{binning.sec}

In order to assign pixels to objects, we use the SExtractor (Bertin \& Arnouts 1996) segmentation map that was produced for the $H_{160}$ detection band of our parent catalog.  Rather than treating the multi-wavelength photometry in each pixel individually, we group pixels following the Voronoi 2D binning technique by Cappellari \& Coppin (2003),  so as to achieve a minimum signal-to-noise level of 10 per bin in the $H_{160}$ band.  The pixel binning forms a crucial step, as significant biases in SED modeling results may arise when fitting stellar population models to SEDs of individual pixels that have marginal or no significance over most or even the entire wavelength range (see also Section\ \ref{integrated.sec}).

\subsection {Resolved SED modeling}
\label{resolved.sec}

We compute rest-frame $2800\AA_{rest}$, $U_{rest}$, and $V_{rest}$ photometry for all of the spatial bins in each of the objects using EAZY (Brammer et al. 2008).  To this end, we keep the redshift fixed at the photometric redshift that was derived from the integrated $U$-to-8 $\mu$m photometry of the galaxy (or the spectroscopic redshift when available).  Briefly, each rest-frame band is interpolated from the observed bands straddling the rest-frame wavelength of interest, using the EAZY templates as a guide (i.e., this approach is only weakly model dependent).

A more heavy reliance on models enters when we want to translate the surface brightness levels in different bands to resolved stellar population properties such as stellar mass, age, or dust extinction.  Degeneracies in stellar population modeling, such as between age and dust, have long been known from analyses of integrated photometry, where sampling of the SED over as wide a wavelength range as possible has proven crucial in improving the constraints on these parameters (see, e.g., Wuyts et al. 2007).  In this paper, we adopt a novel, two-step approach to exploit the resolved color information provided by HST, while still accounting for the available integrated constraints at shorter and longer wavelengths.  In other words, to make full use of both the resolved and integrated photometry, we couple them together in a global function that is then to be minimized.

First, we perform standard stellar population modeling of the $BVizYJH$ SED of each spatial bin individually.  That is, we fit Bruzual \& Charlot (2003, hereafter BC03) models to the 7-band $B$-to-$H$ SED and search for the least-squares solution using the fitting code FAST (Kriek et al. 2009).  We apply similar constraints as in previous work on integrated photometry: allowing ages since the onset of star formation between 50 Myr and the age of the universe, and visual extinctions in the range $0 < A_V < 4$ with the reddening following a Calzetti et al. (2000) law.  We adopt a Chabrier (2003) IMF, assume a uniform solar metallicity, and allow exponentially declining SFHs with efolding times down to $\tau = 300$ Myr (see Wuyts et al. 2011a).  Where relevant, we comment on how alternative assumptions regarding the SFH (minimum age, minimum $\tau$, 2-component SFHs, and delayed $\tau$ models that allow for a phase of increasing SFRs) affect our conclusions.  The chi-square function to be minimized in this first, entirely resolved, step is:

\begin{equation}
\chi^2_{res} = \sum_{i=1}^{N_{bin}} \sum_{j=1}^{N_{res\ bands}} \frac{ (F_{i,j} - M_{i,j})^2 } {(E_{i,j})^2}
\label{chires.eq}
\end{equation}

where $N_{bin}$ is the number of spatial bins, and $N_{res\ bands}$ is the number of passbands observed at high resolution.  The observed flux and flux error in a spatial bin and band are $F_{i,j}$ and $E_{i,j}$ respectively, and $M_{i,j}$ is the corresponding stellar population model, which has mass, age, $\tau$, and $A_V$ as free parameters.  The FAST algorithm calculates the best-fitting normalization (i.e., mass) for each discrete cell in (age, $\tau$, $A_V$) space, and selects the cell for which the fit gives the lowest $\chi^2$ value.  If we want to simultaneously match the resolved and integrated photometric constraints, the function to be minimized is:

\begin{equation}
\chi^2_{res+int} = \chi^2_{res} + \sum_{j=1}^{N_{int\ bands}} \frac { (F_j - \sum_{i=1}^{N_{bin}} M_{i,j})^2 } {(E_j)^2 }
\label{chiresint.eq}
\end {equation}

where $N_{int\ bands}$ represents the number of passbands for which only integrated photometry is available (in the case of GOODS-South, this is $U$, $K_s$, and the 4 IRAC bands).  Since the stellar population models of all spatial bins are linked in $\chi^2_{res+int}$, it is no longer computationally feasible to compute a $\chi^2$ value for each cell in this multi-dimensional space (which has grown by a power of the number of spatial bins).  We therefore follow an iterative approach, in which we adopt the stellar population models of all spatial bins obtained in step one as an initial guess.  From there, we consider the improvement in $\chi^2_{res+int}$ when adjusting the model parameters for one of the spatial bins.  We select the spatial bin for which a maximum improvement could be reached, but only adjust its (age, $\tau$, $A_V$) so as to achieve the nearest (i.e., slightly lower) $\chi^2_{res+int}$ value.  The latter approach was found to be most effective in avoiding large corrections to the model of one bin that serve to compensate deviations from the integrated constraints that are due to imperfect stellar population models of multiple bins.  We repeat this walk through the multi-dimensional parameter space until no improvement in $\chi^2_{res+int}$ is found, or a maximum of 500 iterations is reached.  The adopted maximum number of iterations is sufficiently large for the iteratively adjusted $\chi^2_{res+int}$ to asymptote, and the sum of all bin models yields overall satisfactory fits to the $U$-to-8 $\mu$m integrated photometry.  We present tests of the resolved SED modeling procedure in Appendix A.

\begin {figure*}[htbp]
\centering
\plotone{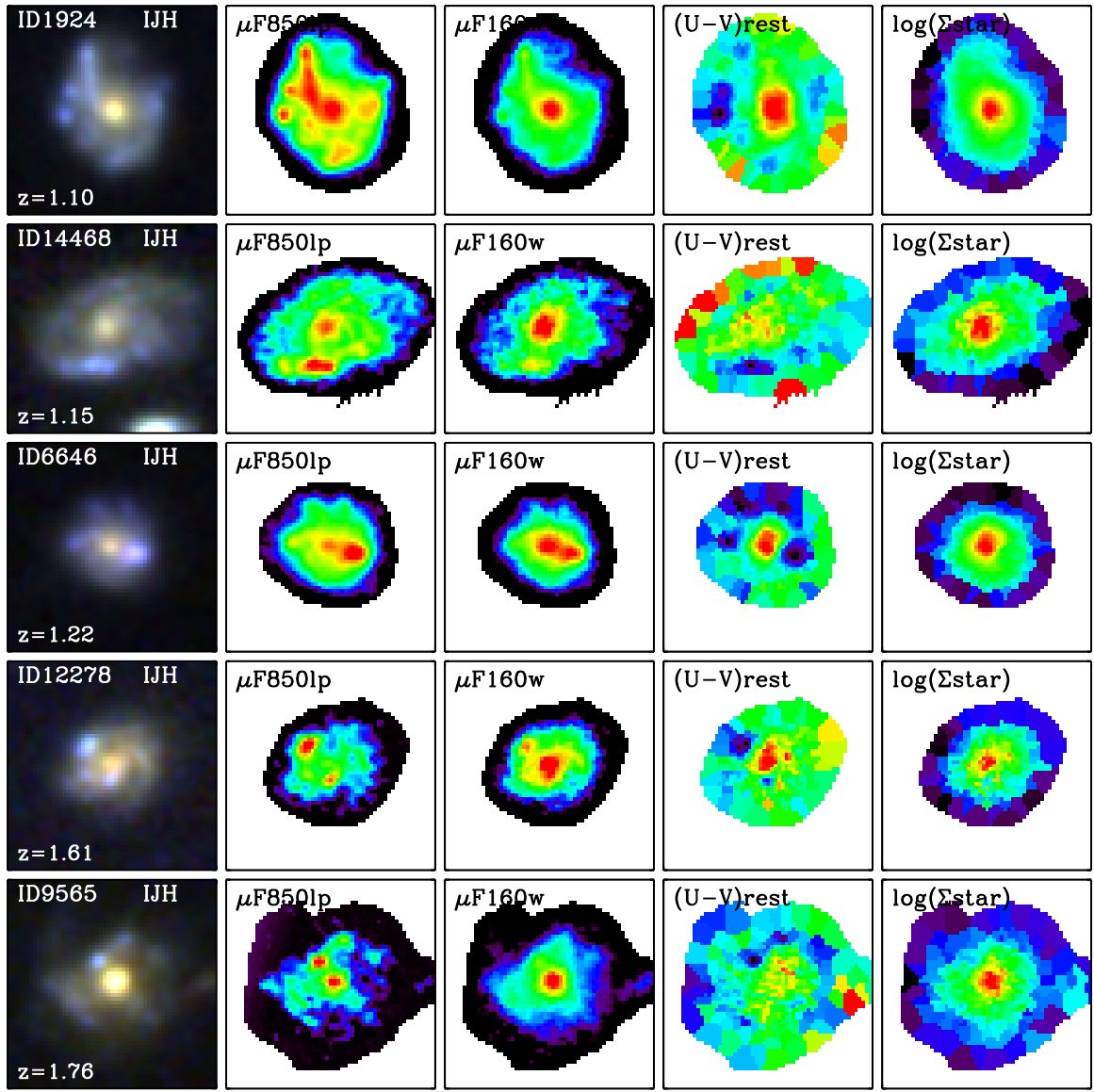}
\caption{Case examples of galaxies at $z \approx 1 - 2$ exhibiting off-center peaks in the surface brightness distribution.  From left to right (f.l.t.r.): observed $I_{775}J_{125}H_{160}$ 3-color postage stamps sized $3\farcs 4\ \times 3\farcs 4$, surface brightness distributions in the observed $z_{850}$ and $H_{160}$ bands, rest-frame $U-V$ color maps, and the distribution of stellar surface mass density.  The off-center regions with elevated surface brightness tend to be blue, and therefore less pronounced (but still present) in $H_{160}$ compared to $z_{850}$.  With a few notable exceptions (ID1683 and ID12328), the stellar mass maps are centrally concentrated, and lack regions with elevated surface mass density at large radii.
\label{examples.fig}}
\end {figure*} 

\subsection {Structural parameters}
\label{structure.sec}

We derive the following non-parametric structural measures: $R_e$, Concentration, Gini, and M20.  In all estimators but the Gini coefficient, a center of the galaxy needs to be defined.  Throughout the paper, we consistently adopt the stellar mass-weighted center, irrespective of the map on which the structural measurement is carried out (i.e., also in determining the semi-major axis half-light radius $R_{e,\ U}$ of the $U_{rest}$ distribution).  The non-parametric half-light radius $R_e$ is defined as follows.  We construct the curve of growth from elliptical apertures centered on the mass-weighted center.  To this end, we measured the position angle of each object pixel to the center, and adopted the mode of the distribution as the position angle of the elliptical apertures.  The axial ratio was derived from the histogram of distances from each object pixel to the center, specifically by taking the ratio of the mode to the extent of the distribution.  Our motivation to define the ellipse parameters based on the segmentation map rather than a surface brightness map, is to be driven by the outer isophotes.  This way, we avoid that individual bright clumps superposed on an otherwise smooth surface brightness distribution affect the aperture shape, which may happen in a light-weighted approach.  The semi-major axis length at which the curve of growth reaches 50\% of its maximum value is defined as $R_e$.  Throughout the paper, quoted half-light and half-mass radii are understood to be measured along the major axis.  We stress that our size definition is different from the effective radius $r_e$ plotted in Figure\ \ref{sample.fig} in that no functional form (e.g., Sersic) is assumed for the profile, but more importantly in that the measurement is carried out at the observed resolution, without accounting for smearing by the PSF.  We argue that this approach is sufficient for our present goal, since we only aim to compare the structural parameters measured on light and mass maps that all have identical resolution.

Concentration is defined as $R_{80} / R_{20}$, with $R_{80}$ and $R_{20}$ being the radii of circles enclosing 80\% and 20\% of the total light (or mass) respectively.  The Gini coefficient (see, e.g., Abraham et al. 2003; Lotz et al. 2004; F\"{o}rster Schreiber et al. 2011a) acts as an alternative measurement of how concentrated a pixel distribution is, without the need to define a center.  Finally, we consider the M20 parameter, first introduced by Lotz et al. (2004), as an indicator sensitive to the presence of multiple nuclei or clumps in the distribution.

In all cases, we measured the above parameters on pixelized images where the light (or mass) of the Voronoi bins was distributed uniformly over its constituent pixels.  Again, since the same binning was used to construct rest-frame light and stellar mass maps, this does not affect the relative comparison.

\setcounter{figure}{1}
\begin {figure*}[htbp]
\centering
\plotone{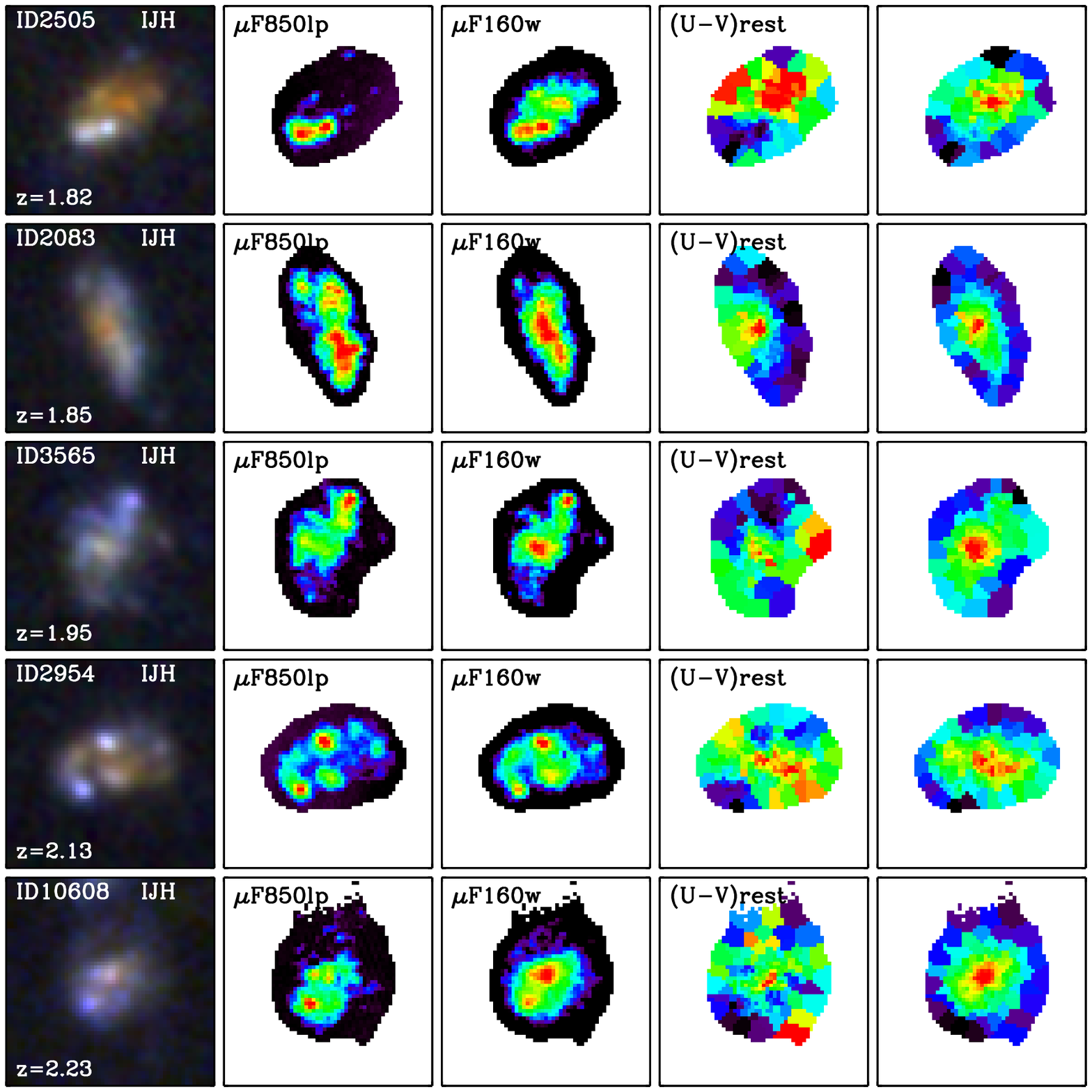}
\plotone{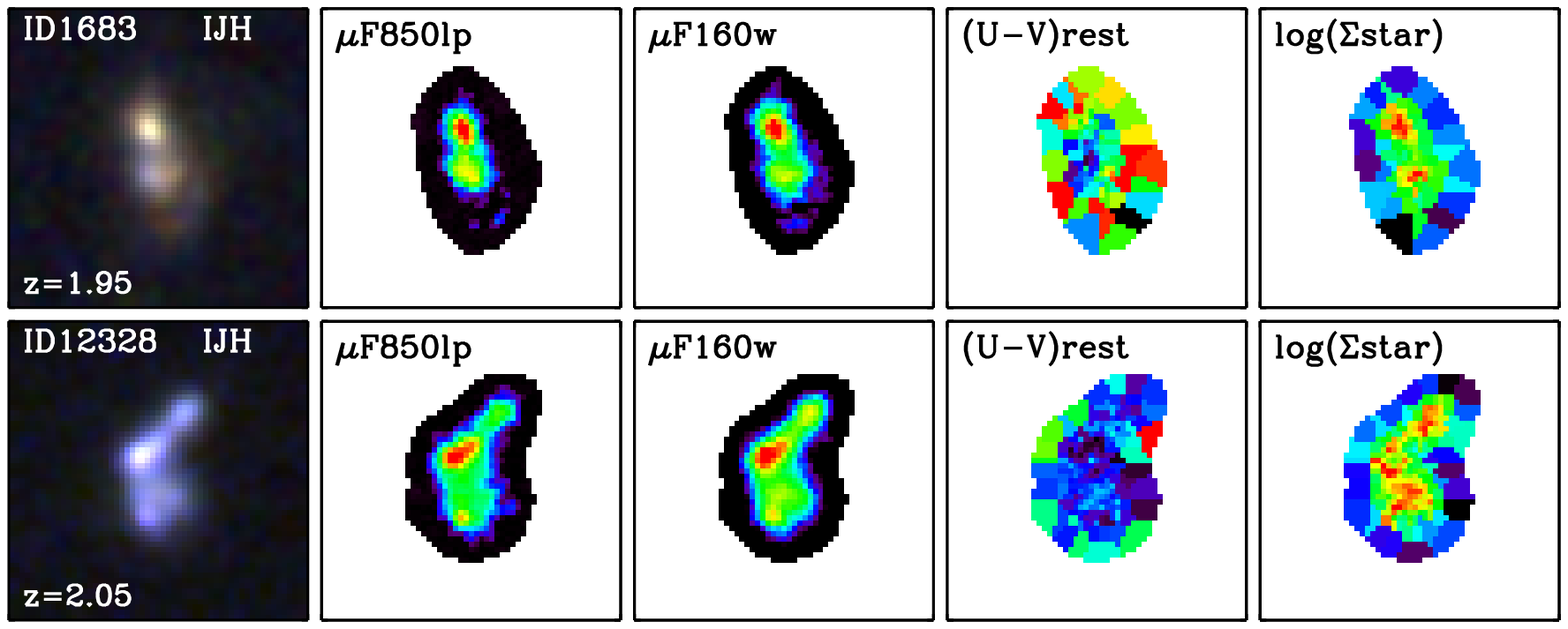}
\caption{
continued
\label{examples.fig}}
\end {figure*}

\section {Case examples}
\label{examples.sec}

With the PSF-matched images and the tools for resolved SED modeling in hand, we now turn to an object by object inspection of a set of case examples in Figure\ \ref{examples.fig}.  The objects span a redshift range from $z = 1.10$ to $z = 2.23$.  They were not drawn randomly from our SFG sample, but instead visually selected for the appearance of strong, clump-like features in the PSF-matched $I_{775}J_{125}H_{160}$ 3-color postage stamps.  However, as we will demonstrate in Section\ \ref{profres.sec}, the key characteristic that these single object images reveal is reflected in the co-added profiles of the complete SFG population as well.

Next to the 3-color postage stamp of each object, Figure\ \ref{examples.fig} shows the surface brightness maps in the observed $z_{850}$ and $H_{160}$ bands.  At $z \sim 2$, these bands correspond roughly to rest-frame 2800\AA\ and rest-frame $V$-band wavelengths.  Here, we only display those pixels that SExtractor assigned to the segmentation map of the respective object.  In each case, the $z_{850}$ map features peaks in the surface brightness distribution located away from the center.  In addition, an elevated central surface brightness with respect to the outer disk is notable in several, but not all, galaxies.  Generally, this nucleus is not brighter in $z_{850}$ than the off-center clump(s).  Inspecting the counterparts of these maps in the $H_{160}$ band, we note that in the majority of the cases the off-center clumps seen in $z_{850}$ remain present, albeit less pronounced, while no new off-center peaks emerge in these examples.  In contrast to the off-center clumps, the nucleus of the galaxy manifests itself more strongly at longer wavelengths.

A reflection of this general trend in the rest frame, is illustrated in the fourth column of Figure\ \ref{examples.fig}: the rest-frame $(U-V)_{rest}$ color maps typically feature bluer colors at the location of clumps that dominate the $z_{850}$ appearance.  This trend is reminiscent of observations by F\"{o}rster-Schreiber et al. (2011b), who found radial trends in color and H$\alpha$ equivalent widths in two $z \sim 2$ SFGs.  Similar trends were also found by Guo et al. (2011b) based on a set of 10 $z \sim 2$ SFGs in the Hubble Ultra Deep Field.  While off-center clumps tend to be blue, the reddest colors are frequently seen in the center.  Since the $(U-V)_{rest}$ color is tightly correlated with the stellar $M/L$ ratio (Marchesini et al. 2007), it comes as no surprise that the stellar mass maps derived from resolved SED modeling show a prominent nucleus, but lack the off-center features that were characteristic for the light profiles (particularly at short wavelengths).  Two rare exceptions to the trend (ID1683 and ID12328) are shown at the bottom of Figure\ \ref{examples.fig}.  Here, the absence of strong spatial variations with color leads to an approximately uniform translation from light to mass.  Consequently, the multiple clumps seen in the surface brightness profiles remain present in the stellar mass maps.  We mention in passing that object ID12328 serves as the prototype of a merging system in the CANDELS visual classification effort (e.g., Kocevski et al. 2012; Kartaltepe et al. 2012).

While a statistical analysis of the complete SFG sample follows in Section\ \ref{profres.sec}, the handful of examples discussed here already reveal the main message of this paper: while the exquisite WFC3 imaging provides us with a better proxy of stellar mass than was available thus far at high resolution from ACS, the $H_{160}$ profiles should not be equated to stellar mass profiles.  Generally, the distribution of stellar mass is more centrally concentrated, and (at least at the resolution of WFC3) does not exhibit as evidently a complex of multiple clumps.

\section {Results}
\label{results.sec}

\subsection {Integrated stellar population properties}
\label{integrated.sec}

\begin{figure}[t]
\centering
\includegraphics[width=0.49\columnwidth]{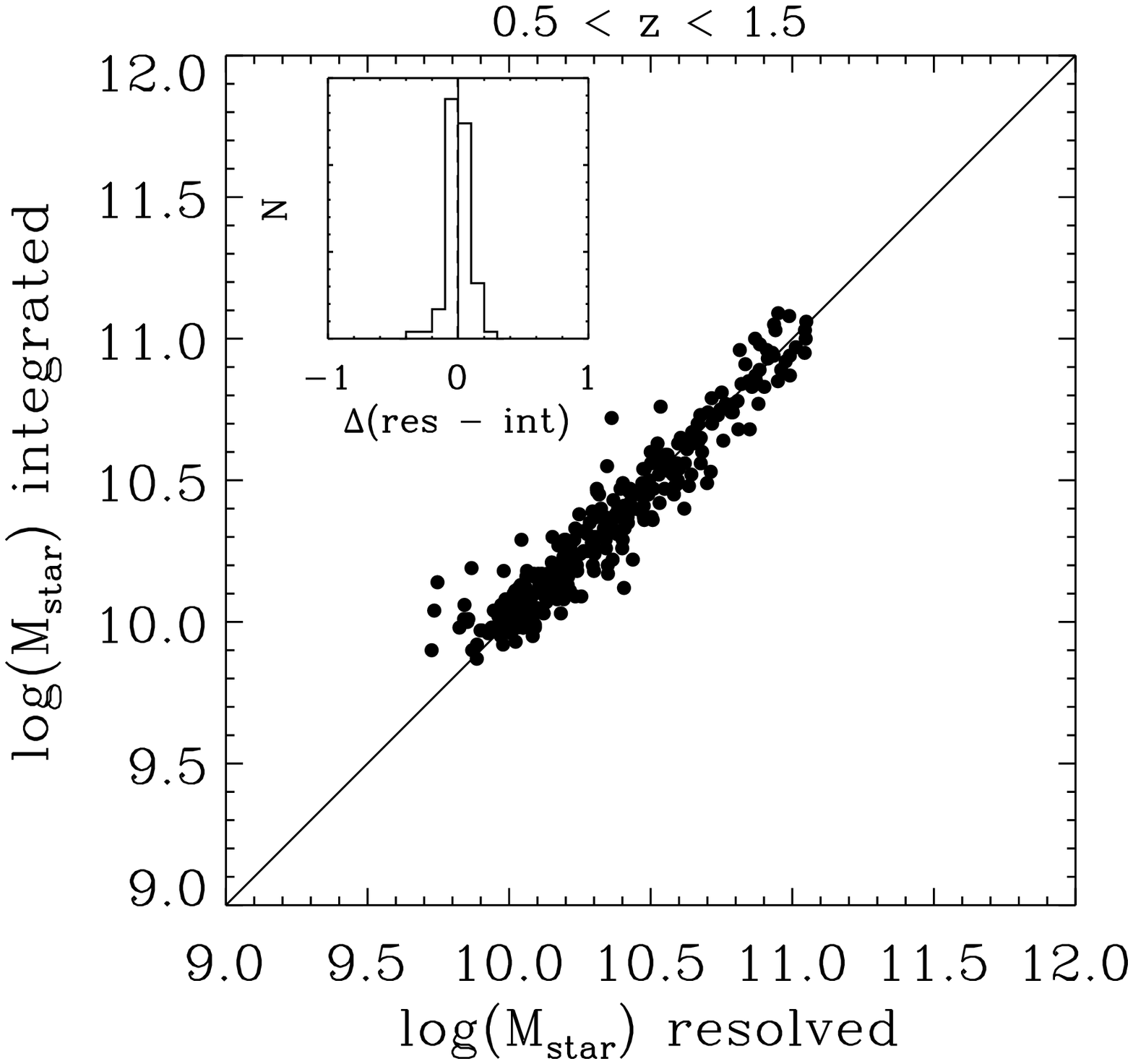}
\includegraphics[width=0.49\columnwidth]{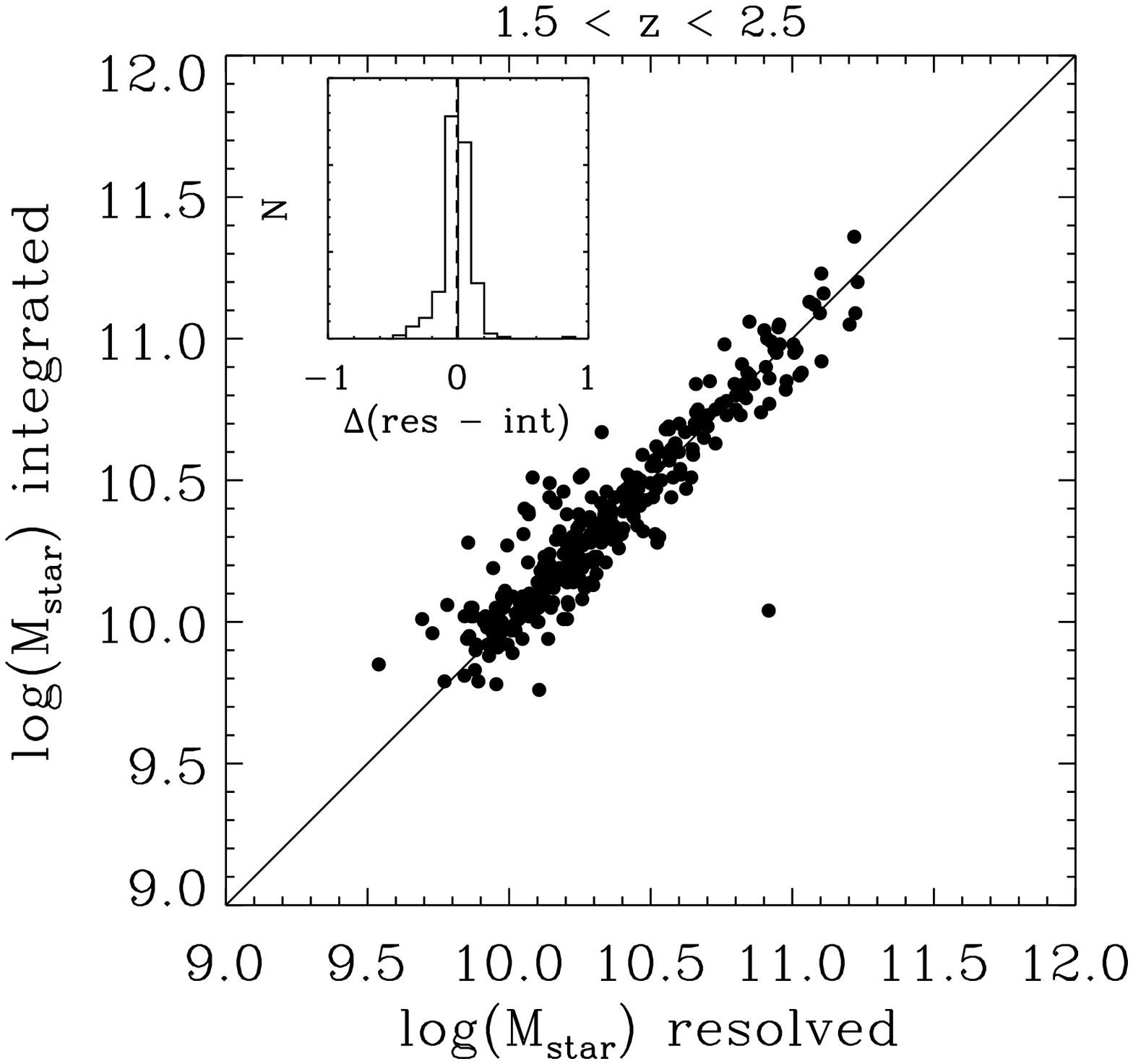}
\includegraphics[width=0.49\columnwidth]{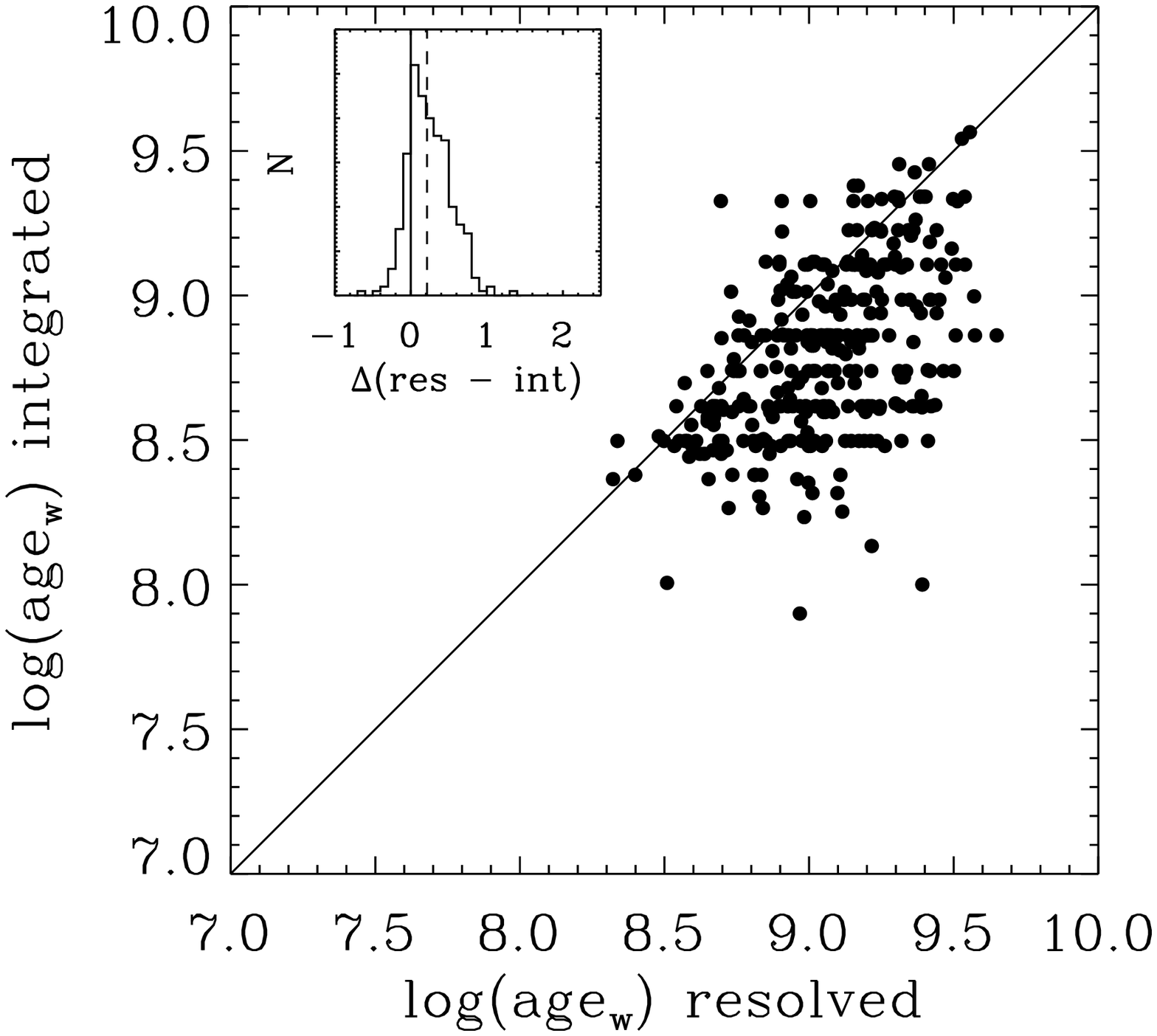}
\includegraphics[width=0.49\columnwidth]{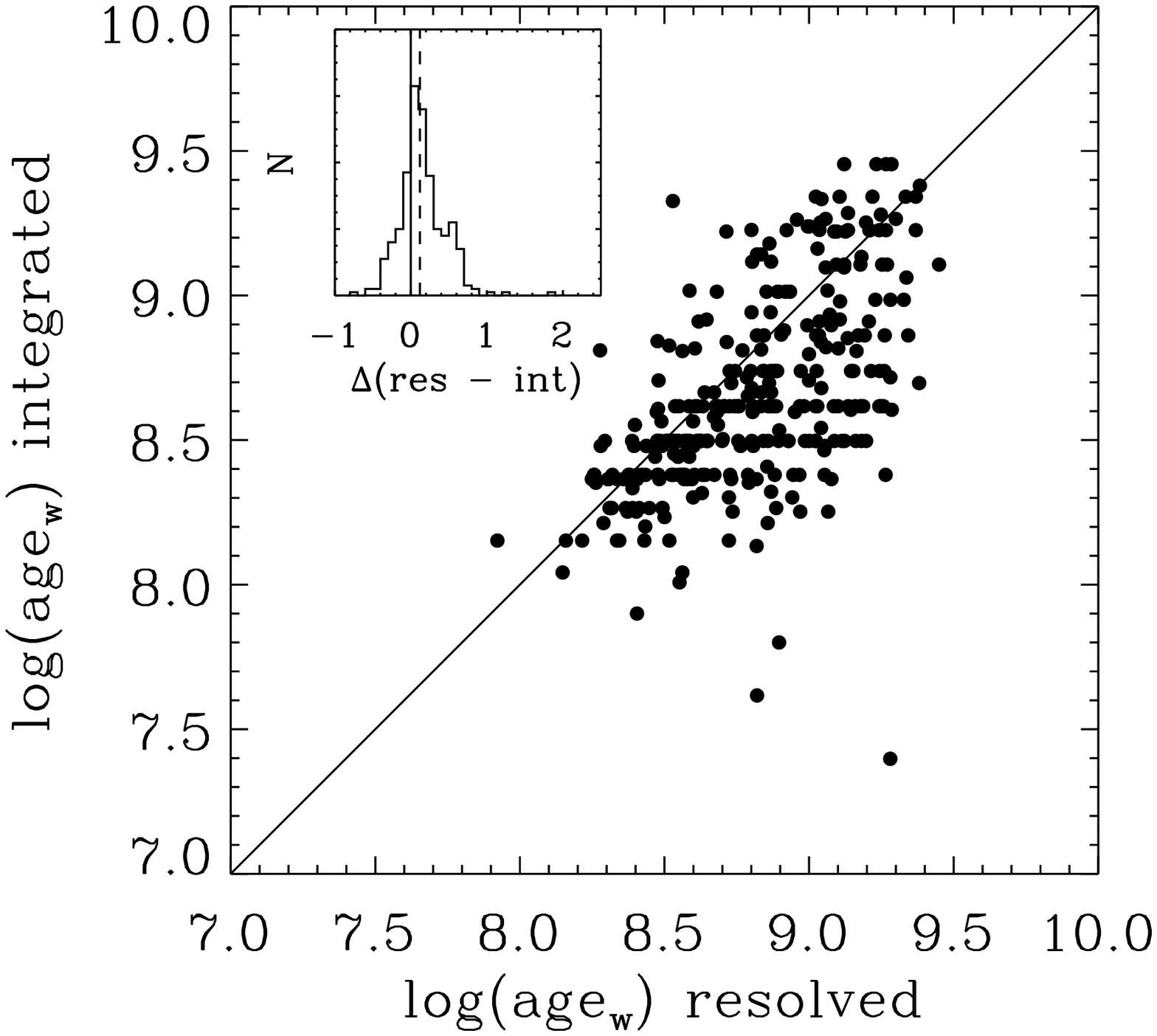}
\includegraphics[width=0.49\columnwidth]{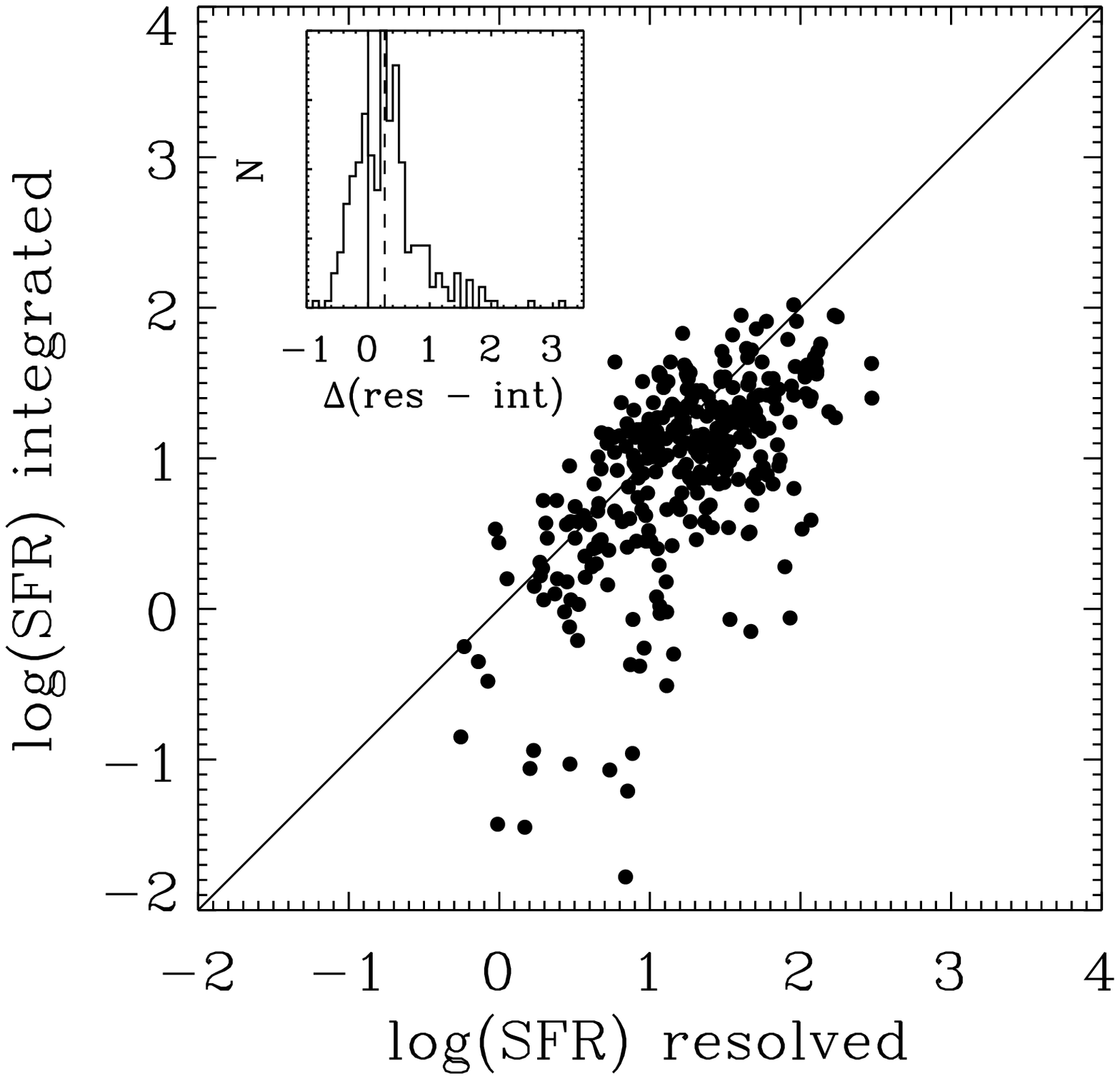}
\includegraphics[width=0.49\columnwidth]{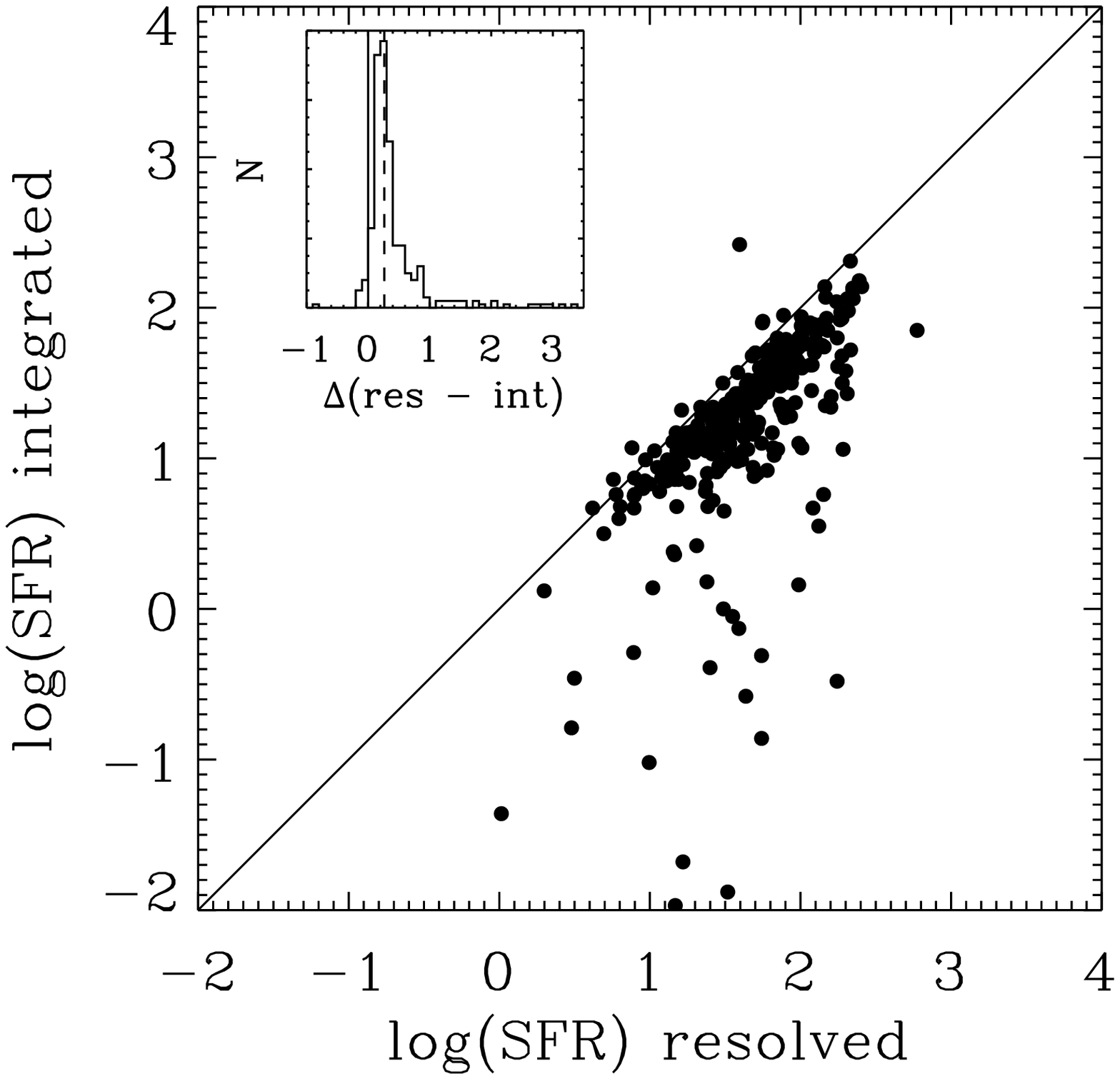}
\includegraphics[width=0.49\columnwidth]{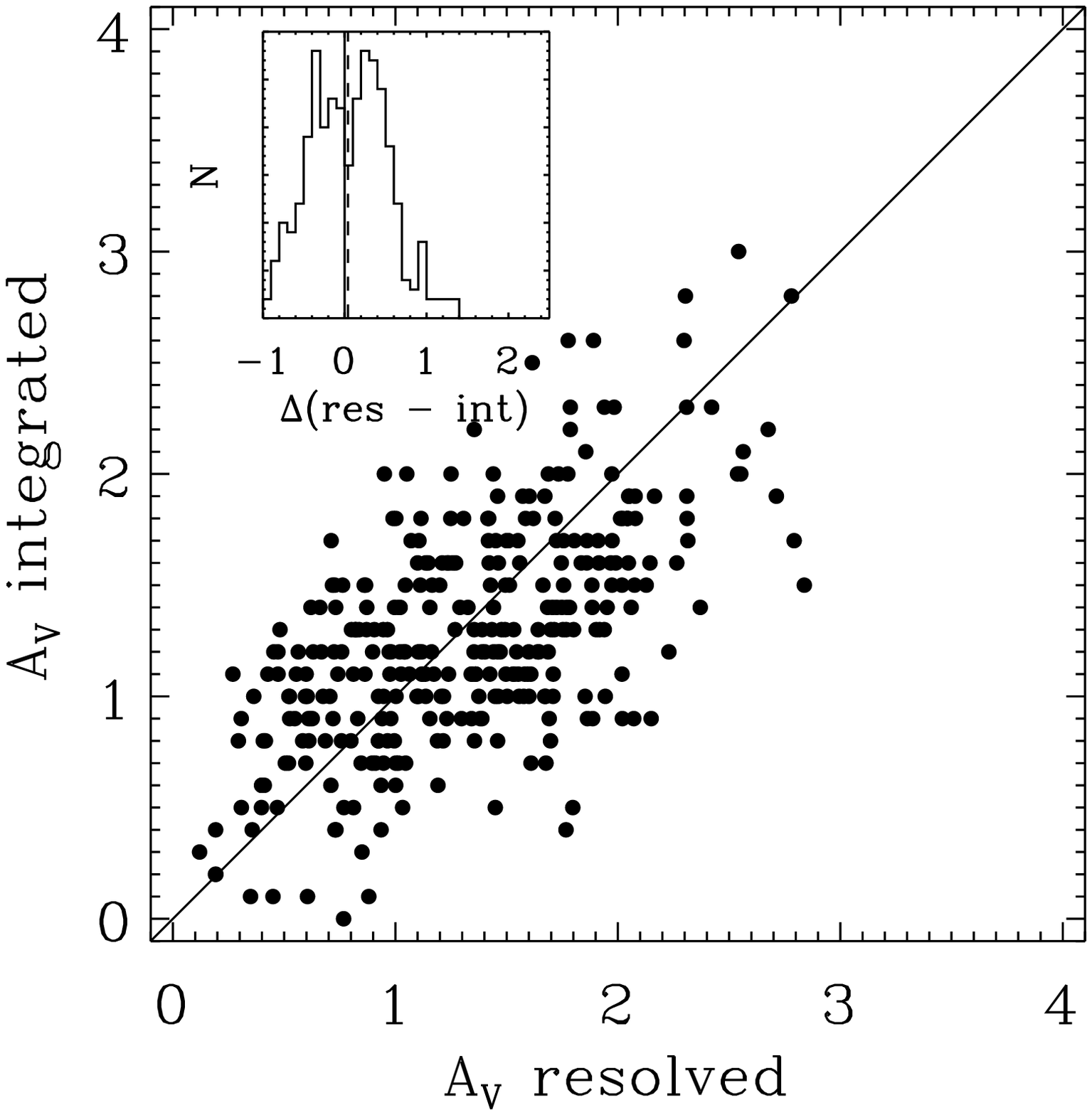}
\includegraphics[width=0.49\columnwidth]{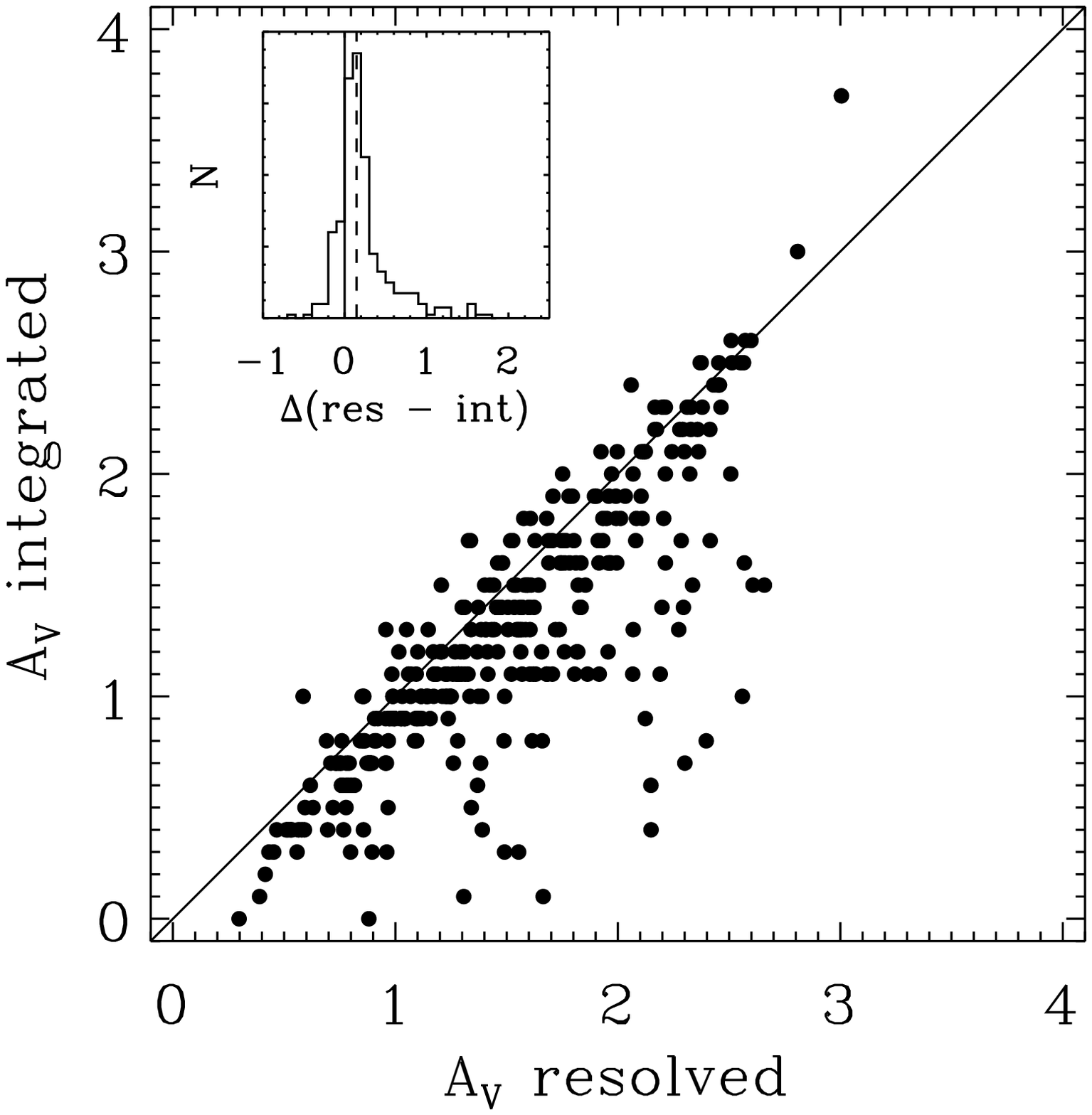}
\caption{
Comparison of global stellar population properties inferred from modeling the integrated $U$-to-8 $\mu$m photometry versus modeling the resolved HST photometry taking into account the constraints from integrated $U$, $K_s$, and IRAC photometry.  Results for our sample at $0.5 < z < 1.5$ are presented in the left-hand panels, those for the $1.5 < z < 2.5$ sample in the right-hand panels.  While the stellar mass can be quantified robustly, severe underestimates of the SFR-weighted age can occur when no information on internal $M/L$ ratio variations is resolved.
\label{int_vs_res.fig}}
\end{figure}

Before delving deeper into the resolved stellar populations, we ask ourselves the question whether one can benefit from the analysis of resolved colors to understand integrated galaxy properties.  After all, the methodology detailed in Section\ \ref{resolved.sec} uses all the information that is typically included in SED modeling of the integrated galaxy light, and more.  In cases where dust lanes or gradients and/or spatial variations in stellar age are present, the more extincted or older subregions may be downweighted in the integrated light, while their presence can, at least to the extent they are resolved at the WFC3 resolution, be recovered when using the resolved color information.

We address this question by contrasting the stellar population properties of galaxies as a whole as derived from SED modeling of the integrated light versus reconstructed from resolved SED modeling (see Figure\ \ref{int_vs_res.fig}).  Throughout this paper, we adopted identical assumptions regarding attenuation law, SFH, etc for the resolved modeling of each spatial bin as for the integrated SED modeling.  Moreover, in order to obtain a fair comparison, we rescaled the integrated photometry for each galaxy by a uniform factor so as to normalize the total $H_{160}$-band flux to the sum of the $H_{160}$-band fluxes of pixels belonging to the object's segmentation map.  The latter scaling was implemented to avoid any spurious trends due to aperture corrections in the multi-wavelength catalog containing the integrated photometry.  The corrections are typically of order 10\%.

The first conclusion to draw from Figure\ \ref{int_vs_res.fig} is that the estimates of the total stellar mass derived from only integrated photometry are remarkably robust when compared to those obtained from resolved SED modeling.  The median offset in $\log(M)$ is 0.00 dex, with a scatter (normalized median absolute deviation) of 0.08 dex.  The good correspondence also remains if we were to ignore the constraints from integrated $U$, $K_s$, and IRAC photometry in the resolved approach (i.e., minimize $\chi^2_{res}$ from Equation\ \ref{chires.eq} instead of $\chi^2_{res+int}$ from Equation\ \ref{chiresint.eq}.  A significantly larger systematic offset (by 0.2 dex for $z \sim 2$ SFGs) would be obtained if no pixel binning to a minimum $S/N$ level were applied.  The reason is that in such a scenario there are many individual pixels in the outskirts that have essentially no color constraints, and to which unreasonably large $M/L$ ratios can be assigned.  This will then lead to erroneously large estimates of their contribution to the stellar mass when applying resolved SED modeling.  Zibetti et al. (2009) carried out a detailed analysis of 9 nearby galaxies (D $<$ 26 Mpc) that is similar in spirit to ours.  These authors find that stellar masses derived from resolved color information exceed the masses inferred from integrated photometry by 0 to 0.2 dex.  We note that in this study, the physical scales probed are an order of magnitude smaller than attainable with WFC3 for distant galaxies.  We caution therefore that color and stellar population variations on spatial scales below the resolution limit may still affect our estimates of the total stellar mass.

More dramatic differences are revealed when considering estimates of the SFR-weighted stellar age, which quantifies the age of the bulk of the stars, and equals the mass-weighted age modulo stellar mass loss (see Equation 4 in Wuyts et al. 2011a).  In the resolved case, the SFR-weighted age of a galaxy is computed equivalently:
\begin {equation}
age_{w,\ resolved} = \frac {\sum_{i=1}^{N_{pix}} \int_0^{t_{obs}} \! SFR_i(t) (t_{obs} - t) \, \mathrm{d}t } 
                                               {\sum_{i=1}^{N_{pix}} \int_0^{t_{obs}} \! SFR_i(t) \, \mathrm{d}t }
\end {equation}
While for galaxies that from integrated photometry are inferred to be the oldest, the age estimate does not depend sensitively on the adopted approach, the recovered age of the younger systems (or at least those inferred to be the youngest from their integrated photometry) can increase by up to an order of magnitude in the most extreme cases when taking into account the resolved photometric constraints.  Outshining of the underlying older population by regions with young, newly formed stars, is likely responsible for the observed offsets (see also Maraston et al. 2010, Wuyts et al. 2011a).  

The median offset in inferred SFRs for the entire SFG sample is 0.27 dex.  However, the distribution shows a tail towards larger underestimates of the SED modeled SFR when no resolved photometric information is used, exceeding an order of magnitude.  Welikala et al. (2011) find a qualitatively similar tail towards high $\log (SFR_{resolved} / SFR_{integrated})$ for a sample of optically selected SFGs at $z \sim 1$.  While we adopt default $\tau$ models for the SED modeling of each spatial bin, the resulting galaxy-integrated SFH does not necessarily follow this simple functional form (see also Abraham et al. 1999; F\"{o}rster Schreiber et al. 2003).  Appendix B illustrates that, for the SFGs in our sample, resolved SED modeling reveals a typical effective SFH in which stars initially formed at a slow pace, and over time the stellar mass build-up proceeded at an increased rate.  Several recent studies arrived to similar conclusions following independent lines of evidence (e.g., Renzini 2009; Maraston et al. 2010; Lee et al. 2011; Papovich et al. 2011).

Finally, we find the inferred visual extinction to be reasonably well behaved in the median (0.13 mag more attenuation given by the resolved approach), but with significant outliers on an individual object basis.  Here, the effective visual extinction in the resolved case is derived from the ratio of the total attenuated and inferred intrinsic $V_{rest}$ luminosity:
\begin {eqnarray}
A_{V,\ resolved} = -2.5 \log \left ( \sum_{i=1}^{N_{pix}} L_{V,\ attenuated,\ i} \right ) \nonumber \\
+2.5 \log \left ( \sum_{i=1}^{N_{pix}} L_{V,\ intrinsic,\ i} \right )
\end {eqnarray}
The deviations between integrated and resolved SED modeling are correlated for different stellar population parameters, with $\Delta \log(M)$ following a positive correlation with $\Delta \log (age_w)$, and $\Delta \log (SFR)$ and $\Delta A_V$ being anticorrelated to $\Delta \log (age_w)$.  The trends described in this Section are seen in both the $z \sim 1$ and $z \sim 2$ SFG population.  Any differences between the two redshift intervals may stem from intrinsic differences between the stellar populations and/or their spatial distribution, but could potentially also be affected by observational aspects (e.g., at lower redshift the $B_{435}$-to-$H_{160}$ wavelength range with resolved information spans a bluer part of the rest-frame SED).  Given the overall consistency between the comparisons at $z \sim 1$ and $z \sim 2$, we defer a more in depth analysis of these potential second order effects to future work, and proceed with the core objective of this paper: dissecting the resolved stellar populations of SFGs out to $z \sim 2.5$.

\begin{figure}[htbp]
\centering
\plotone{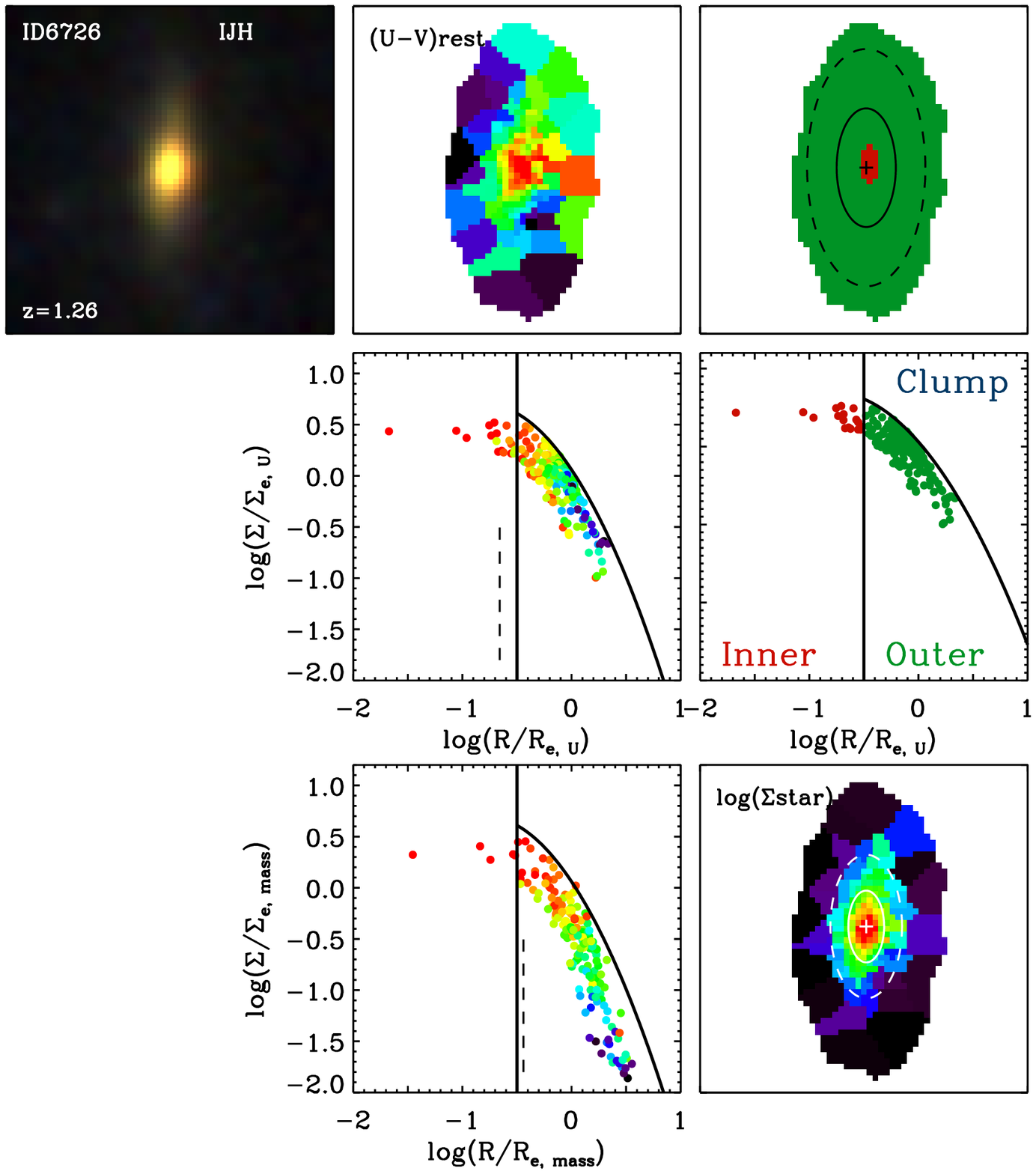}
\plotone{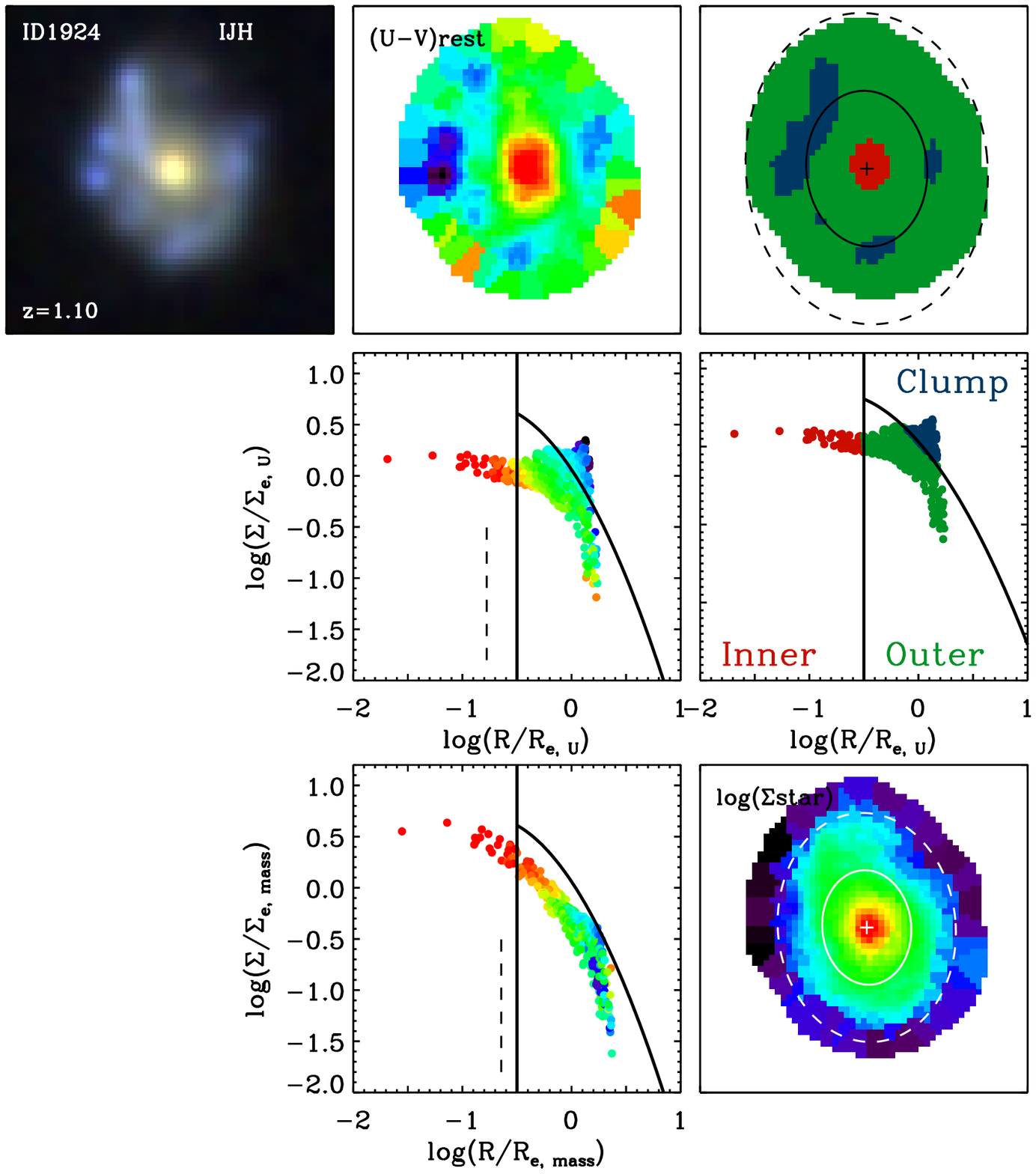}
\caption{
Examples of 2D maps and 1D normalized profiles of $z \approx 1-2$ SFGs without (ID6726) and with (ID1924, ID2954) clumpy features in their rest-frame U-band surface brightness distribution.  For each object, the top row shows the observed-frame $IJH$ postage stamp, the $(U-V)_{rest}$ color map, and a map indicating whether pixels are assigned to the center, outer disk, or clump regime.  Solid black ellipses contain half the rest-frame $U$-band light.  Dashed ellipses have twice the major axis length.  The middle row shows the normalized rest-frame $U$ light profiles, color coded by $(U-V)_{rest}$ and pixel type (inner / outer / clump) respectively.  Solid lines separate the center, disk, and clump sectors.  The vertical dashed line marks the WFC3 resolution.  Normalized stellar mass profiles color coded by $(U-V)_{rest}$, and the corresponding 2D stellar mass maps are presented in the bottom row for each object.  We overplot solid and dashed white ellipses with semi-major axis lengths of $R_{e,\ mass}$ and $2 R_{e,\ mass}$ respectively.  All three examples shown here feature negative color gradients and a more compact stellar mass than light distribution.  In the case of ID1924 and ID2954, off-center regions with elevated surface brightness ('clumps') are identified.  They exhibit bluer colors than the underlying disk.
\label{prof_explanation.fig}
}
\end{figure}

\setcounter{figure}{3}
\begin{figure}[t]
\centering
\plotone{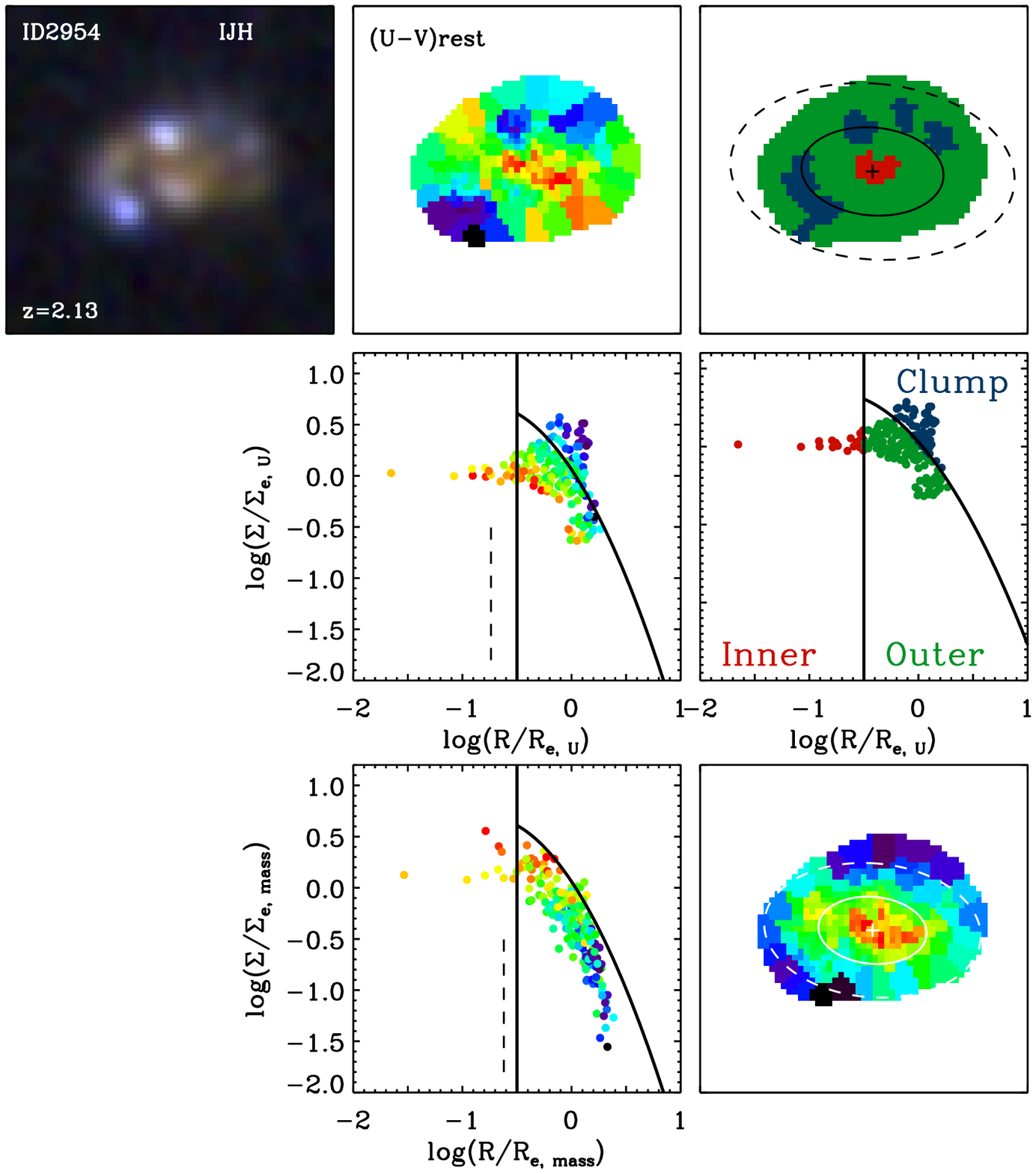}
\caption{
continued
\label{prof_explanation.fig}
}
\end{figure}

\begin{figure*}[t]
\centering
\includegraphics[width=\textwidth]{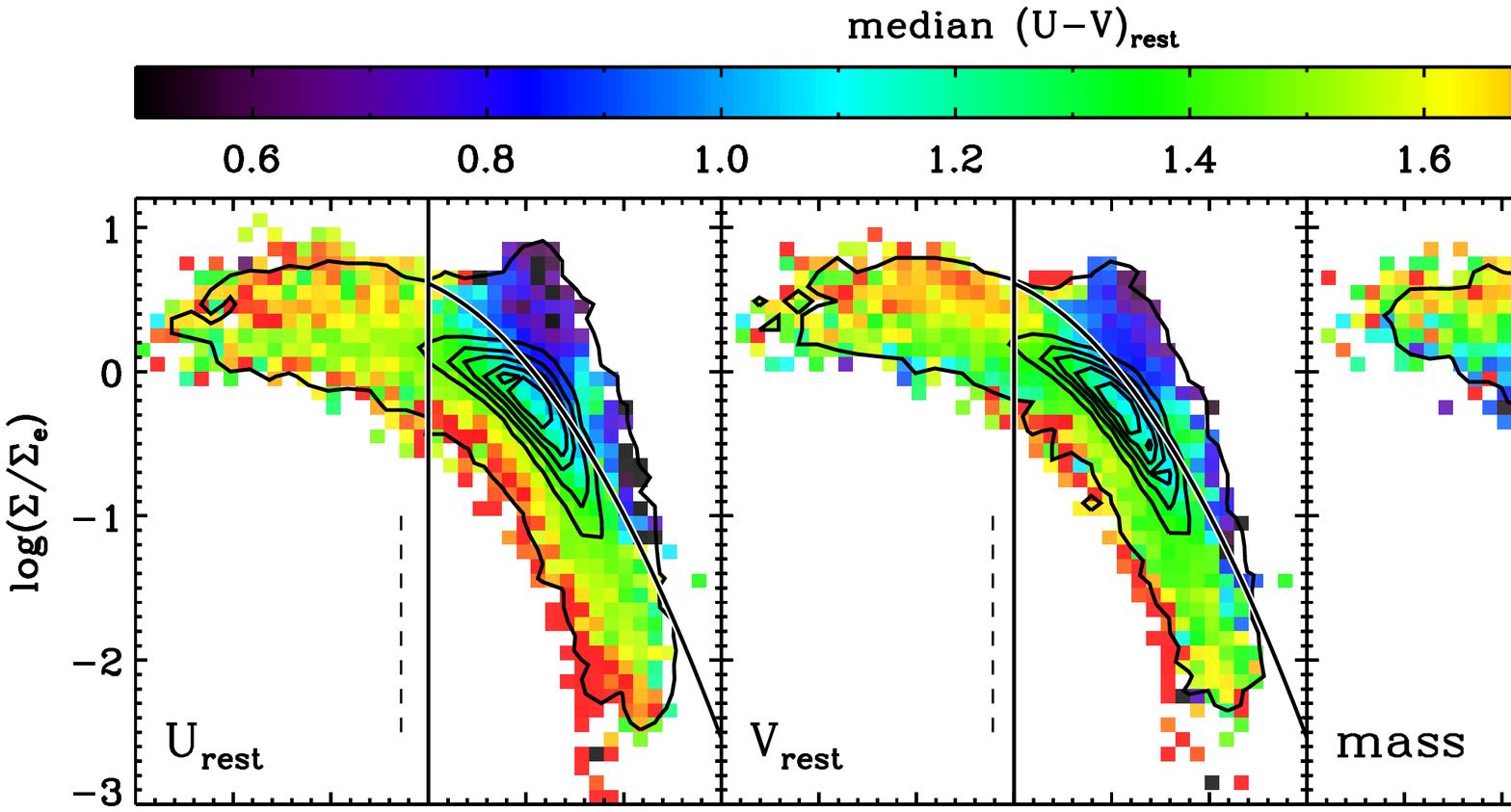}
\includegraphics[width=\textwidth]{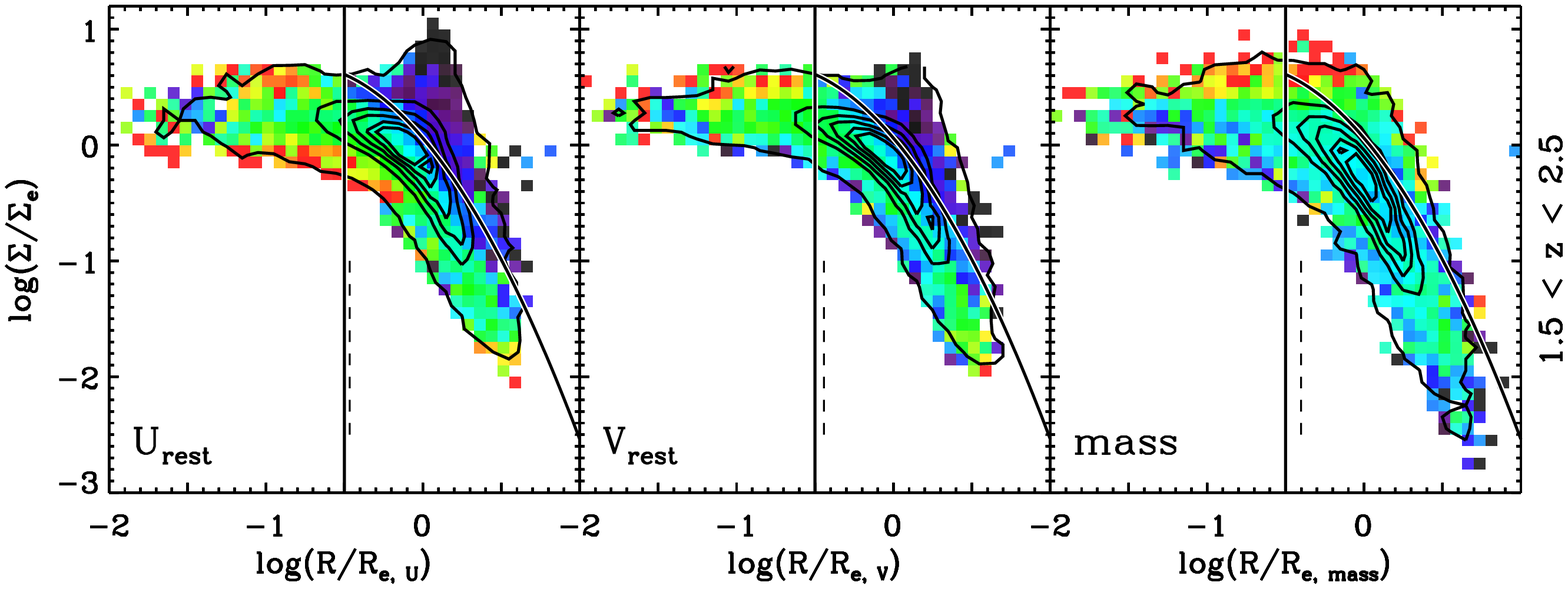}
\epsscale{1.0}
\caption{
Co-added normalized profiles for the SFGs at $0.5 < z < 1.5$ ({\it top row}) and $1.5 < z < 2.5$ ({\it bottom row}), uncorrected for smearing by the PSF.  F.l.t.r., the surface brightnesses/densities and half-light/mass radii were measured on the $U_{rest}$, $V_{rest}$, and stellar mass maps, always adopting the center of stellar mass as a reference for measurements of radii.  The profiles are color-coded by the median rest-frame $(U-V)_{rest}$ color within each $(\Sigma/\Sigma_e,\ R/R_e)$ bin.  The vertical dashed line indicates the typical resolution, and solid lines separate the center, disk and clump regime as in Figure\ \ref{prof_explanation.fig}.  Bright, off-center clumps notable at short wavelengths, and to a lesser extent present even at the longest wavelengths attainable at high resolution, tend to be bluer than the underlying galaxy disk.  The reddest colors are found in the center.
\label{respop3_rfUV.fig}}
\end{figure*}

\begin{figure*}[t]
\centering
\includegraphics[width=0.48\textwidth]{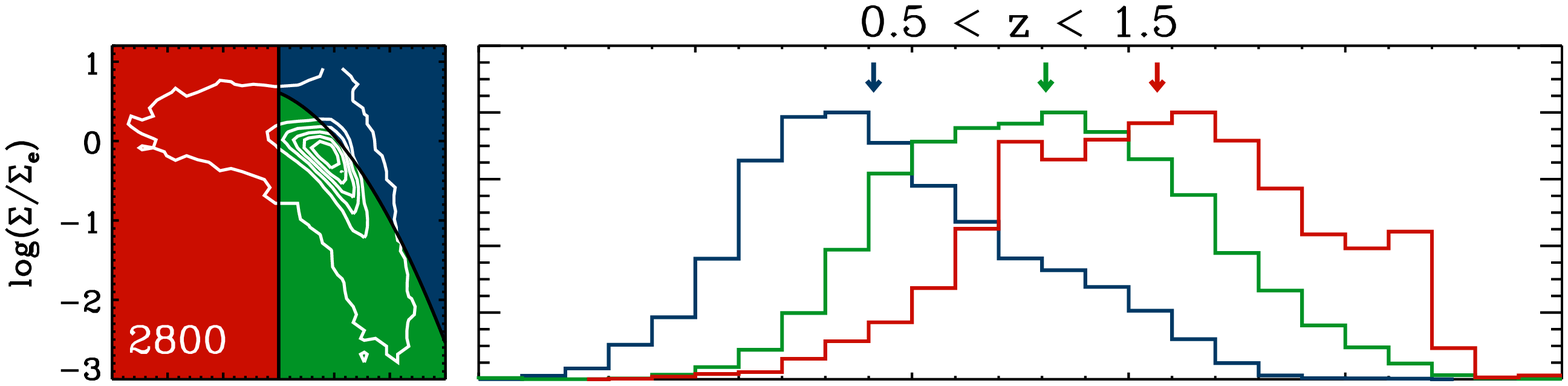}
\includegraphics[width=0.48\textwidth]{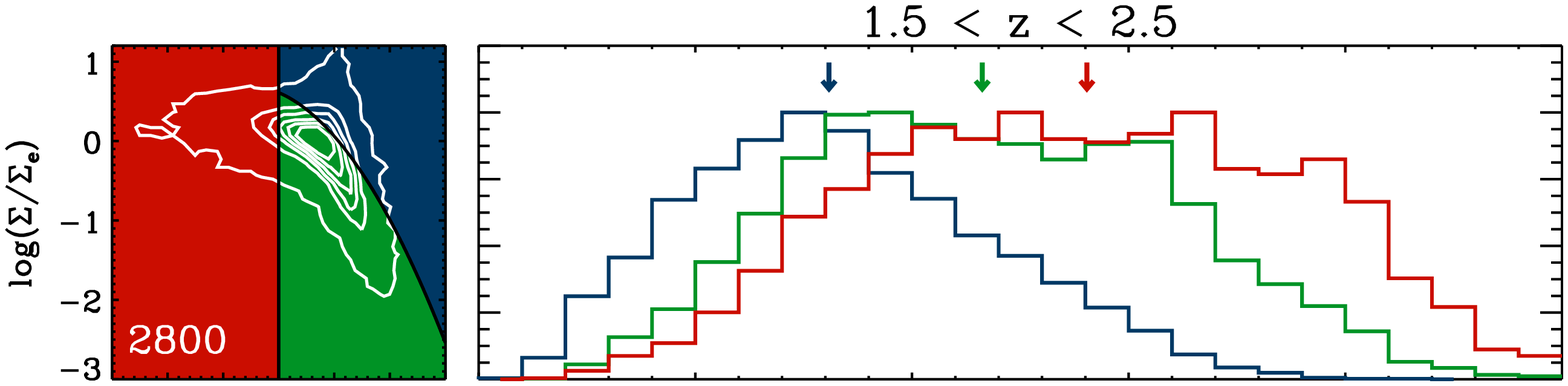}
\includegraphics[width=0.48\textwidth]{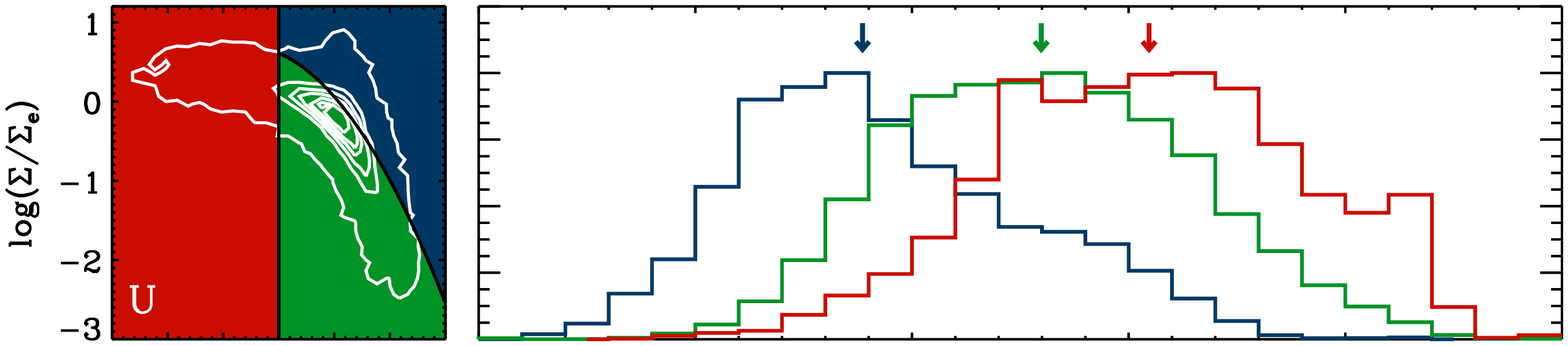}
\includegraphics[width=0.48\textwidth]{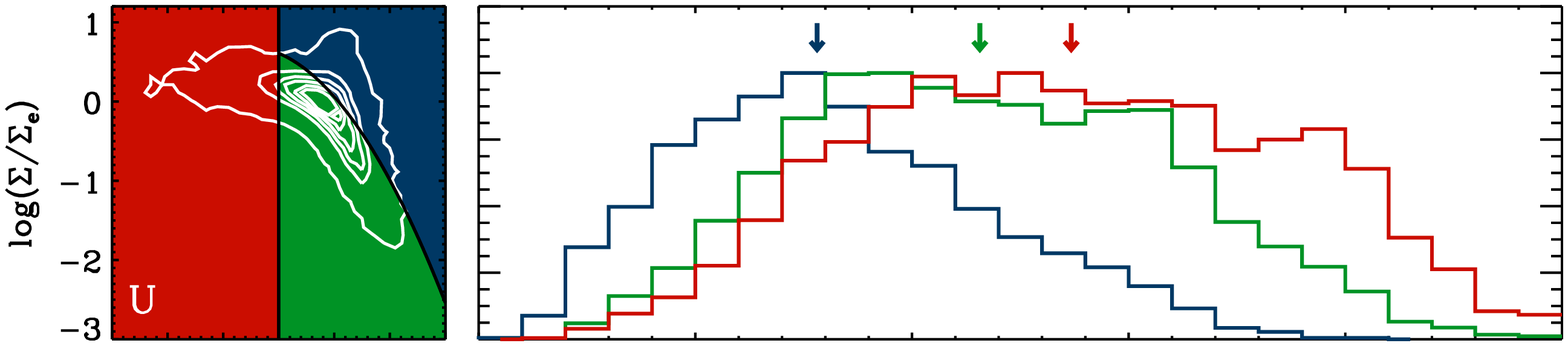}
\includegraphics[width=0.48\textwidth]{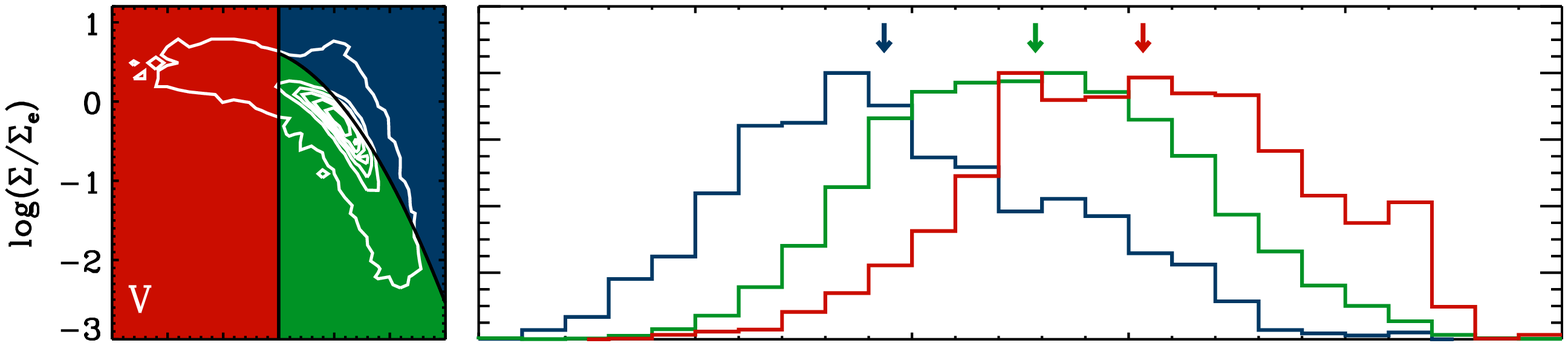}
\includegraphics[width=0.48\textwidth]{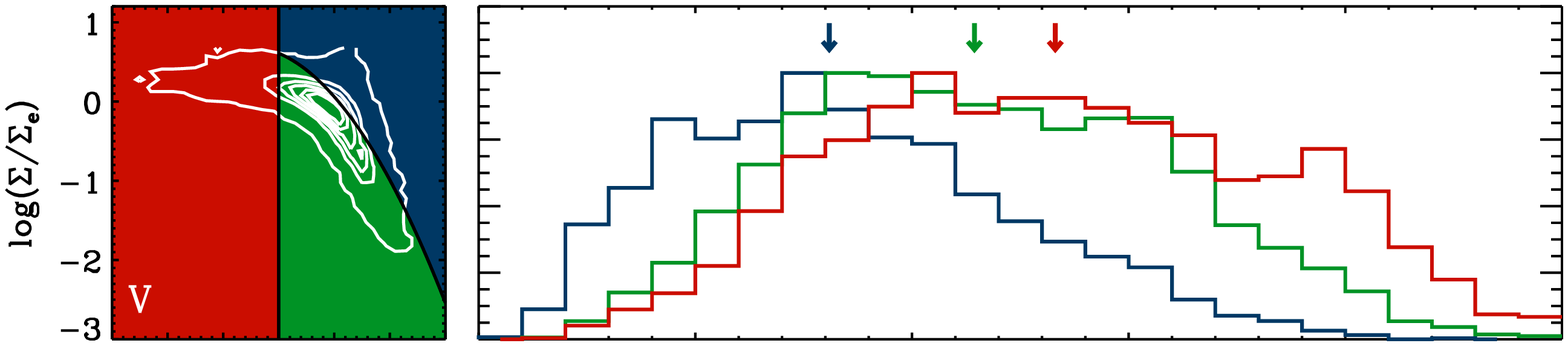}
\includegraphics[width=0.48\textwidth]{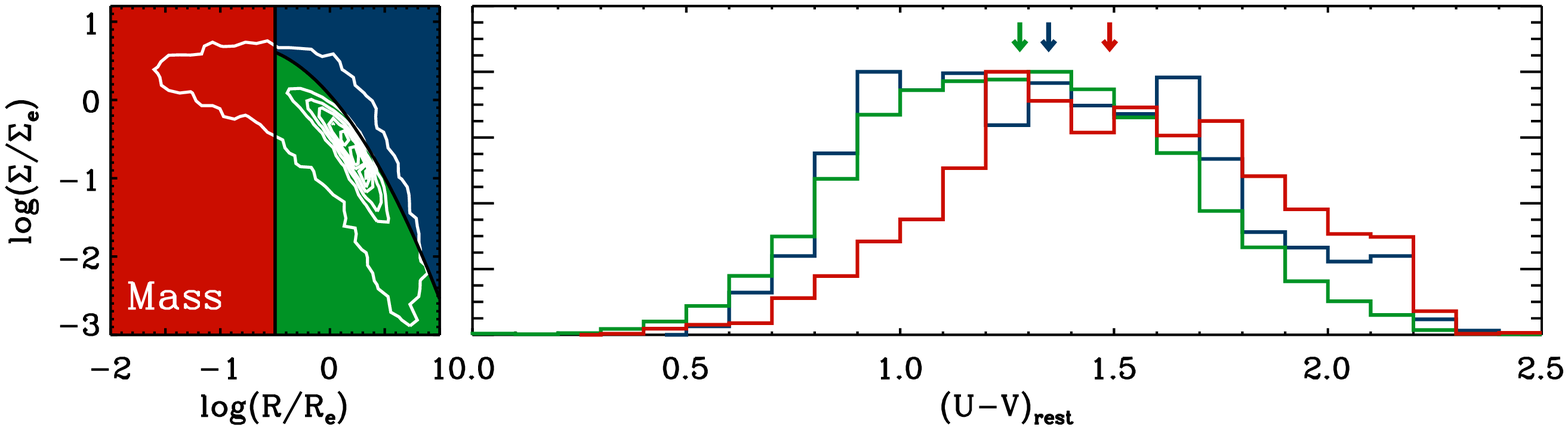}
\includegraphics[width=0.48\textwidth]{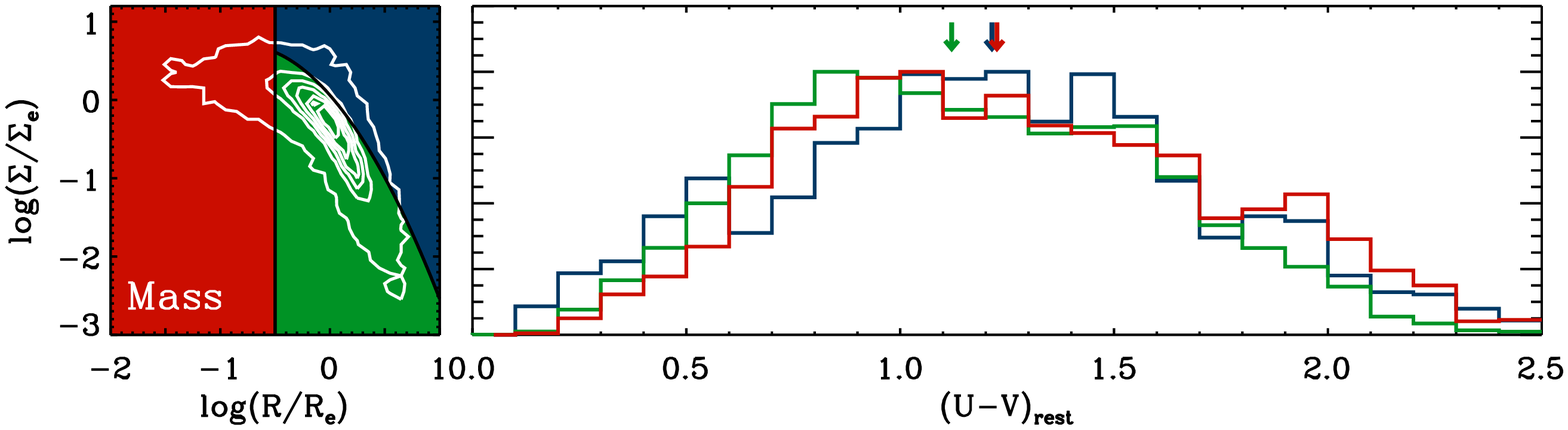}
\epsscale{1.0}
\caption{
Peak-normalized histograms of the rest-frame $(U-V)_{rest}$ color distribution in the center (red), outer disk (green), and clump (blue) regions of SFGs at $0.5<z<1.5$ and $1.5<z<2.5$.  From top to bottom rows, spatial bins are assigned to the center/disk/clump regime based on the normalized rest-frame $2800\AA$, $U$, $V$, and stellar mass profiles respectively.  Medians of the distribution are indicated with arrows for each of the three galaxy components.  While a large spread in $(U-V)_{rest}$ colors is seen in all cases, the color distributions of the three galaxy components as identified on the surface brightness maps are significantly offset with respect to each other.  The galaxy centers are generally characterized by redder colors than the outer disk.  Clumps selected by their elevated surface brightness on the other hand tend to be bluer than the underlying disk.  The latter difference disappears, or if anything reverses, when identifying clumps as regions with elevated surface mass density on the stellar mass maps.
\label{hist_rfUV.fig}}
\end{figure*}

\subsection{Resolved color and stellar population profiles}
\label{profres.sec}

\subsubsection{Constructing Co-added Normalized Profiles}
\label{construct.sec}

In this Section, we quantify the dependence of resolved stellar population properties on surface brightness and stellar surface mass density, while simultaneously probing trends with galactocentric radius.  We do this for our complete sample of SFGs above $10^{10}\ M_{\sun}$ by constructing co-added normalized profiles.  Addressing variations in color, surface mass density, SFR, age and extinction in the 2-dimensional profile space $(\Sigma,\ R)$ allows us to isolate dependencies on local conditions (e.g., correlations between age and surface brightness at a given galactocentric radius) from global galactic gradients (e.g., the dependence of extinction on galactocentric radius).

The first step in constructing the co-added normalized profiles, is defining a center.  Throughout this paper, we consistently adopt the stellar mass-weighted center as reference for all radial measurements:

\begin{equation}
(x_{center};\ y_{center}) = \left ( \frac{ \sum_{i=1}^{N_{pix}} x_i M_i } {\sum_{i=1}^{N_{pix}} M_i};\  \frac{ \sum_{i=1}^{N_{pix}} y_i M_i } {\sum_{i=1}^{N_{pix}} M_i} \right )
\end{equation}

where $M_i$ stands for the stellar mass in pixel $i$.
The motivation for this choice is that the center of stellar mass is more likely to coincide with the kinematic center, where the potential well is deepest, than any light-weighted center.  Having established a reference center, we next determine the semi-major axis length of an ellipse that contains half the light (mass) as outlined in Section\ \ref{structure.sec}, and compute the average surface brightness (surface mass density) within that half-light (half-mass) radius.  The third step is to normalize the surface brightness of each spatial bin in a galaxy by the average surface brightness $\Sigma_e$ within the half-light radius $R_e$ of that galaxy.  Likewise, the distance to the center in major axis units is normalized by the (major-axis) half-light radius.  Normalized stellar mass profiles are constructed following equivalent procedures.  Note that a given pixel will have a different normalized distance to the center when measured on a $U_{rest}$, $V_{rest}$, or stellar mass map, since the half-mass radius is not necessarily (and will typically not be) equal to the half-light radius, and half-light radii themselves depend on wavelength.  Our approach of normalizing the galactocentric distance for a given light profile by the half-light radius measured in that band (as opposed to, e.g., always normalizing by the half-mass radius) has the advantage that objects with identical surface brightness distributions will be mapped to identical positions in the normalized $(\Sigma / \Sigma_e,\ R / R_e)$ space, irrespective of the extent of their underlying mass distribution.  This facilitates a consistent classification of spatial bins into inner, outer, and clump regions based on the waveband of interest (Equations\ \ref{center.eq} to\ \ref{clump.eq} in Section\ \ref{internal_color.sec}).  By means of illustration, we show in Figure\ \ref{prof_explanation.fig} the normalized profiles in $(\Sigma / \Sigma_e,\ R / R_e)$ space alongside the corresponding 2D maps for three individual examples.  In the final step, we stack the normalized profiles of every galaxy.  I.e., we place every spatial bin of every SFG in one diagram of $\Sigma/\Sigma_e$ versus $R/R_e$.  Radial and local dependencies of any stellar population parameter X are now characterized for the entire SFG population by taking the median over all X of spatial bins that ended up at a given $(\Sigma/\Sigma_e;\ R/R_e)$.  In all of the above, we do not attempt to correct for smearing by the WFC3 PSF, and it is understood that our ability to trace spatial variations in stellar populations is limited by its finite width ($0\farcs 18$, see also Appendix C).

\begin{deluxetable*}{lrrrrrrrrrrrr}
\tablecolumns{13}
\tablewidth{\linewidth}
\tablecaption{Contribution of off-center 'clumps' to the integrated stellar populations of SFGs}
\tablehead{
\colhead{Property used for clump identification\tablenotemark{a}} & \colhead{$f_{\rm clumpy}$\tablenotemark{b}}
 & \multicolumn{5}{c}{$\Sigma_{\rm clumps} / \Sigma_{\rm all\ SFGs}$\tablenotemark{c}}
 & \colhead{}
 & \multicolumn{5}{c}{$\Sigma_{\rm clumps} / \Sigma_{\rm clumpy\ SFGs}$\tablenotemark{d}} \\
 \\
\cline{3-7} \cline{9-13} \\
\colhead{} & \colhead{} & \colhead{$L_{2800}$} & \colhead{$L_U$} & \colhead{$L_V$} & \colhead{mass} & \colhead{SFR}
& \colhead{}
& \colhead{$L_{2800}$} & \colhead{$L_U$} & \colhead{$L_V$} & \colhead{mass} & \colhead{SFR} \\
}
\startdata
\multicolumn{13}{c}{$0.5 < z < 1.5$} \\
\cline{1-13} \\

$2800_{\rm rest}$  & 0.79 &               0.17 & 0.14 & 0.10 & 0.05 & 0.15 &  &            0.20 & 0.16 & 0.12 & 0.06 & 0.17 \\
$U_{\rm rest}$        & 0.57 &               0.12 & 0.09 & 0.07 & 0.03 & 0.09 &  &            0.19 & 0.15 & 0.11 & 0.06 & 0.14 \\
$V_{\rm rest}$        & 0.27 &               0.05 & 0.04 & 0.03 & 0.02 & 0.04 &  &            0.16 & 0.14 & 0.12 & 0.07 & 0.12 \\
mass                        & 0.15 &               0.01 & 0.01 & 0.01 & 0.02 & 0.01 &  &            0.08 & 0.08 & 0.09 & 0.15 & 0.05 \\

 \\
\multicolumn{13}{c}{$1.5 < z < 2.5$} \\
\cline{1-13} \\

$2800_{\rm rest}$  & 0.74 &               0.19 & 0.17 & 0.13 & 0.07 & 0.18 &  &            0.25 & 0.22 & 0.17 & 0.09 & 0.22 \\
$U_{\rm rest}$        & 0.60 &               0.16 & 0.13 & 0.10 & 0.05 & 0.13 &  &            0.24 & 0.21 & 0.16 & 0.09 & 0.19 \\
$V_{\rm rest}$        & 0.42 &               0.11 & 0.09 & 0.07 & 0.04 & 0.09 &  &            0.22 & 0.20 & 0.17 & 0.12 & 0.19 \\
mass                        & 0.41 &               0.04 & 0.04 & 0.04 & 0.07 & 0.04 &  &            0.10 & 0.10 & 0.11 & 0.16 & 0.09 \\

\enddata
\tablenotetext{a}{Rest-frame waveband or physical property based on which normalized profiles were constructed and off-center spatial bins in the 'clump' regime were identified.}
\tablenotetext{b}{Fraction of SFGs that are 'clumpy' in the respective light or mass map used for clump identification (see a).  A galaxy is defined as being 'clumpy' in, e.g., the $V_{rest}$ surface brightness distribution, if at least 5\% of the total $V_{rest}$ luminosity of that galaxy is contributed by spatial bins in the 'clump' regime.}
\tablenotetext{c}{Fractional contribution of spatial bins in the 'clump' regime to the integrated luminosity, mass, and SFR of all SFGs in the redshift interval of interest.}
\tablenotetext{d}{Fractional contribution of spatial bins in the 'clump' regime to the integrated luminosity, mass, and SFR of 'clumpy' SFGs (see b).}
\label{clump.tab}
\end{deluxetable*}

\subsubsection{Internal Color Variations}
\label{internal_color.sec}

An example of such co-added normalized profiles is shown in Figure\ \ref{respop3_rfUV.fig}.  The top row presents the results for $z \sim 1$ SFGs, while the bottom row shows the equivalent results for the $z \sim 2$ SFGs.  For each redshift interval, co-added normalized profiles (i.e., with dimensionless units on both axes) were constructed based on the $U_{rest}$, $V_{rest}$, and stellar mass maps.  The color coding indicates the median $(U-V)_{rest}$ color of all Voronoi-binned pixels in a given $(\Sigma/\Sigma_e;\ R/R_e)$ bin.  In each panel, a vertical dashed line corresponds to the normalized half-light radius of the WFC3 PSF for the typical redshift and size of galaxies in the sample.  Any radial trends within this resolution limit should not be considered as physical.  

The co-added 1D profiles reveal a number of trends that we alluded to already when discussing the multi-wavelength appearance of a few case examples in Section\ \ref{examples.sec}.  We start by focussing on the $U_{rest}$ profiles.  Within an effective radius, the radial profiles are relatively flat, at least in part due to beam smearing.  At radii $R >> R_e$, the surface brightness drops.  At all radii, the co-added profiles exhibit a considerable scatter in $\Sigma/\Sigma_e$.  The range in $\Sigma/\Sigma_e$ at a fixed radius reaches a maximum around $R \approx R_e$, spanning a dynamic range of roughly 2.5 - 3 dex.  At this radius, surface brightness levels are reached that even exceed the surface brightness of the inner regions.  In the remainder of this paper, we will refer to these off-center regions with elevated surface brightness relative to the underlying disk as 'clumps'.  We use this nomenclature for ease of description and inspired by the morphological appearance of the galaxies shown in Figure\ \ref{examples.fig}, but stress however, that regions of excess surface brightness with other geometric shapes, such as spiral arms, would show up similarly in the 1D profile plots.  The reason that clumps with the highest surface brightness levels line up at $R \approx R_e$ can be understood from the following extreme example.  In a case where a single off-center clump dominates the entire light budget at $U_{rest}$ wavelengths, an ellipse centered on the center of mass needs to grow out to the galactocentric radius of that clump before it encompasses half the $U_{rest}$-band light, yielding $R_e \approx R_{\rm clump}$.  At all azimuthal angles away from the clump, low surface brightness levels corresponding to the outer underlying disk will be measured at a radius $R_e$, leading to the observed wide range in $\Sigma/\Sigma_e$ at the effective radius.

Interestingly, the color of subregions within the SFGs cannot adequately be described by a simple function of distance relative to the galaxy center.  Outside $\log(R/R_e) = -0.5$, a strong trend is revealed in which the $(U-V)_{rest}$ color correlates with the $U_{rest}$ surface brightness at a given radius.  Blue clumps are superposed on a redder underlying disk.  This characteristic behavior is seen at $z \sim 2$ as well as $z \sim 1$.  Turning to the middle panels of Figure\ \ref{respop3_rfUV.fig}, we note that a substantial range in surface brightness levels is still present around $R \sim R_e$ at $V_{rest}$ wavelengths, albeit slightly less pronounced.  Again, we find evidence that off-center regions with elevated surface brightness ('clumps') are in the median bluer than the underlying galaxy disk.  Within $\log(R/R_e) = -0.5$, we find typically redder colors than in the outer disk or clumps, and the relation with normalized surface brightness seems to be reversed.  Finally, when considering the mass profiles, the presence of off-center clumps is not evident as it was for the light profiles.  Outside the half-mass radius, little dependence of color on surface mass density is noted at a given radius.  Within the half-mass radius, the optical color correlates positively with the local surface mass density.

The co-added normalized profiles in Figure\ \ref{respop3_rfUV.fig} only present the median color characteristics of clumps relative to the underlying disk and galaxy nucleus.  They do not contain information on the distribution of colors among clumps, among galaxy nuclei, or among outer disk regions.  We quantify this further by splitting the normalized profiles in center, disk, and clump sectors (solid lines in Figure\ \ref{respop3_rfUV.fig}):

\begin{align}
[center]\ \ &  X < -0.5 \label{center.eq} \\
[disk]\ \ &  X > -0.5\ \&\ Y < 0.06 - 1.6 X - X^2 \label{disk.eq} \\
[clump]\ \ & X > -0.5\ \&\ Y > 0.06 - 1.6 X - X^2 \label{clump.eq}
\end{align}

where $X \equiv \log(R/R_e)$ and $Y \equiv \log(\Sigma/\Sigma_e)$.  The separating line between disk and clump regions (Equation\ \label{disk.eq} -\ \label{clump.eq}) was defined by eye, upon inspection of both the normalized stacked profiles, and the equivalent diagrams for individual objects in combination with their appearance in postage stamps (as illustrated for case examples in Figure\ \ref{prof_explanation.fig}).  In Figure\ \ref{hist_rfUV.fig}, we display the $(U-V)_{rest}$ color distributions of Voronoi-binned pixels assigned to each of these three regimes, as well as the median of the distribution, indicated with an arrow.  We use maps at three rest-frame wavelengths (2800\AA, $U$, and $V$), as well as the stellar mass maps to identify center, clump, and disk pixels.  A large spread in colors is seen in both nuclear, disk, and clump pixels.  Nevertheless, a clear trend emerges in that the light profiles feature red centers, green disks, with blue clumps superposed.  Considering the same cuts in $(\Sigma/\Sigma_e;\ R/R_e)$ space for the normalized mass profiles, we do not find significant evidence for color differences between the clump and outer disk regime.  Furthermore, the trend of redder central colors is somewhat reduced.  The latter observation can be understood from red pixels with $\log(R/R_{e,\ \rm light}) < -0.5$ being located at $\log(R/R_{e,\ \rm mass}) > -0.5$.  As we will discuss in Section\ \ref{light_mass.sec}, half-mass radii are in the median smaller by 0.1-0.2 dex than half-light radii measured at $U_{rest}$ wavelengths.  Larger differences can occur for those galaxies where a strong color gradient (a red core) is present.

\subsubsection{Contributions of Clumps to the Integrated Light and Mass of SFGs}
\label{contribution.sec}

Not all SFGs have a significant fraction of their light or mass contributed by pixels in the 'clump' regime (which, we remind the reader, could also include spiral or ring-like features with excess surface brightness/density).  In Table\ \ref{clump.tab}, we list the fraction of SFGs that are classified as clumpy in a given band, where we define 'clumpy' as having at least 5\% of the galaxy's total luminosity in the respective band originating from pixels in the clump regime.  Equivalently, the stellar mass distribution of a galaxy is considered to be clumpy when pixels in the clump regime contribute at least 5\% of the total stellar mass.  The fraction depends sensitively on which waveband is used to identify the clumps, dropping progressively from 79\% (74\%) for $2800_{rest}$-selected clumps at $z \sim 1$ (2) to 27\% (42\%) for $V_{rest}$-selected clumps.  The fraction of galaxies with clumps decreases even further, down to 15\% (41\%) at $z \sim 1$ (2), when applying the clump selection on the stellar mass maps.  We stress that this dependence on the property used for clump identification stems from real, intrinsic properties of the resolved stellar populations, and is not driven by wavelength-dependent resolution effects that may smear out or blend clumps, because all of our analysis is carried out on images that were PSF-matched to the WFC3 resolution.  We do note, however, that higher resolution deeper data might resolve more clumps (especially fainter and smaller ones, if present).  The fractional contribution of clumps to the integrated galaxy light of our mass-complete sample of SFGs shows a similar dependence on the waveband used for clump selection, decreasing with increasing wavelength.  Meanwhile, for a given selection band, the fractional contribution of clumps to the integrated galaxy light is a decreasing function of wavelength.  Even when considering the contribution of clumps selected at 2800\AA\ to the integrated $L_{2800}$ luminosity of only those galaxies classified as clumpy in that band, the fractional contribution is limited to 20-25\%.  In terms of contribution to the overall stellar mass budget of SFGs, clumps only account for 2-7 percent, depending on the redshift interval and property used for clump identification.  We note that here, no attempt was made to subtract the underlying disk emission at the location of a clump, as is done in some previous studies of smaller samples of $z \sim 2$ SFGs (e.g., F\"{o}rster Schreiber et al. 2011b; Guo et al. 2011b).  While beyond the scope of this paper, we point out that such corrections would only reduce the fractional contribution to the total stellar mass in SFGs.  Likewise, we cannot exclude that some of the surface brightness peaks do not correspond to clumps formed in situ in the disk, but rather to a minor merger where the satellite is seen in projection on the massive SFG.  Again, this effect, if present, biases the fractional contributions listed in Table\ \ref{clump.tab} to higher values.  When considering the contribution of clumps to the outer disk regions only (i.e., excluding pixels with $\log (R/R_e) < -0.5$), the fractional contributions increase only modestly, by 0.01 to 0.02.

Several factors inhibit a meaningful quantitative comparison of the fractional contributions listed in Table\ \ref{clump.tab} to estimates from the literature (Elmegreen et al. 2005, 2009; F\"{o}rster Schreiber et al. 2011; Guo et al. 2011b).  First, none of these authors consider mass-complete samples of SFGs.  Instead, they study the contribution of clumps to the class of clumpy galaxies specifically, selected by their clumpy morphology directly and/or sampling indirectly the UV bright population of galaxies through the requirement of a spectroscopic redshift.  Second, while we do not include the center of the galaxy as a clump, these studies do include the contribution of bulge-like clumps, if present.  Finally, the clump identification by Elmegreen et al. (2005, 2009) and Guo et al. (2011b) was carried out at the ACS resolution, and the clump masses reported in the earlier work were in most cases based on ACS photometry only.  All of the above reasons, together with sheer statistics, likely contribute to the fact that the fractional contribution of clumps to the total SFG population ($\Sigma_{clumps} / \Sigma_{all\ SFGs}$ in Table\ \ref{clump.tab}) is at face value somewhat lower (by factors $\sim 1.5$ to 3) than fractions from the literature.  However, in common to all of these studies is the observation that clumps contribute significantly less in terms of stellar mass than in terms of SFR or UV luminosity.

\begin{figure*}[t]
\centering
\includegraphics[width=\textwidth]{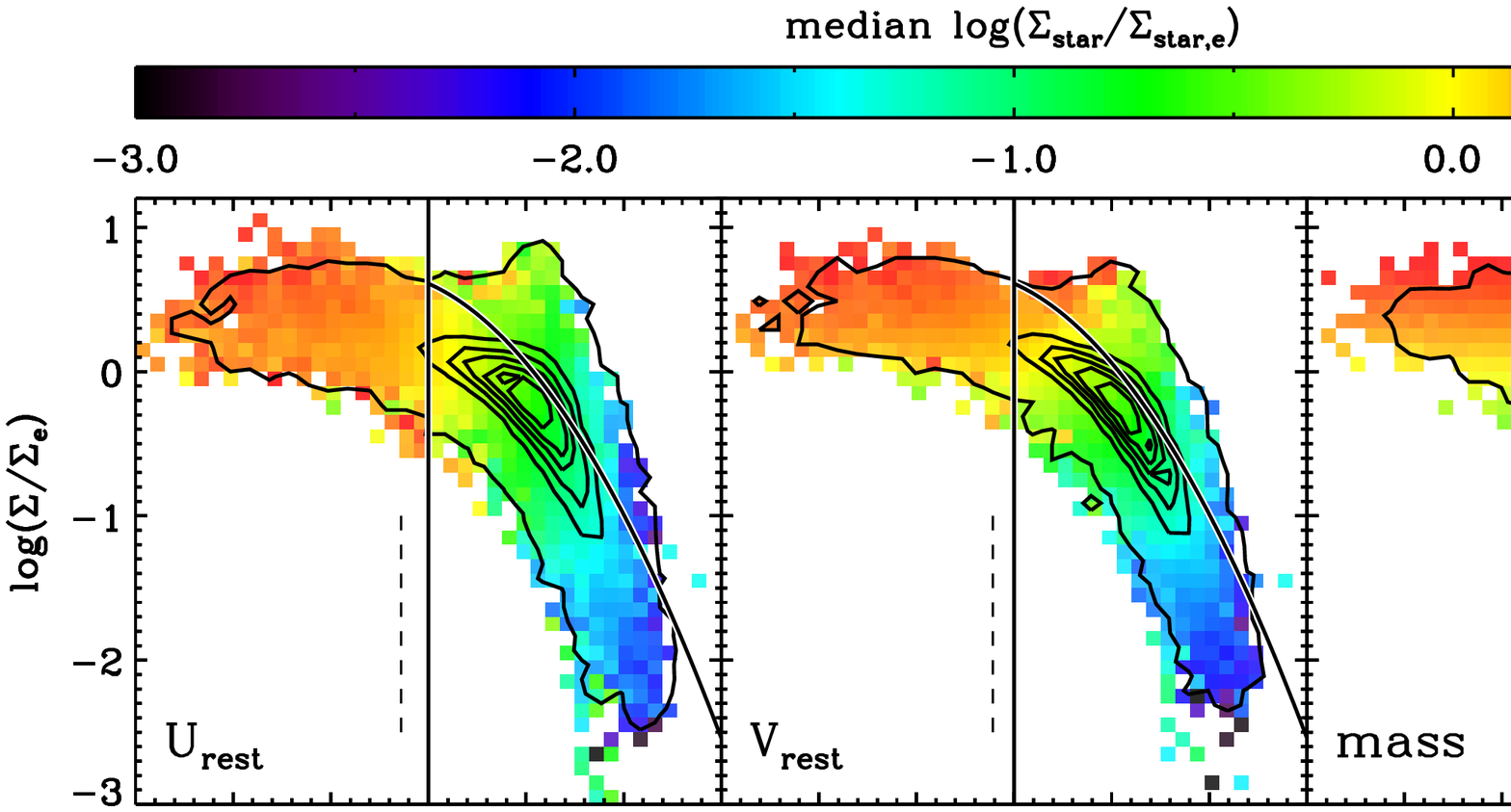}
\includegraphics[width=\textwidth]{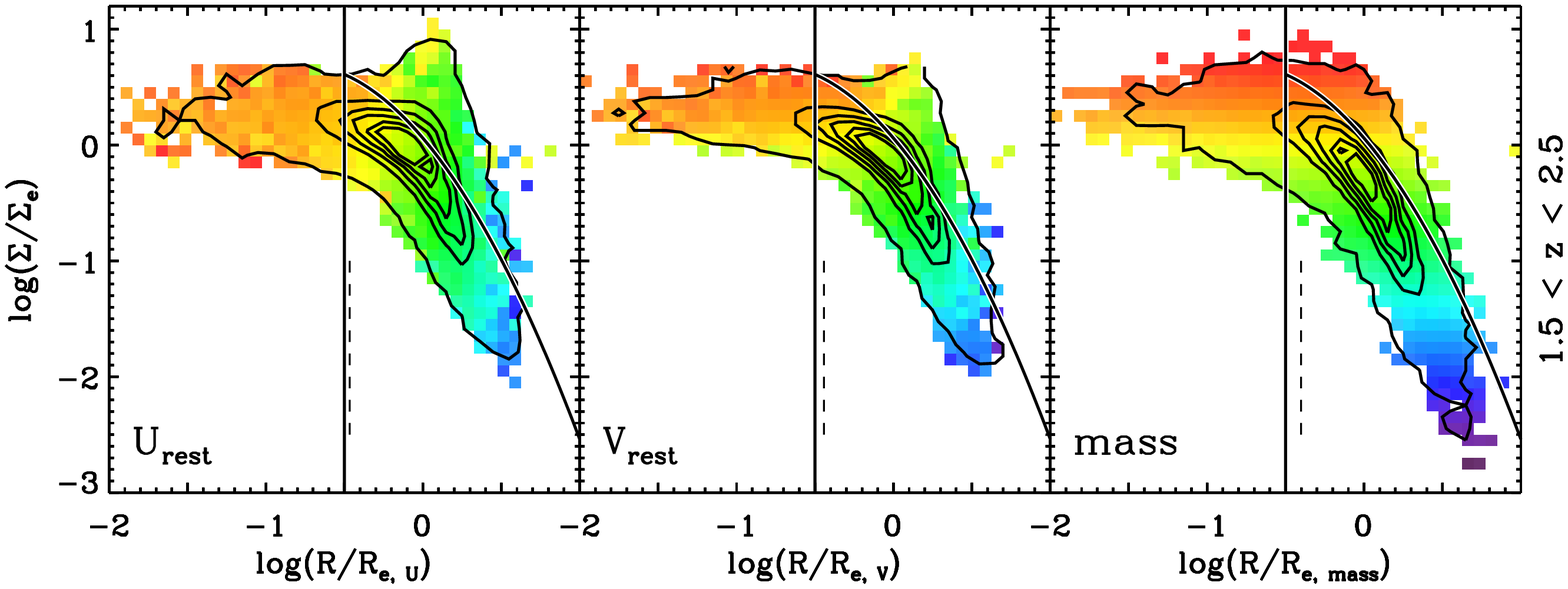}
\epsscale{1.0}
\caption{
Idem to Figure\ \ref{respop3_rfUV.fig}, but with the color-coding indicating the normalized stellar surface mass density.  When considering shorter wavelengths (i.e., poorer proxies of stellar mass), little variation in the stellar surface mass density is seen at a given radius, despite a substantial range in surface brightness levels.
\label{respop3_Sigma.fig}}
\end{figure*}

\begin{figure*}[t]
\centering
\includegraphics[width=\textwidth]{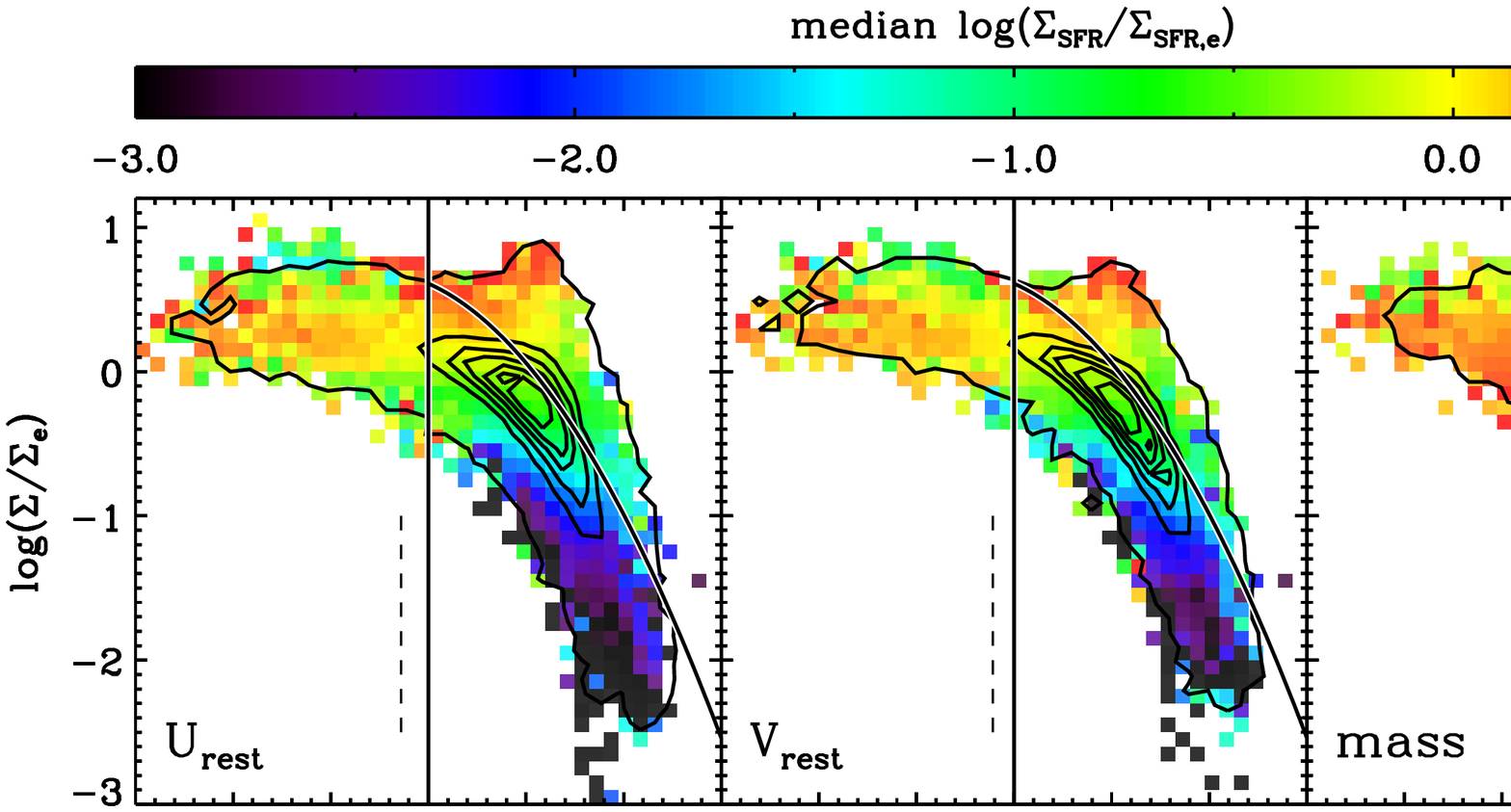}
\includegraphics[width=\textwidth]{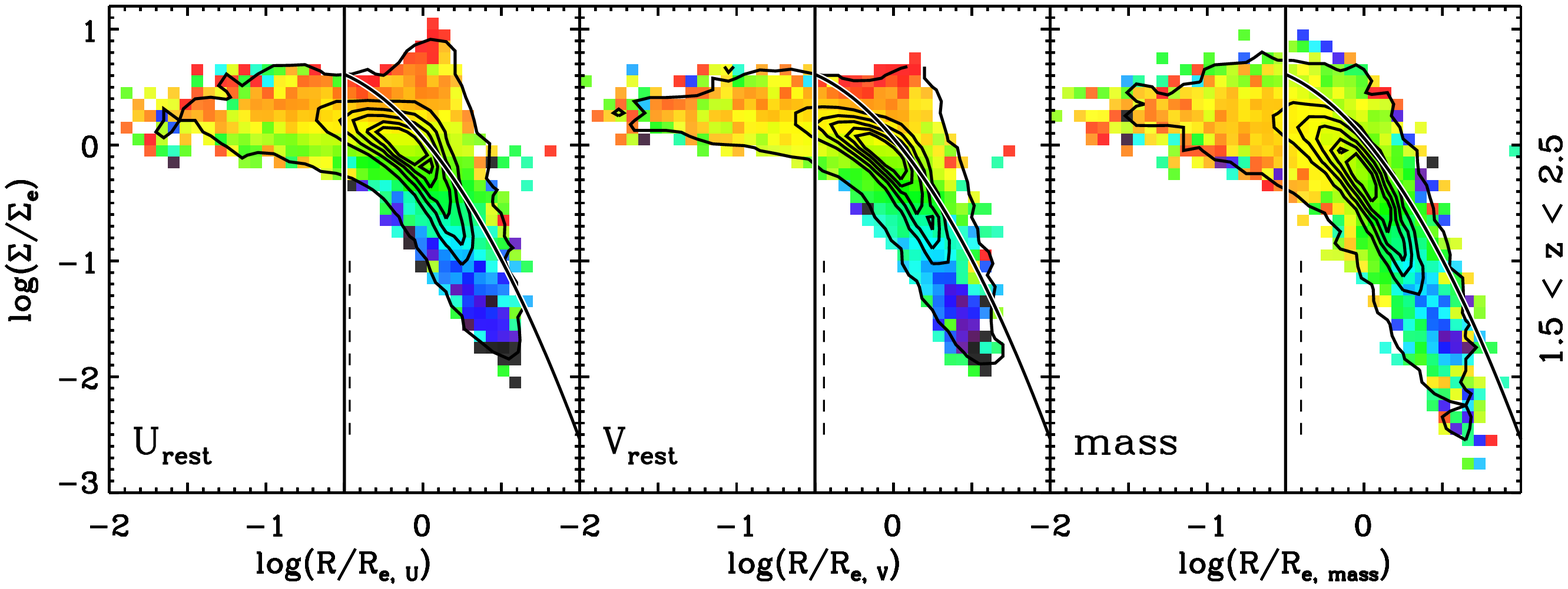}
\epsscale{1.0}
\caption{
Idem to Figure\ \ref{respop3_rfUV.fig}, but with the color-coding indicating the normalized star formation surface density.  In the light profiles, the star formation surface density is a function of surface brightness, with no evidence for a radial dependence.  In the mass profile, however, the star formation surface density correlates more strongly with radius than with variations in surface mass density at a given radius.
\label{respop3_SFR.fig}}
\end{figure*}

\subsubsection{Surface Mass Density}
\label{surfdens.sec}

As alluded to already in our discussion of case examples, the resolved color information gives us a handle on spatial variations in the $M/L$ ratio, and can therefore be exploited to derive stellar surface mass densities.  We show its variation along and across the co-added normalized profiles in Figure\ \ref{respop3_Sigma.fig}.  We remind the reader that all available photometric information was used in constraining the surface mass density, and not just the one $(U-V)_{rest}$ color that was discussed in the previous Section. In the mass profiles (right-hand panels), the isocontours run by construction horizontally, as the property indicated with color coding is identical to that plotted along the y axis of the diagram.  It is therefore more interesting to inspect its behavior in the light profiles (left-hand and middle panels).  Particularly at the shorter $U_{rest}$ wavelengths, it is striking that the surface mass density correlates more strongly with galactocentric distance than with surface brightness (i.e., the isocontours now run vertically).  The implication is that, while a large spread in surface brightness levels is observed at a given radius (particularly around $R \sim R_{e,\ \rm U}$), clumps present in the light distribution do not necessarily correspond to an excess in the surface mass density with respect to the underlying disk.  This is reflected quantitatively in Table\ \ref{clump.tab}, where we see that clumps selected from the $U_{rest}$ surface brightness profile, account for only 3-5\% of the integrated stellar mass in SFGs.

An interesting question is whether, and how much mass could be hidden if the actual SFHs of clump regions differed from the smooth, 1-component $\tau$ models assumed in our analysis.  We address this question by exploring the impact of 2-component SFHs following equivalent procedures to Papovich et al. (2001).  Briefly, we fit the superposition of a maximally old burst and a younger $\tau$ model to the SED of each spatial bin, with the relative mass fraction of the young and old component as a free parameter.  To maximize the effect of outshining of the old population by a young burst, we allow ages since the onset of star formation for the young component down to 10 Myr.  Furthermore, since we are interested in an upper limit on the amount of mass that can potentially be hidden, we do not adopt the least squares solution, but rather the maximum total stellar mass corresponding to a fit that is within $\Delta \chi^2_{red} = 1$ of the best-fit solution.  We find that under these extreme conditions, the inferred stellar masses of spatial bins in the clump regime increase by a factor $\sim 3$.  In contrast, a similar exercise for the center and outer disk regime leads to factors of 2 and 2.3, respectively.  The fact that less mass can be hidden in these regions that are characterized by redder colors, was noted earlier for 2-component modeling of integrated galaxy SEDs (Papovich et al. 2001; Wuyts et al. 2007).  We conclude that, while a substantial amount of mass can be hidden if the actual SFHs deviate from smooth $\tau$ models, and slightly more so in clumps than in the underlying disk, it is rather unlikely that the extreme SFHs considered here apply to all clumps.  Moreover, if this would be the case, it may be more meaningful to interpret the old component as background stars from the underlying disk that coincide in projection with the star-forming clumps, in which case it should not be included in the fractional mass contributions by clumps listed in Table\ \ref{clump.tab}.

\begin{figure*}[t]
\centering
\includegraphics[width=\textwidth]{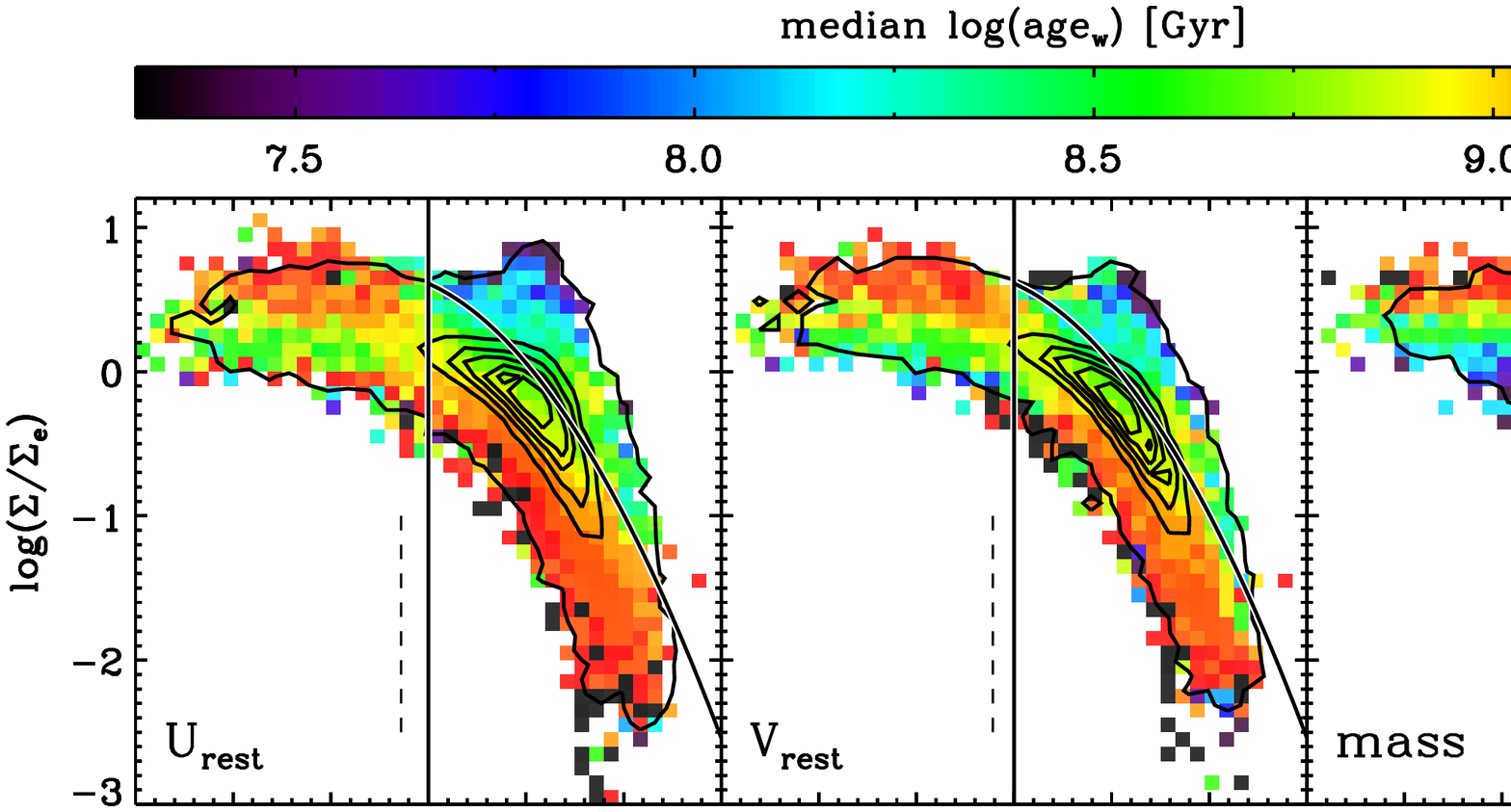}
\includegraphics[width=\textwidth]{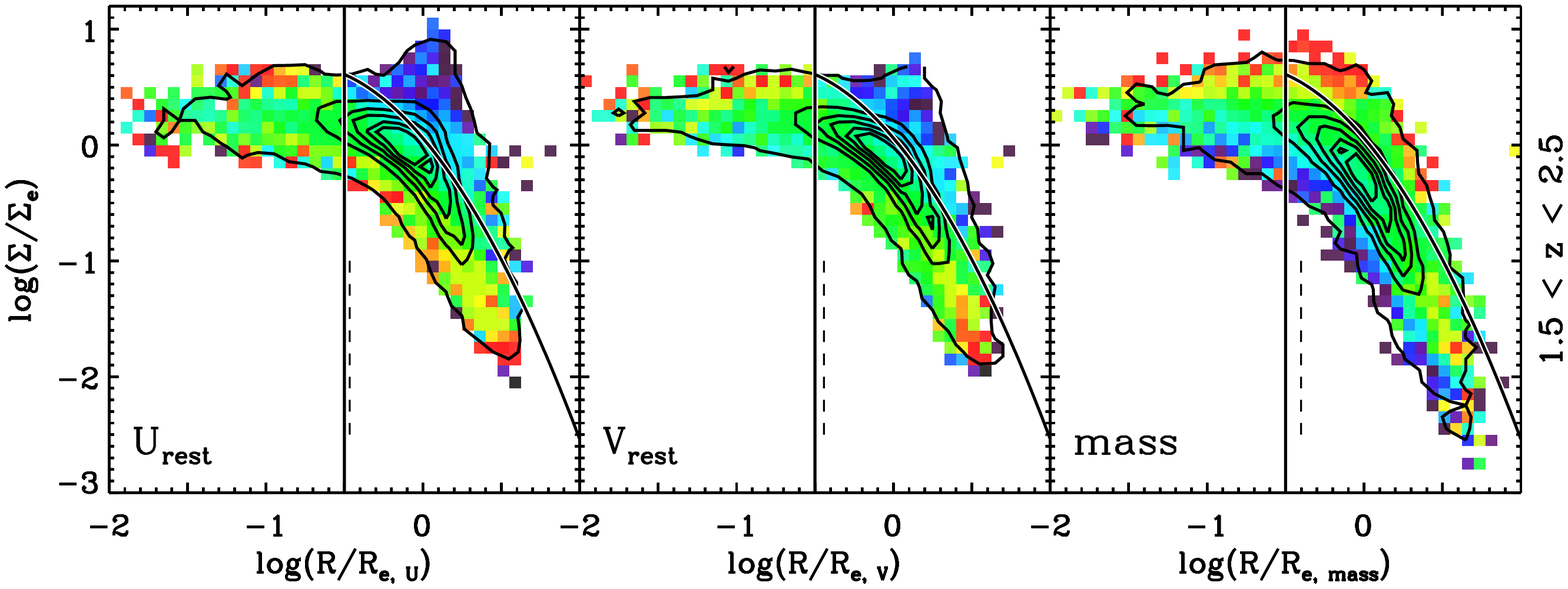}
\epsscale{1.0}
\caption{
Idem to Figure\ \ref{respop3_rfUV.fig}, but with the color-coding indicating the SFR-weighted age of the stellar population.  Clumps selected from surface brightness profiles are significantly younger ($< 200 Myr$) than the underlying galaxy disk.  In contrast, regions with elevated surface mass densities at a given radius have older stellar populations than regions with a lower surface mass density at the same galactocentric radius.
\label{respop3_agew.fig}}
\end{figure*}

\subsubsection{Surface Density of Star Formation}
\label{SFRsurfdens.sec}

In contrast to what is seen in terms of surface mass density, clumps do stand out significantly compared to the underlying disk in terms of the amount of ongoing star formation per unit area.  This is expressed most evidently in the co-added normalized light profiles of our $z \sim 2$ SFG sample (bottom left and middle panel of Figure\ \ref{respop3_SFR.fig}).  First, we find evidence for a radial gradient in $\Sigma_{SFR}/\Sigma_{SFR,\ e}$, increasing towards the center.  Second, superposed on top of this smooth profile are clumps that exceed the star formation surface density of the underlying disk, reaching an excess of more than an order of magnitude for the most extreme clumps.  In the case of these most extreme clumps, their $\Sigma_{SFR}$ even exceeds the amount of star formation per unit area taking place in the galaxy center.  The stellar mass profiles (right-hand panels) exhibit a smooth radial gradient, but any evidence for variations in star formation surface density with stellar mass surface density at a given radius is marginal.

\begin{figure*}[t]
\centering
\includegraphics[width=\textwidth]{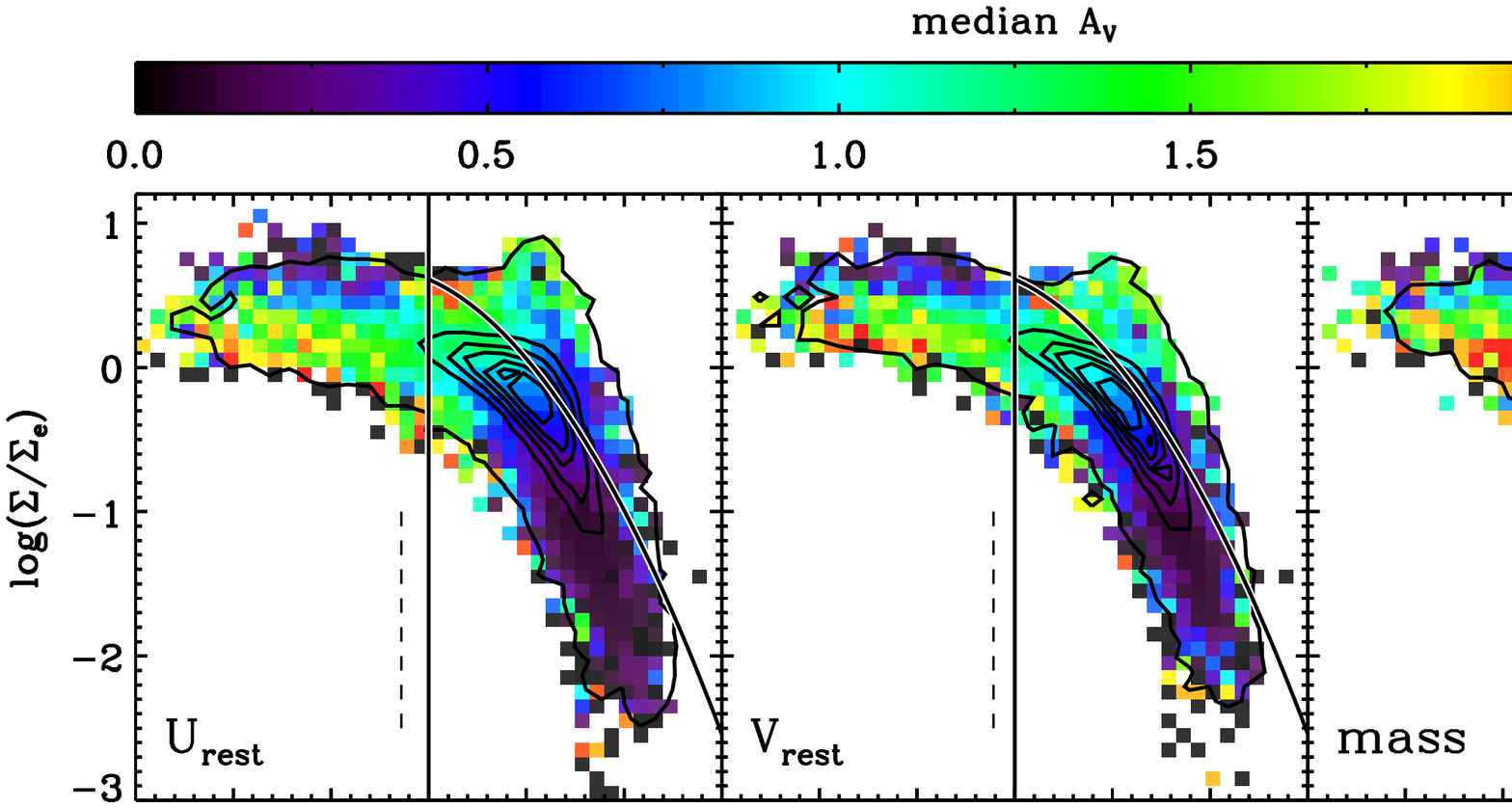}
\includegraphics[width=\textwidth]{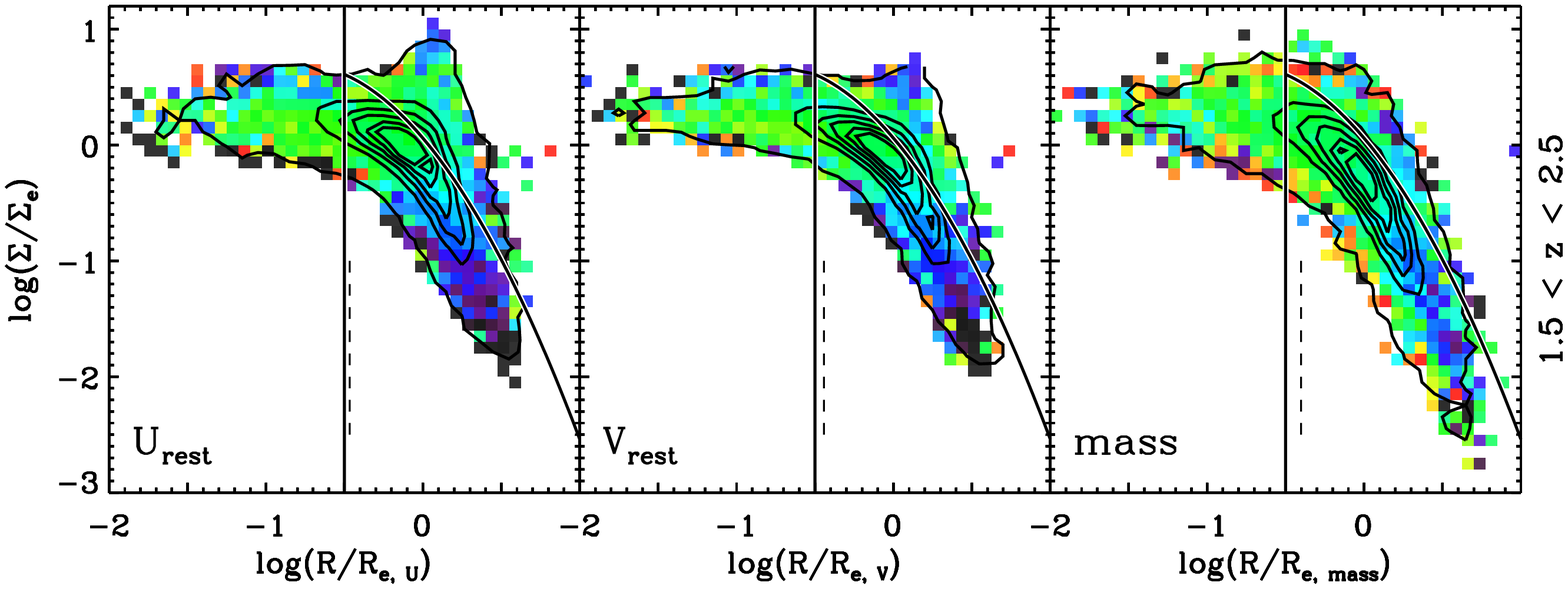}
\epsscale{1.0}
\caption{
Idem to Figure\ \ref{respop3_rfUV.fig}, but with the color-coding indicating the visual extinction $A_V$.  The centers of SFGs are in the median more strongly attenuated than the outskirts.
\label{respop3_Av.fig}}
\end{figure*}

Combining the results on the surface mass density and surface density of star formation, we unavoidably come to the conclusion that the inferred situation cannot be sustained over long timespans.  When left in place, any prolongued excess in $\Sigma_{SFR}$ will build up an excess in $\Sigma_{star}$.  Either the clumps (or spiral or ring-like features) seen here will reside elsewhere (e.g., in the center) by the time they reach a more evolved stage, or they are dispersed on a short timescale before they manage to build up a significant excess in $\Sigma_{star}$.  We return to these scenarios in Section\ \ref{discussion.sec}.

\subsubsection{Age and Dust Profiles}
\label{lagew.sec}

Given the above speculations on the longevity of clumps (or other off-center features), we now investigate the inferred variation in the SFR-weighted age (i.e., the age of the bulk of the stars) along and across the light and mass profiles (Figure\ \ref{respop3_agew.fig}).  It is well known from integrated SED modeling that age-dust degeneracies inhibit constraining the stellar age as tightly as the stellar mass, particularly when only a narrow wavelength baseline is available.  Here, the fact that we also account for the integrated photometric constraints from shorter ($U$) and longer ($K_s + IRAC$) observed wavelengths than the $B_{435}$-to-$H_{160}$ range available at high resolution, can help.  For integrated photometry, the power of a long wavelength baseline in reducing age-dust degeneracies has been demonstrated convincingly through the now widely used $UVJ$ diagram (e.g., Wuyts et al. 2007, 2009; Williams et al. 2009).  In this diagram, evolved populations (with red $(U-V)_{rest}$ and blue $(V-J)_{rest}$) are efficiently distinguished from unobscured star-forming populations (blue $(U-V)_{rest}$, blue $(V-J)_{rest}$) and dusty starbursts (red $(U-V)_{rest}$, red $(V-J)_{rest}$) respectively.  Consider now a star-forming galaxy with an outer disk that is blue in $(U-V)_{rest}$, and a core that is red in $(U-V)_{rest}$.  To address whether the center is red because of dust or because of an evolved population, we can consider the integrated photometry at longer wavelengths.  If the integrated $(V-J)_{rest}$ is red, this cannot be driven by the unobscured star formation in the outer disk.  Consequently, the red center will be interpreted as a dusty core.  Conversely, the center is unlikely to host dusty star formation if  the integrated $(V-J)_{rest}$ color is blue.  In practice, as detailed in Section\ \ref{methodology.sec}, we again use all photometric information available to construct the color-coded profiles in Figure\ \ref{respop3_agew.fig}, and not just one or two colors.

The pattern of age variations with galactocentric radius and surface brightness is very reminiscent of those described for the $(U-V)_{rest}$ color (which is only weakly model dependent) in Section\ \ref{internal_color.sec}.  In other words, much of the optical color variations are interpreted as variations in the age of the bulk of the stars.  At $z \sim 2$, clumps selected at $U_{rest}$ and $V_{rest}$ wavelengths following Equation\ \ref{clump.eq}, have a median SFR-weighted age of 130 Myr and 150 Myr respectively, a factor 2.6 respectively 2.2 younger than the underlying disk.  At $z \sim 1$, both the disk and the clumps are characterized by older ages.  Pixels in the $U_{rest}$-selected clump regime have a median inferred age of 370 Myr during this epoch, and are younger by a factor 2.5 than the underlying disk.  A slightly smaller factor of 2.3, and older SFR-weighted age of 400 Myr is obtained for $V_{rest}$-selected clumps at $z \sim 1$.

To explore the robustness of our results with respect to the assumptions made regarding SFHs (see Section\ \ref{resolved.sec}), we tested the impact of three parameters separately: the minimum age since the onset of star formation (by default set to 50 Myr), the minimum e-folding time (by default 300 Myr), and the function form (by default $\tau$ models).  Lowering the minimum age since the onset of star formation to 5 Myr has the effect of decreasing the typical inferred clump ages by $\sim 0.15$ dex.  Likewise, lowering the minimum $\tau$ to 30 Myr (i.e., allowing more rapidly dropping SFHs) leads to clump ages that are younger by 0.1 - 0.3 dex.  On the other hand, delayed $\tau$ models (with $SFR(t) = t\ \exp(-t/\tau)$, i.e., featuring rising SFRs for $t < \tau$) produce slightly older ages, by $\sim 0.1$ dex.  In most cases, however, the inferred ages of the center and outer disk regions change by a similar amount.  Consequently, the relative stellar population properties of the three galaxy components are more robust.

Not all color variations are assigned to spatial variations in the stellar age.  This becomes evident when color-coding the light and mass profiles by the inferred visual extinction $A_V$ (see Figure\ \ref{respop3_Av.fig}).  A clear trend emerges where the emission from stars inside the half-light (half-mass) radius is significantly more extincted than the light emerging from the outer disk.  The median $A_V$ of pixels inside $R_e$ is 1.3, while the emission originating from stars outside 2 half-light radii suffers $\sim 0.7$ mag less extinction.  As we will discuss in Section\ \ref{toymodel.sec}, the presence of a dust gradient is to be expected in a scenario where star-forming disks grow inside-out.  If star formation has a longer history in the central regions, its associated stellar mass loss will have been more efficient at enriching the interstellar medium (ISM), producing a thicker metal and thus dust column that causes the obscuration.

\subsection{Galaxy structure in light and mass}
\label{light_mass.sec}

\begin{figure*}[t]
\centering
\includegraphics[width=0.48\textwidth]{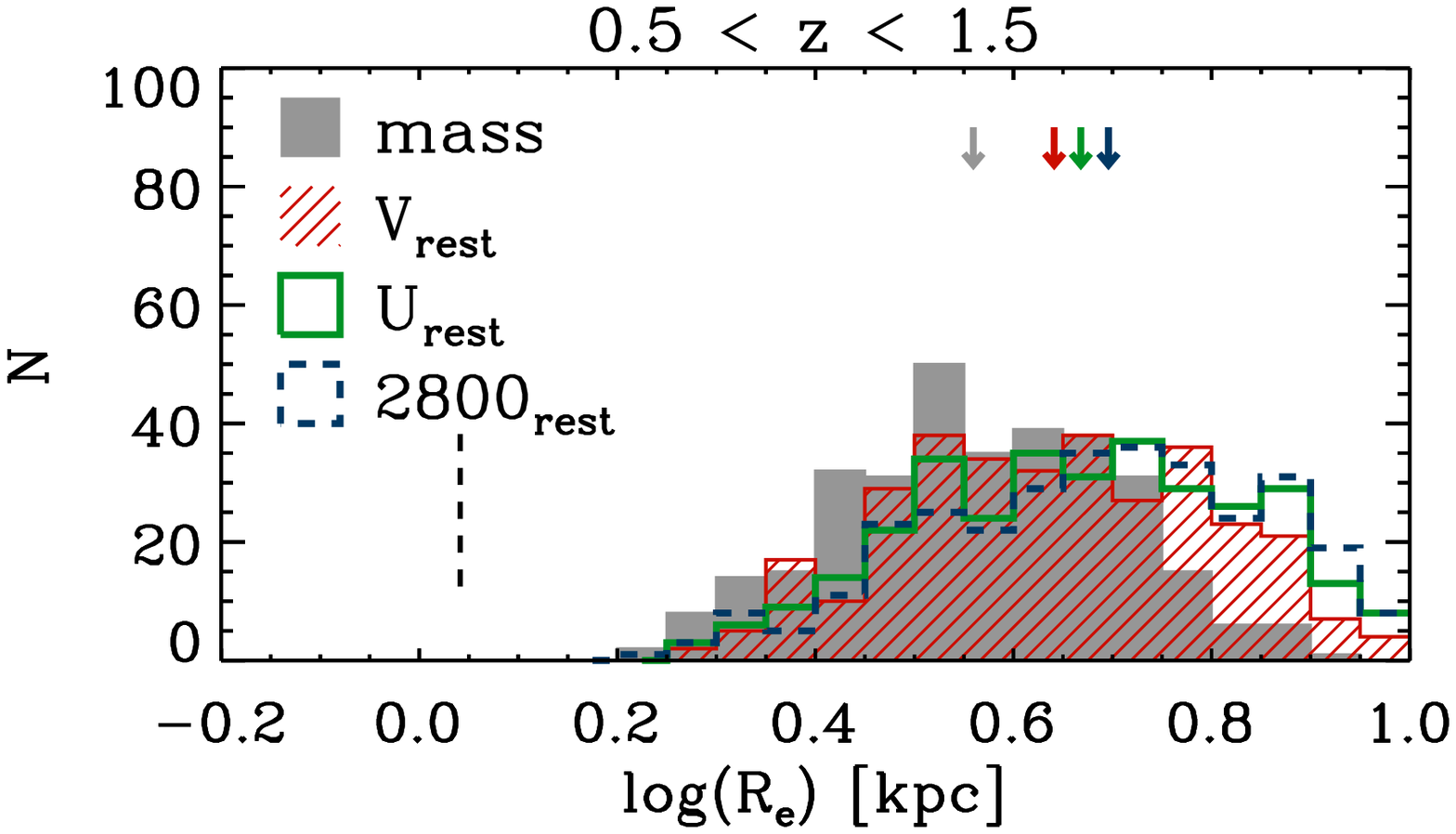}
\vspace{0.1in}
\includegraphics[width=0.48\textwidth]{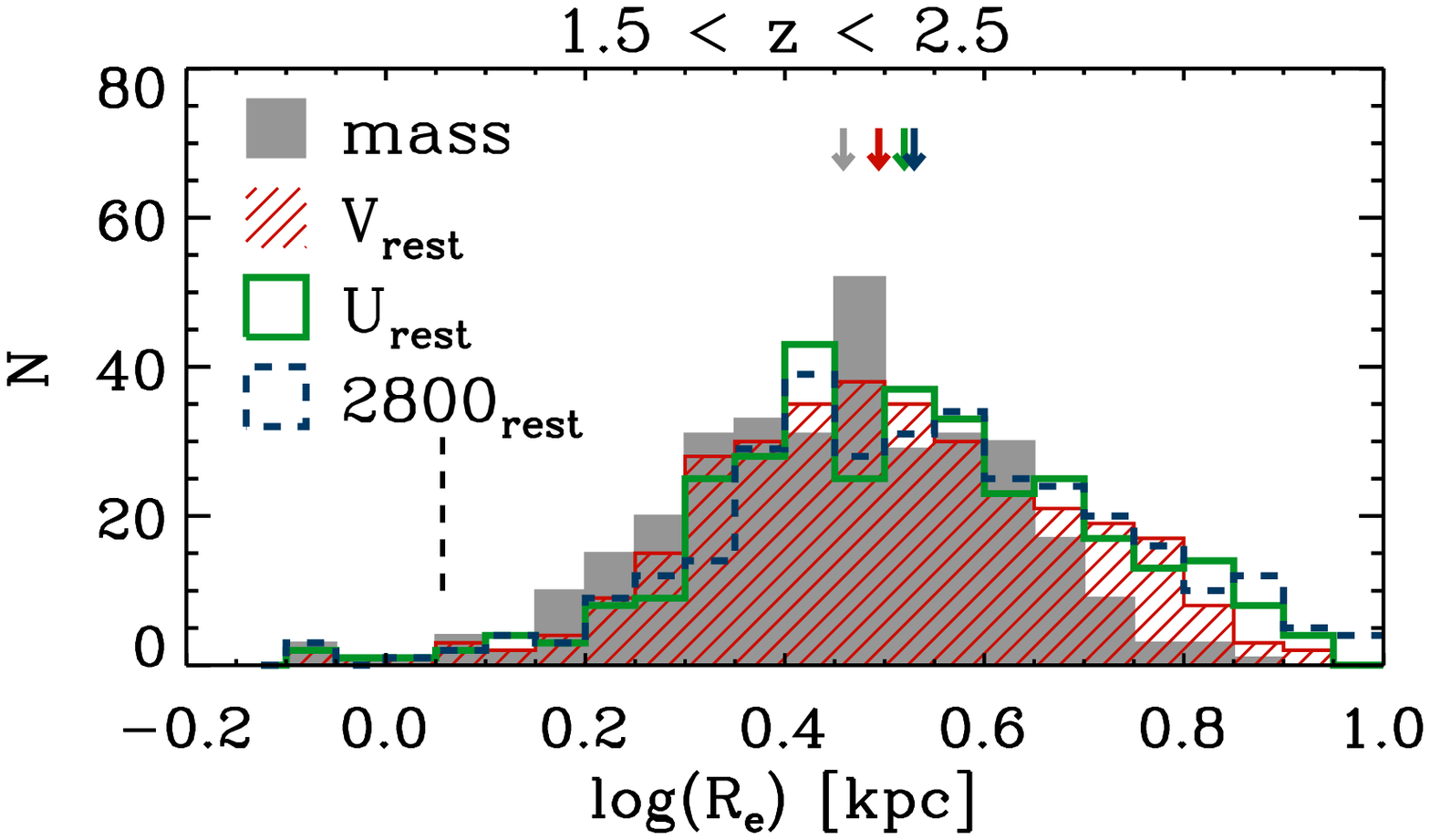}
\vspace{0.1in}
\includegraphics[width=0.48\textwidth]{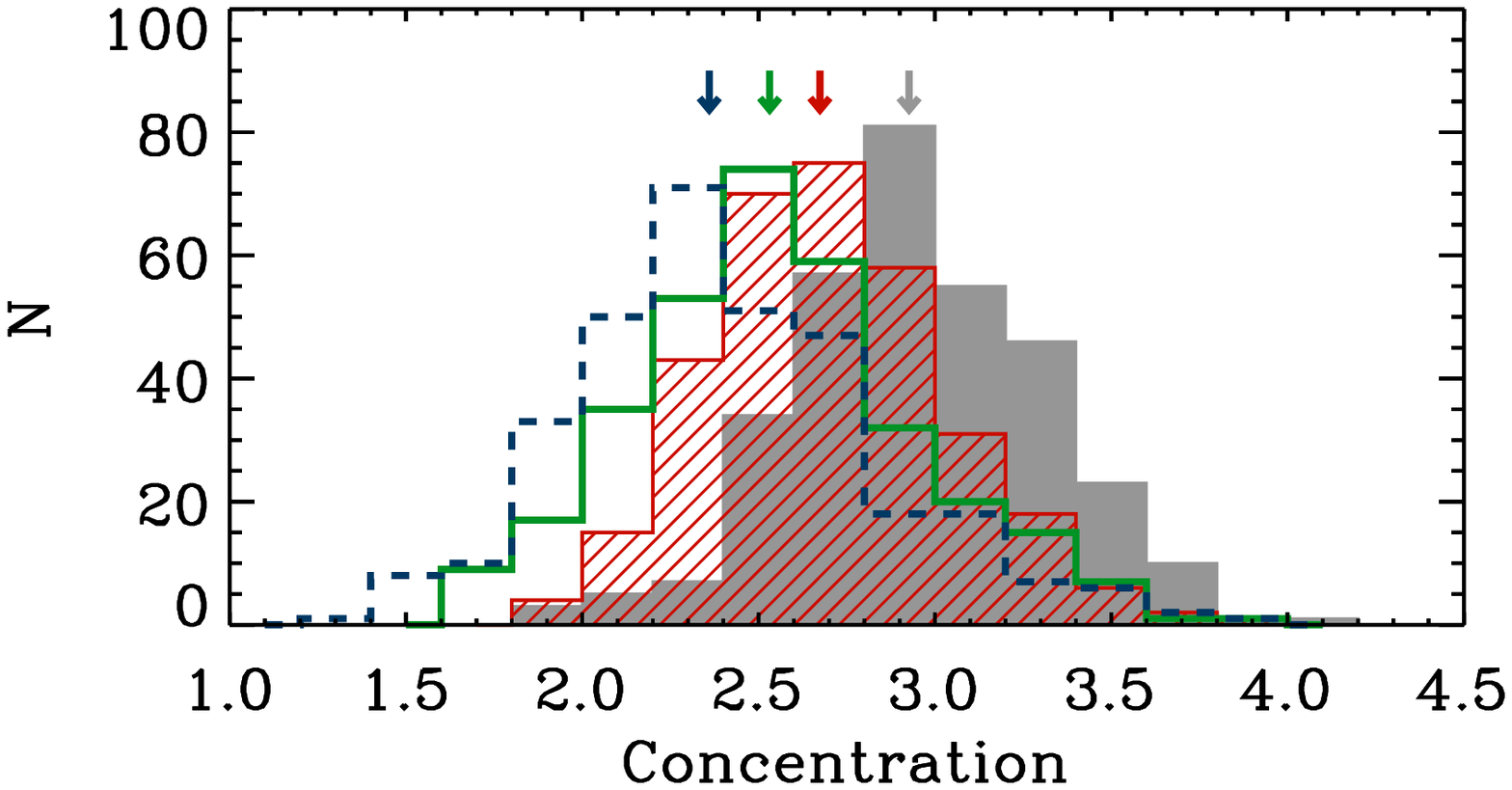}
\vspace{0.1in}
\includegraphics[width=0.48\textwidth]{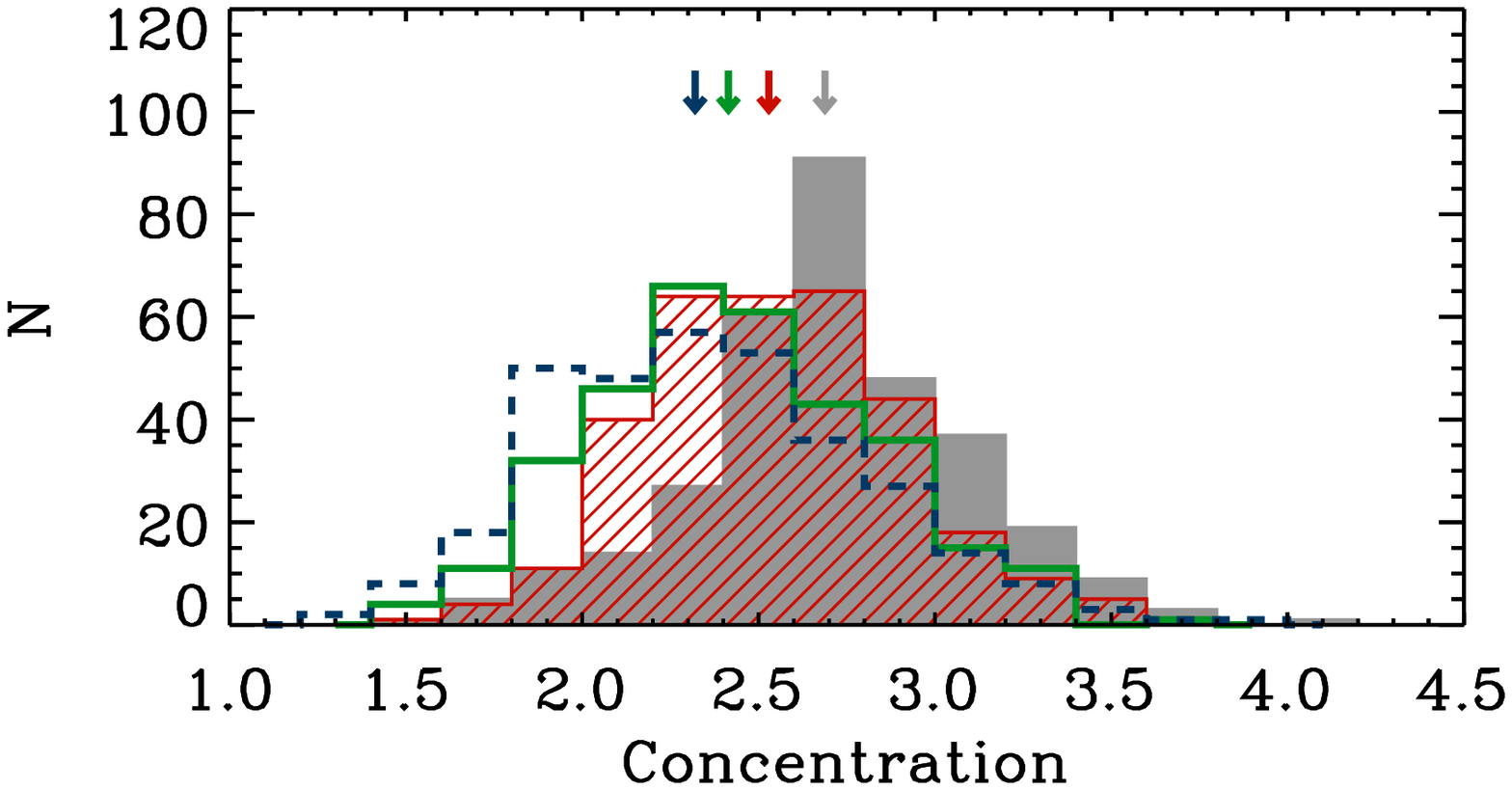}
\vspace{0.1in}
\includegraphics[width=0.48\textwidth]{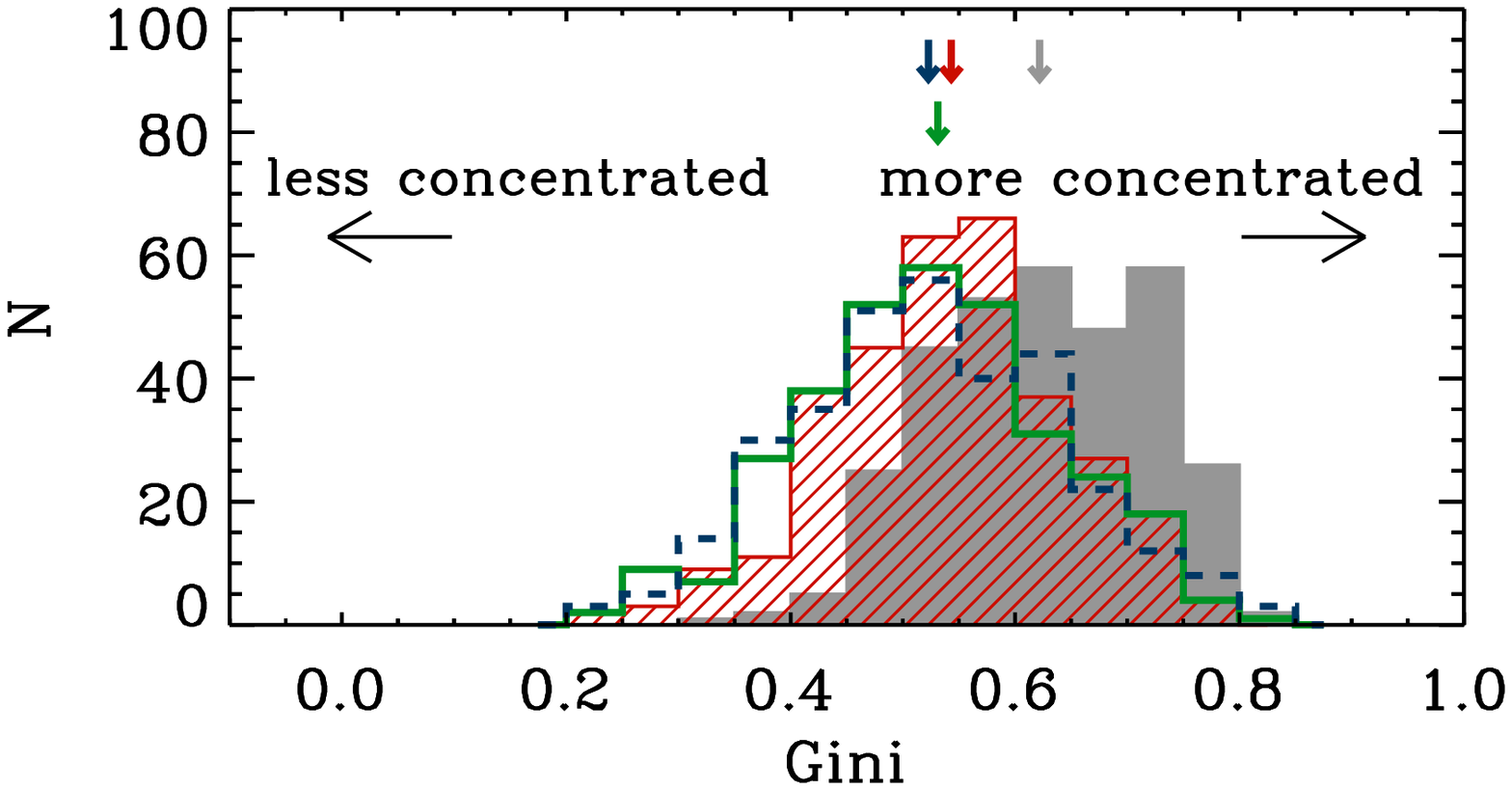}
\vspace{0.1in}
\includegraphics[width=0.48\textwidth]{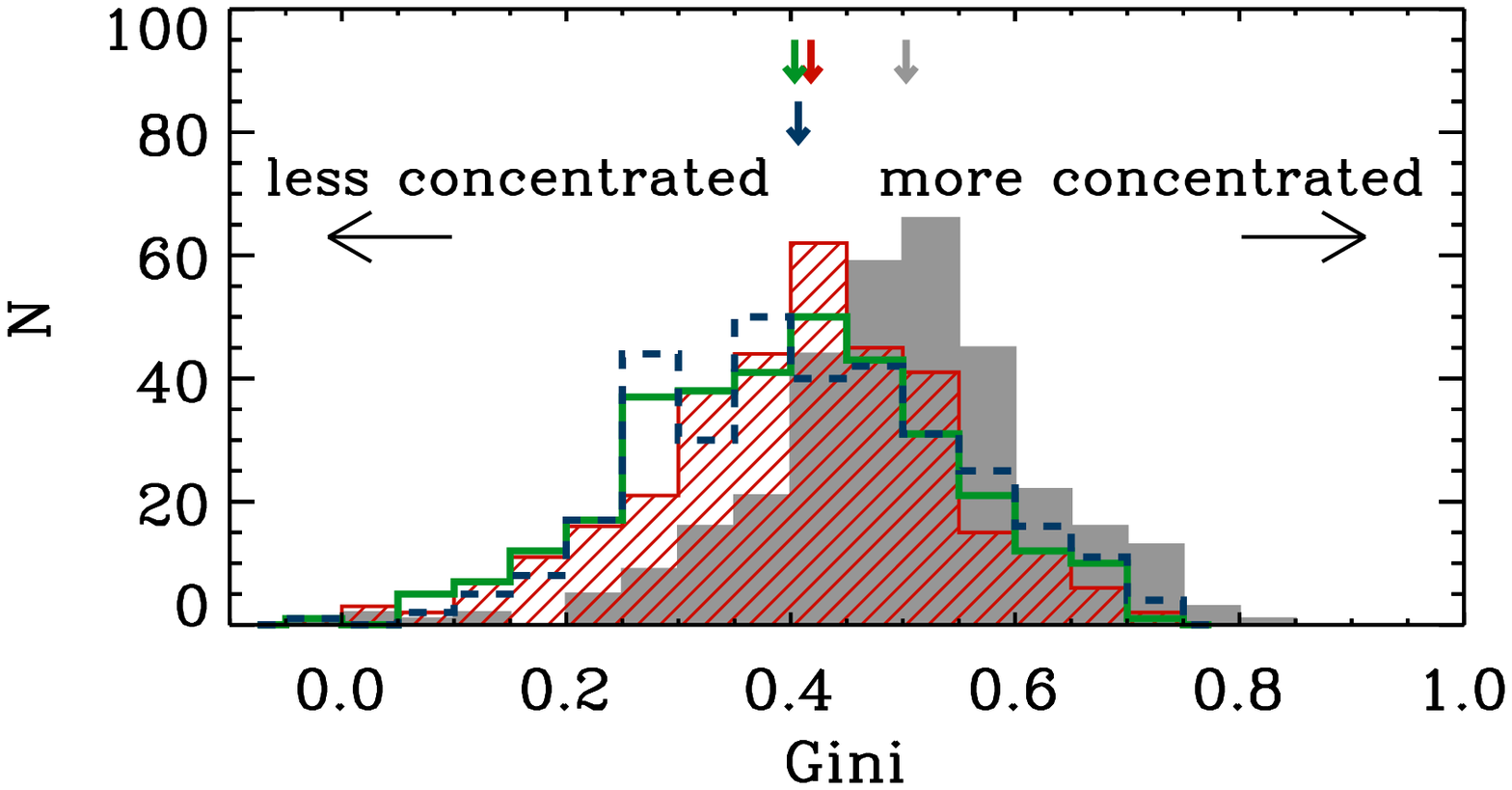}
\vspace{0.1in}
\includegraphics[width=0.48\textwidth]{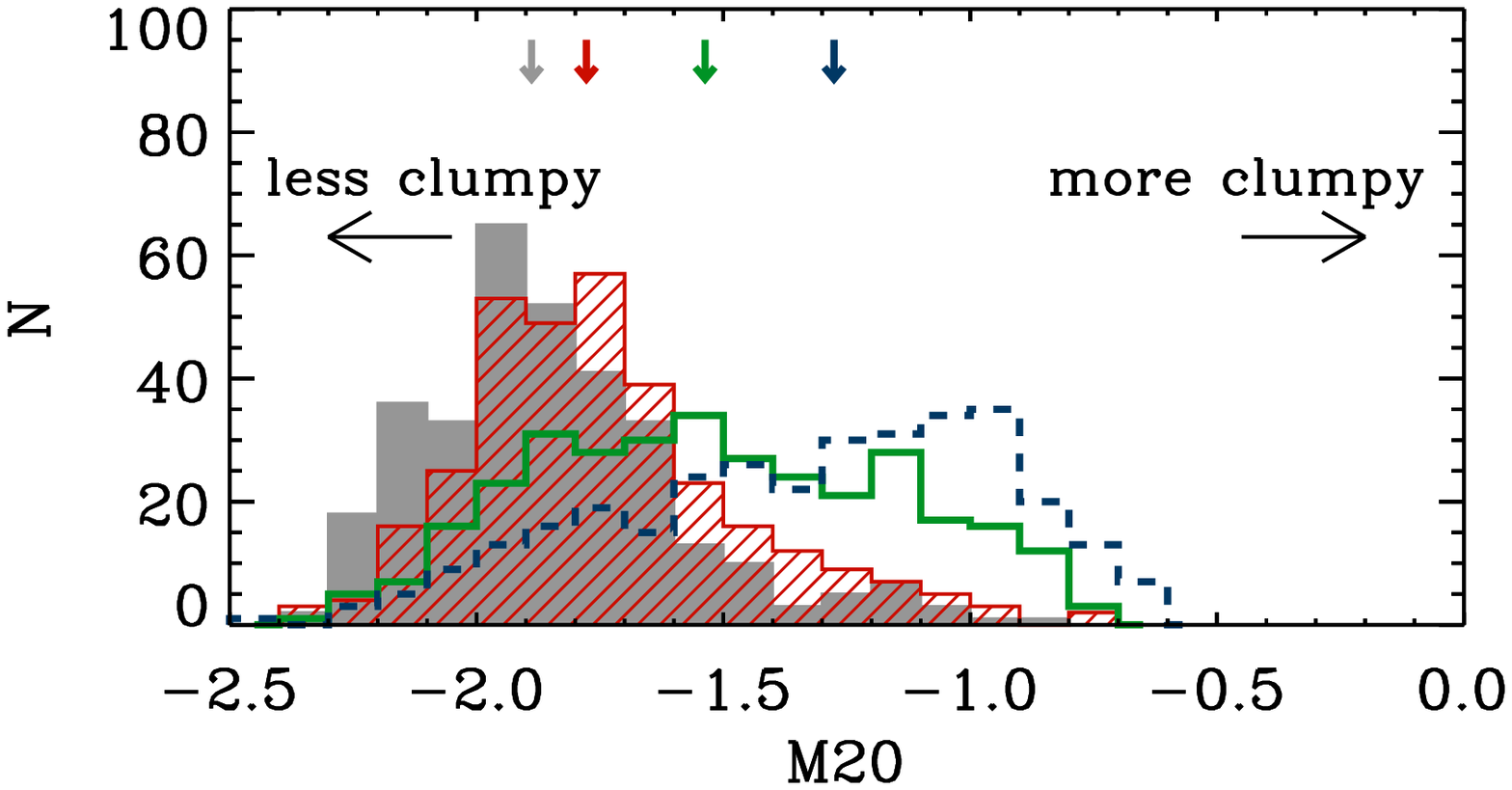}
\vspace{0.1in}
\includegraphics[width=0.48\textwidth]{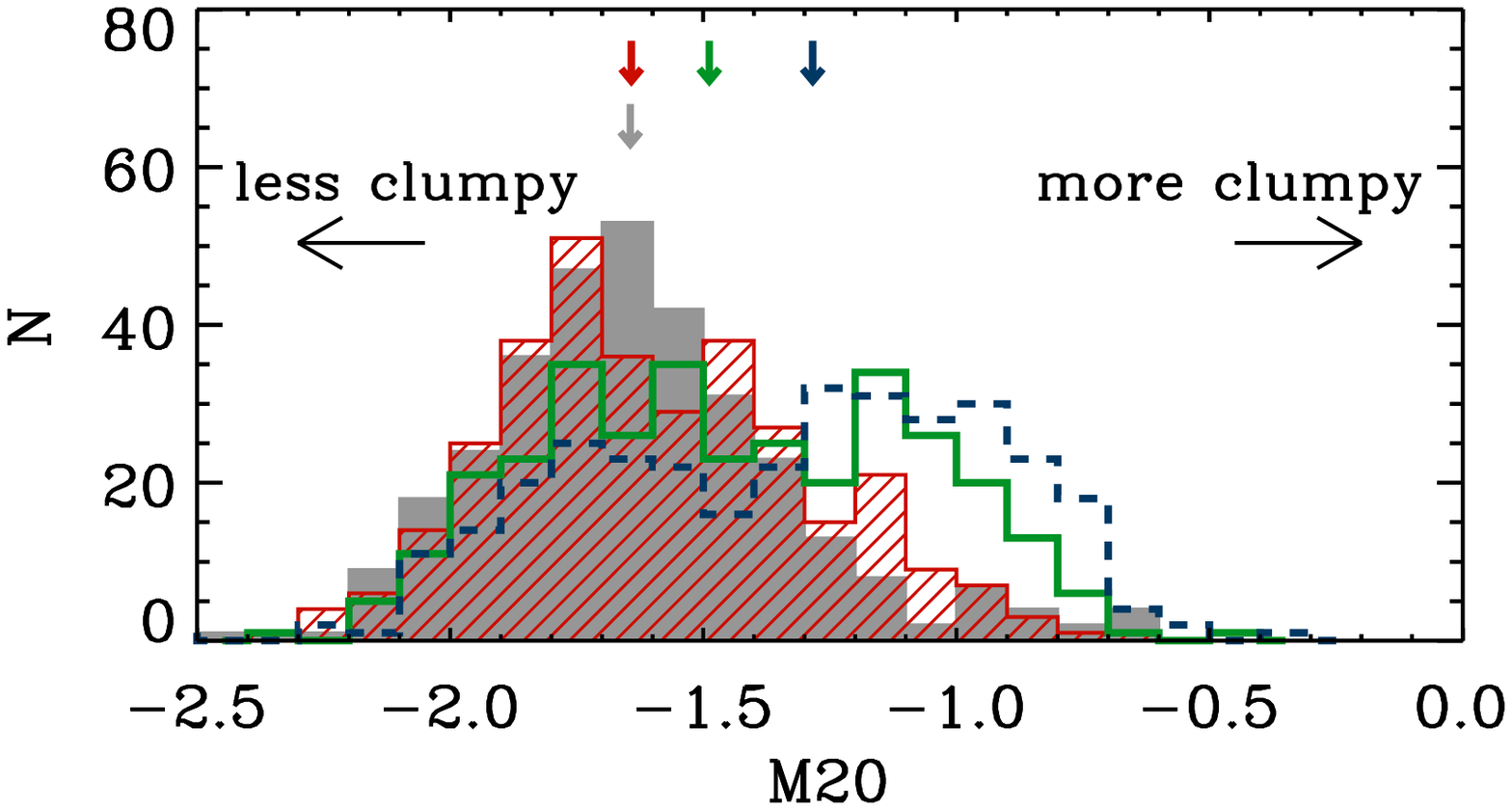}
\caption{
Structural parameters of galaxies at $0.5 < z < 1.5$ ({\it left-hand panels}) and $1.5 < z < 2.5$ ({\it right-hand panels}), as measured on maps of the rest-frame $2800\AA$, $U$, $V$, and stellar mass distribution.  The vertical dashed line indicates the physical scale corresponding to the half-light radius of the WFC3 PSF at the median redshift of the sample.  A change in structure is noted from $2800\AA_{rest}$ over $U_{rest}$ and $V_{rest}$ to stellar mass, in that the median galaxy size and M20 coefficient reduce, while measures of concentration and Gini increase.
\label{structure.fig}}
\end{figure*}

Over the past decade, much effort has been invested in studying the structural properties of high-redshift galaxies, and their evolution with cosmic time.  Advances in this field have been enabled by the advent of near-infrared high-resolution cameras onboard HST (first NICMOS, and now WFC3), which in combination with the optical ACS camera in principle allow one to trace galaxy structure at a fixed rest-frame wavelength over a broad range of lookback times.   However, to our knowledge no measurements of size, concentration, Gini and M20 have been carried out to date on stellar mass maps that were reconstructed from multi-wavelength high-resolution imaging.  We present the results of such an analysis in this Section.

The distribution of half-mass radii, concentrations, Gini and M20 coefficients as determined on the stellar mass maps following the procedures outlined in Section\ \ref{structure.sec} is contrasted with the equivalent measurement carried out on the rest-frame $2800\AA$, $U_{rest}$ and $V_{rest}$ surface brightness maps in Figure\ \ref{structure.fig}.  The comparison should be regarded as a relative one.  As such, it is fair because all three maps were derived from the same $BVizYJH$ images that were PSF-matched to the WFC3 resolution, and subject to the Voronoi pixel binning routine described in Section\ \ref{binning.sec}.  In detail, we carried out the structural measurements on pixelized maps in which the light or mass of a given Voronoi bin was distributed evenly over its constituent pixels.  Nevertheless, the M20 coefficient may in some cases be biased to lower values due to the binning.  More generally, the absolute values of the structural parameters should not be interpreted as purely physical because we did not do any attempt to correct for beam smearing by the PSF.  Beam smearing leads to larger sizes, smears out central cusps, and can blend clumpy features that would otherwise be picked up by the M20 coefficient (see Appendix C).  However, it does so to the same degree for the mass, $V_{rest}$, $U_{rest}$, and $2800\AA_{rest}$ maps, and it is therefore not a relevant concern for the point that we aim to make here.  Namely, the distribution of half-mass radii is shifted by 0.1 to 0.2 dex with respect to that of the half-light radii.  The concentration of the mass profiles on the other hand is boosted with respect to that of the light profiles, by 10 - 15\% in the median in comparison to the concentration of the $U_{rest}$ light.  The Gini coefficient, as an independent measure of concentration that does not rely on the definition of a galaxy center, paints a consistent picture.  While the Gini distribution of the $2800\AA_{rest}$ to $V_{rest}$ light profiles does not differ significantly, a larger fraction of the SFGs is characterized as highly concentrated (i.e., bulge-dominated) in terms of their stellar mass distribution, with the median of the mass-based Gini coefficients being larger by 0.2.  Finally, the M20 coefficient, which is frequently used to signal the presence of multiple peaks in the surface brightness distribution of galaxies, clearly extends to larger values (i.e., a higher degree of clumpiness) when quantified at short wavelengths, rather than at long wavelengths or from the inferred stellar mass distribution.

We conclude that an assessment of the physical structure of high-redshift SFGs benefits from a multi-wavelength approach.  Even structural properties derived from rest-frame $V_{rest}$-band light profiles, which thanks to CANDELS and other surveys now become readily available for statistically significant samples, can differ significantly from those of the distribution of stellar mass.  The latter is a physically more relevant parameter, which can be compared more straightforwardly with predictions from models.  Resolved SED modeling can serve as a powerful tool to reach this goal.  Alternatively, modelers should apply a realistic treatment of the intrinsic $M/L$ ratios of the stellar populations in combination with radiative transfer in order to confront their model predictions with monochromatically derived structural properties.  In recent years, the methodology to bridge the gap between the physically modeled galaxy properties and their appearance in light has been developed by, among others, Jonsson (2006).  Applications of the SUNRISE radiative transfer code (Jonsson 2006; Jonsson et al. 2010) for an apples to apples comparison between observed and simulated (structural) galaxy properties at $z \sim 2$ have been presented by Wuyts et al. (2009; 2010).  Similar to what is found for SFGs in this paper, Wuyts et al. (2010) conclude on the basis of simulations combined with radiative transfer that quiescent $z \sim 2$ galaxies as well may be more compact in terms of their mass distribution than they appear in emission, due to the presence of a $M/L$ ratio gradient (see Guo et al. 2011a for observational support).

\section {Discussion}
\label{discussion.sec}

Throughout the paper, when analyzing the normalized co-added profiles of high-redshift SFGs, we addressed both radial variations in color and stellar population parameters, and variations with surface brightness/density at a given radius.  In Section\ \ref{nature_fate.sec}, we first discuss the implications of the latter trends for the nature and fate of clumps.  Next, Section\ \ref{toymodel.sec} sketches a cosmological framework for galaxy growth that can serve as context for the observed radial trends.

\subsection {The Nature and Fate of Clumps}
\label{nature_fate.sec}

\begin{figure}[t]
\centering
\plotone{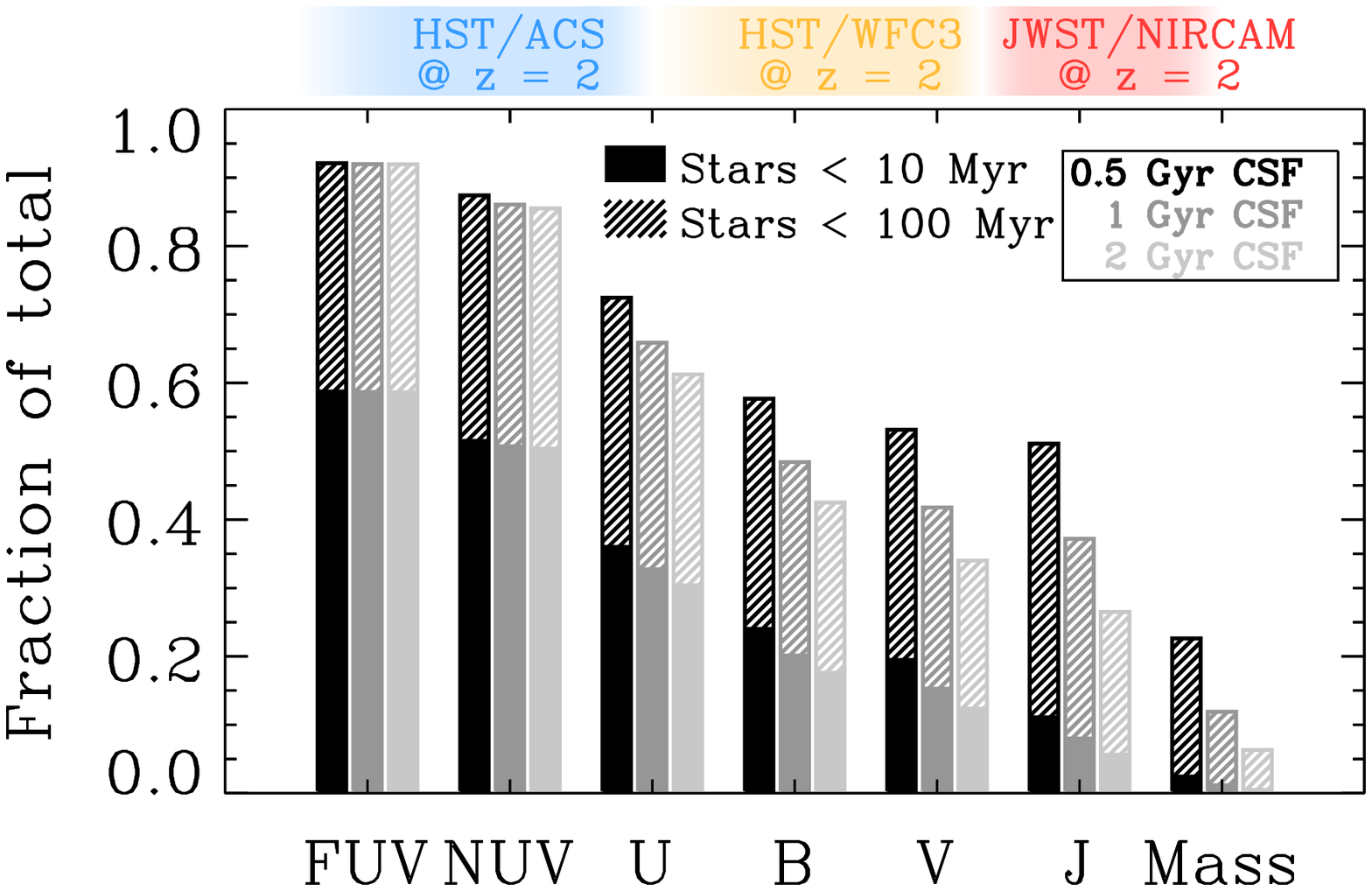}
\caption{
Fractional contribution of stars younger than 10 Myr and 100 Myr to the total emission at rest-frame FUV to $J$-band wavelengths, and to the total mass present in stars.  Gray shades indicate BC03 models with constant SFHs that started 0.5 Gyr, 1 Gyr and 2 Gyr prior to the epoch of observation.  While the rest-frame $V$-band provides a better proxy for stellar mass than the UV part of the SED, it is still substantially biased towards the youngest generation of stars.  Even with JWST, which will provide rest-frame near-infrared imaging of $z \sim 2$ galaxies at HST resolution, monochromatic light distributions may deviate significantly from the underlying mass profiles.  Resolved color information is the key to constraining spatial $M/L$ variations due to non-uniform SFHs (or attenuation) across a galaxy, and recovering the true distribution of stellar mass.
\label{contribution.fig}}
\end{figure}

In order to interpret our observational results on resolved colors and stellar populations of SFGs out to $z = 2.5$, it is useful to consider an idealized case of a galaxy that has been forming stars at a constant rate.  At first, let us assume that the SFH does not vary spatially.  If the constant star formation has been going on for 1 gigayear, the relative fraction of the mass in stars present at the epoch of observation that was contributed during the last 10 Myr (100 Myr) is small: 1.4\% (12\%).  Here, we accounted for the fact that the stars formed earlier on have ejected a larger fraction of their mass back into the ISM, using the time dependence of stellar mass loss for a Chabrier IMF as given by BC03.  It is a well-known fact that the emerging light on the other hand, is more dominated by the strong emission of hot, massive and short-lived O and B stars, and increasingly so at shorter wavelengths (see, e.g., Maraston et al. 2010).  This effect is illustrated quantitatively in Figure\ \ref{contribution.fig}.  The rest-frame UV emission (probed by ACS at $z \sim 2$) from stars younger than 10 Myr outweighs the entire integrated UV luminosity from all stars formed during the 990 Myr beforehand.  In fact, the last 100 Myr of star formation is responsible for roughly 90\% of the UV emission for the assumed SFH.  Figure\ \ref{contribution.fig} demonstrates quantitatively how the rest-frame optical wavebands, probed by WFC3 at $z \sim 2$, are significantly less dominated by the youngest generation of stars than the UV.  However, it is important to realize that the fractional contribution of stars younger than 10 Myr (100 Myr) in the $V_{rest}$ band is still a factor of 11 (3.5) larger than to the total stellar mass, amounting to 16\% (42\%) of total.  In the absence of spatial variations in the SFH, the distribution of $V_{rest}$ light will still relate one-to-one to that of the stellar mass.  If, however, galaxies grow for instance inside-out, this is no longer true.  Color corrections, such as the resolved SED modeling procedure presented in this paper, then need to be applied in order to determine the physical structure.  The same is true if clumps with excess star formation are superposed on the underlying disk, as implied by the results presented in this paper.  It can also be appreciated (or perhaps not appreciated) from Figure\ \ref{contribution.fig} that even in the era of JWST, which will provide rest-frame near-infrared imaging at HST resolution, taking into account color information will be essential to reconstruct the actual distribution of stellar mass from the observed light profiles.

Exploiting the numbers that are graphically summarized in Figure\ \ref{contribution.fig}, we now consider what happens to our idealized galaxy model when a local excess by a factor 5 in the gas surface density occurs 10 Myr prior to the epoch of observation.  Physically, such deviations from a uniform gas surface density across the disk are likely to occur by gravitational instabilities in high-redshift SFGs (Bournaud et al. 2008; Genzel et al. 2008; Dekel et al. 2009b), because they are so gas rich (see, e.g., Tacconi et al. 2010).  Assuming the power law slope of $N \approx 1.4$ of the Kennicutt-Schmidt relation (Kennicutt et al. 2007), the excess in $\Sigma_{gas}$ translates to a boost in $\Sigma_{SFR}$ during the last 10 Myr by a factor 9.5.  The amount of stellar mass that was formed during the clump's life consequently amounts to 12\% of the total stellar mass present at the location of the clump at the epoch of observation.  In light, however, the enhancement in surface brightness with respect to the underlying, unperturbed disk, extends to a factor 2.3, 3.8, and 5.3 at $V_{rest}$, $U_{rest}$, and $2800_{rest}$ wavelengths respectively.  

For a clump that originated 100 Myr ago, the surface mass density will be in excess by a factor 2 with respect to the underlying disk, while the excess increases to factors of 4.5, 6.6, and 8.3 for the $V_{rest}$, $U_{rest}$, and $2800_{rest}$ surface brightness levels respectively.  The $(U-V)_{rest}$ color of the clump in this toy model will be 0.4 mag bluer than the underlying disk.\footnote{In detail, real galaxies may deviate from our dust-free toy model in that they follow a correlation between $\Sigma_{SFR}$ and the local visual extinction (compare Figures\ \ref{respop3_SFR.fig} to\ \ref{respop3_Av.fig}), which would weaken the excess in surface brightness and blue color of the clumps, or require a higher enhancement in $\Sigma_{gas}$ to produce the same effect.}  Keeping in mind that, for the sketched SFH at the clump location (constant star formation with a recent burst superposed) the outshining effect (see, e.g., Papovich et al. 2001; Shapley et al. 2005; Wuyts et al. 2007) may inhibit a full recovery of the stellar mass present in old stars, we obtain a consistent picture with our observational results.  Clumps become progressively less prominent with increasing wavelength, and leave little to no imprint on the distribution of stellar mass.  The latter aspect would change if the star-forming clumps would survive at their radius and continue building stars at the elevated rate for several 100 Myr longer.  Since this is not observed, we can conclude that their fate is sealed on a timescale of 100 - 200 Myr, consistent with the median SFR-weighted age of clumps in $z \sim 2$ SFGs inferred from resolved SED modeling in Section\ \ref{lagew.sec}, as well as recent findings based on smaller samples (Elmegreen et al. 2009; F\"{o}rster Schreiber et al. 2011b; Genzel et al. 2011; Guo et al. 2011b).

This fate can take two possible forms.  A first possible interpretation is that the clumps represent a short-lived star-forming phase, that comes to an end through feedback from the same star formation event that constitutes its existence.  Indeed, strong outflows emanating from individual clumps have already been observed as a broad component in the line emission of the ionized gas of non-AGN $z \sim 2$ SFGs (Genzel et al. 2011; Newman et al. 2012).  These authors inferred mass loading factors of several times the SFR from their IFU observations, implying short lifetimes before the clumps are dispersed.  Furthermore, even in the deepest integral-field data, Genzel et al. (2011) find no prominent kinematic imprint of a local minimum in the potential at the location of the clumps, consistent with our observational results.  Theoretical support for such short-lived star-forming clumps has been presented on the basis of smoothed particle hydrodynamics (SPH) cosmological zoom-in simulations (Genel et al. 2012) as well as complementary SPH simulations with pc-scale resolution of isolated high-redshift, gas-rich disks (Hopkins et al. 2011).  The latter include feedback in the form of momentum imparted on sub-GMC scales from stellar radiation pressure, radiation pressure on larger scales from light that escapes the star-forming regions, HII photoionization heating, as well as the heating, momentum deposition, and mass loss by supernovae and stellar winds from O and AGB stars, implemented with the time-dependence given by stellar evolution models.  The combination of these different processes, and interplay between them, results in a highly efficient feedback which disrupts the clumps on timescales of $\sim 10 - 100$ Myr.  Genel et al. (2012) come to similar conclusions (and consistent timescales) for their cosmological zoom-in simulations that do contain strong momentum-driven winds.  Both studies find a reduction in the bulge formation rate through inward migration and clump coalescence by a factor of several compared to simulations that only include the thermal heating by supernovae (e.g., Bournaud et al. 2008; Ceverino et al. 2010, 2012).  This does not necessarily mean that the build-up of bulges through internal processes should be discarded.  Several recent theoretical studies argued that, even if the clumps do not remain bound, torques in unstable disks will lead to gas inflow (Krumholz \& Burkert 2010; Bournaud et al. 2011; Cacciato et al. 2012).

This brings us to the second possible fate of off-center clumps that are typically formed only recently.  More evolved clumps may be rare in the outer disk, not because they were disrupted before they reached a later stage in their evolution, but because they migrated inward by that time.  A prediction from such a migration scenario that has been discussed as an efficient means to form bulges (e.g., Bournaud et al. 2007; Genzel et al. 2008; Dekel et al. 2009b; Ceverino et al. 2010), is that the age, and hence color of clumps should show a negative correlation with galactocentric radius.  F\"{o}rster Schreiber et al. (2011b) and Guo et al. (2011b) reported on observations that are at least qualitatively consistent with the predicted trend.  The redder colors of the nuclei (which, in the nomenclature adopted in this paper, are not included in what we refer to as clumps) compared to the off-center clumps could be interpreted in the same regard.  However, as we will discuss in Section\ \ref{toymodel.sec}, and was already pointed out by Genel et al. (2012), the observed trend could be expected also from an inside-out growth scenario of star-forming disks that lack radial migration.  A constraint from the present study, and specifically the young age of off-center clumps, is that inward migration, if at play, should happen fast.  While the 400-500 Myr migrational timescales predicted by earlier simulations of isolated galaxies undergoing violent disk instabilities (Bournaud et al. 2007; see also Noguchi 1999) may be in tension with the observational constraints, more recent state-of-the-art cosmological simulations that determine the migration rate in a cosmological steady state predict faster migration times, of about 1 - 1.5 orbital times, or typically 200 - 250 Myr at $z \sim 2$ (Ceverino et al. 2010, 2012; Cacciato et al. 2012).  Similar timescales are obtained on the basis of analytical arguments (Dekel et al. 2009b; Bournaud et al. 2011; Cacciato et al. 2012).  In fact, the range of clump ages of 100 - 200 Myr reported by Ceverino et al. (2010) corresponds very well to the estimates derived in this paper.  One could even speculate that the inferred increase by a factor $\sim 3$ in typical clump age between $z \sim 2$ and $z \sim 1$ stems from an increased migrational timescale:

\begin {equation}
t_{mig} \sim 1.5\ t_{orbit}\ (f_{gas}/0.5)^{-2},
\end {equation}

where the orbital time scales as $t_{orbit} \sim (1+z)^{-3/2}$ and the gas fraction roughly as $f_{gas} \sim (1+z)$ (Dekel et al. 2009b), although an increased fraction of background light from the older and more prominent underlying disk at $z \sim 1$ may contribute as well.

Aside from implications for individual star-forming clumps, the efficiency of feedback processes implemented in numerical simulations may also impact global galaxy properties, such as the integrated gas mass fraction, or the overall stellar profile.  Adopting the definition of (molecular) gas mass fraction from Tacconi et al. (2010):
\begin {equation}
f_{gas} = \frac{M_{gas}}{M_{gas} + M_{*, tot}},
\end {equation}
where $M_{*, tot}$ is the total stellar mass (i.e., in bulge + disk), we derive average gas mass fractions for the simulated $z = 2.3$ galaxies in Ceverino et al. (2010, 2012) of 10\%, with a range from 4\% to 18\%.  For comparison, Tacconi et al. (2010) report an average $f_{gas} \sim 0.44$ for a sample of 10 normal SFGs at the same mean redshift.  The gas mass fractions in the simulations by Genel et al. (2012; $\sim 43\%$) and Hopkins et al. (2011) correspond more closely to what is observed, although it should be stressed that in the latter case this is entirely by construction, since those simulations do not follow the evolution in a cosmological context.   The simulated $z \sim 2.3$ galaxies by Ceverino et al. (2010, 2012) have $36-76\%$  more stars in the bulge than in the disk component.  Recent observations using, among other data sets, deep CANDELS WFC3 imaging suggest that the rest-frame optical surface brightness profiles of typical SFGs at $z \sim 2$, similar to SFGs today, are well characterized by exponential disks (Wuyts et al. 2011b; Bell et al. 2011).  Quiescent galaxies on the other hand have cuspier surface brightness profiles, also similar to what is seen in the nearby universe.  In other words, highly efficient bulge formation, through clump migration and additional gas inflows by torquing, should either go hand in hand with quenching of the star formation, or should be compensated by continuous replenishment and growth of the gas disk to keep the overall profile disk dominated.  An apples-to-apples comparison of observed galaxy profiles, and simulated ones that are mock-observed in the same waveband, will be presented by Mozena et al. (in prep).

We conclude that our observations are consistent with a so-called "Christmas tree" model (Lowenthal et al. 1997) in which the global star formation history of SFGs may be smooth, but on local scales episodic fluctuations in the star formation activity (e.g., clumps) occur.  Phenomenologically, this is akin to spiral galaxies locally, that show an enhancement in star formation and hence bluer colors at the location of the spiral arms compared to the interarm regions, and look smoother at longer wavelengths.  Physically, the processes responsible for the onset and end of the excess in $\Sigma_{SFR}$ is different: a spiral density wave passing by in the case of local disks, compared to gravitational instabilities and subsequent stellar feedback in high-redshift gas-rich disks.  In this sense, the evolution at high redshift could be regarded as more analogous to the lifecycle of individual GMCs (Murray 2011 and references therein).  At the same time, the bluer colors of star-forming clumps with respect to the underlying disk, and even quantitatively their inferred ages, are also consistent with a scenario of rapid clump migration within violently unstable disks.  A synergetic approach combining constraints from individual clump properties with global constraints from the overall stellar distribution and integrated gas mass fraction will be required to adequately discriminate between the two scenarios.

\subsection {Inside-out Disk Galaxy Growth}
\label{toymodel.sec}

\begin{figure}[t]
\centering
\plotone{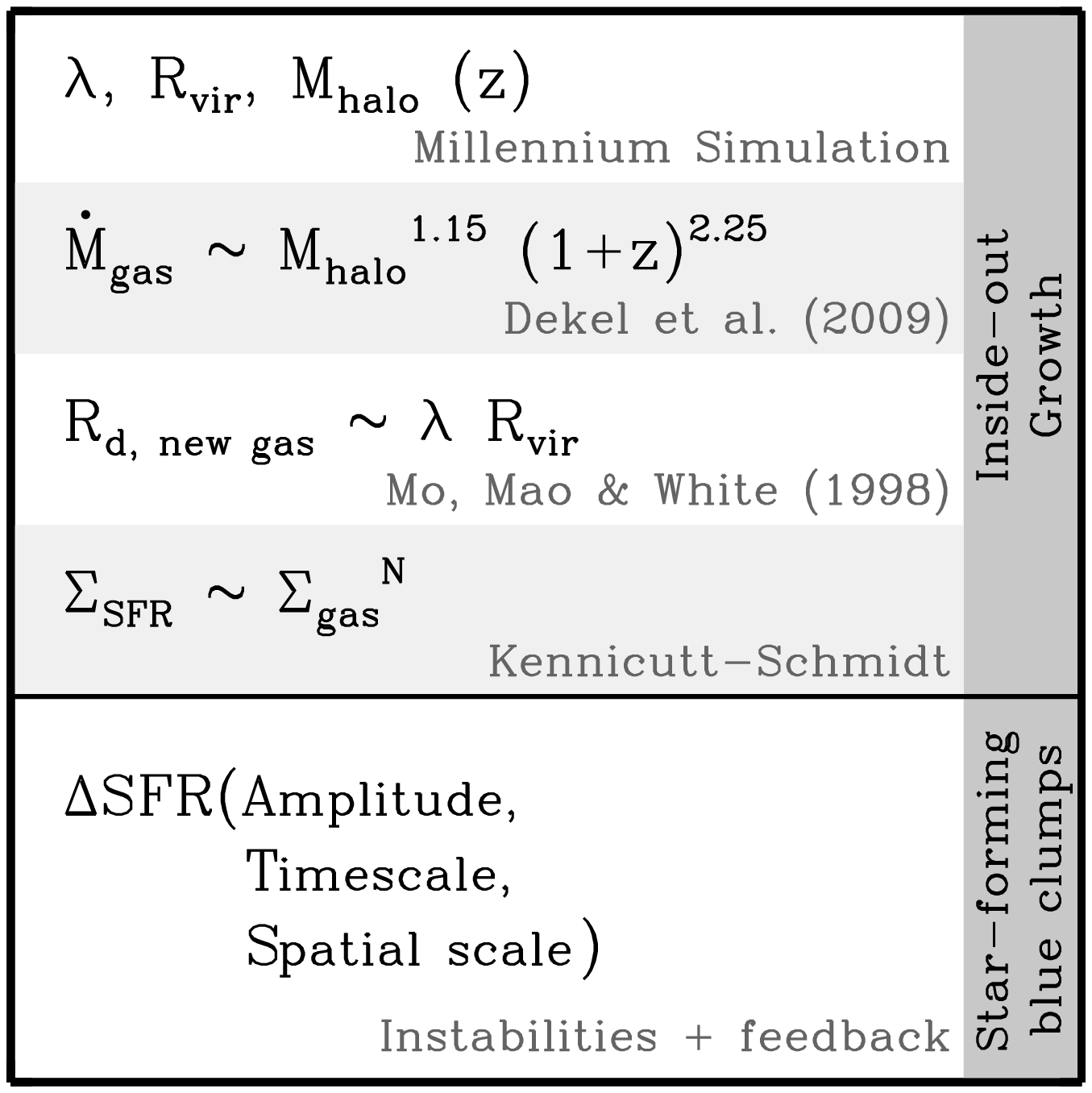}
\caption{
Schematic overview of a simple toy model for disk growth and spatial color variations without radial migration.  Newly accreted gas settles in an exponential disk with a scalelength proportional to the virial radius at that epoch.  Due to halo growth, this scalelength is larger than that of previously accreted gas which has been forming stars, causing an inside-out growth of the stellar disk.  Short-lived, spatial fluctuations in the SFR are required to explain the relation between color and surface brightness variations  at a given radius, which is most pronounced at short wavelengths, and absent in the stellar mass distribution.
\label{scheme.fig}}
\end{figure}

We now tie in the above discussion on star-forming blue clumps with the overall growth of SFGs over cosmic time, using a simple model summarized in Figure\ \ref{scheme.fig}.  Our model does not include torques or inward clump migration within the galaxy disk, even though such processes may well be at play in reality.  It should therefore only be considered as a test to what degree a minimum model would produce trends that are in line with the observations.  Here, we describe qualitatively the trends expected from this model.  In an upcoming paper, we relate its predictions quantitatively to the observed resolved characteristics of SFGs, as well as scaling relations of their global properties.  Our model, which shares several ingredients with that by Bouch\'{e} et al. (2010) and Dutton et al. (2010), is rooted in the cosmic web, with the evolution of the dark matter halos taken from the Millennium Simulation (Springel et al. 2005).  Dekel et al. (2009a) formulated an analytic approximation of the gas accretion rate onto a given halo, which scales slightly above linearly with the halo mass, and with redshift as $\sim (1+z)^{2.25}$ (see also Neistein et al. 2006; Genel et al. 2008).  While at a given mass, accretion rates decrease with cosmic time, the growth in halo mass is typically such that the gas accretion on an individual halo stays roughly constant down to the epoch of our observed $z \sim 2$ SFGs.  Analogous to Fu et al. (2010), we let the newly accreted gas at each timestep settle in an exponential disk according to the Mo, Mao \& White (1998) formalism.  Given that the virial radius of a halo will grow over time, and the distribution of spin parameters $\lambda$ follows the universal log-normal distribution proposed by Bullock et al. (2001), which shows no substantial dependence on halo mass, gas accreted at later times will settle in a disk with a larger scalelength.  At each timestep, we convert gas into stars following the Kennicutt-Schmidt relation by Kennicutt et al. (2007), which leads the integrated SFR to catch up with the rate of incoming gas by $z \sim 2$ for galaxies in the mass range of our observed sample.  The earlier phase, in which the internal gas reservoir of galaxies is built up, has been discussed in the context of the so-called "bath-tub" model (Bouch\'{e} et al. 2010; Krumholz \& Dekel 2011; Dav\'{e} et al. 2012).  As stars are formed from gas disks with increasingly larger scalelengths, the stellar disk will naturally grow inside-out, in line with the observed growth of SFGs over cosmic time (Franx et al. 2008; Wuyts et al. 2011b).  The resulting age gradient translates into a negative gradient in the intrinsic colors, and into intrinsic half-light radii that are slightly larger than the respective half-mass radii.  Similar conclusions, in accord with the empirical evidence presented in this paper, were drawn from the models by Dutton (2009) and Dutton et al. (2011).  Since the enrichment of the ISM with heavy elements released by stellar mass loss has started earlier in the center, the color gradient and offset in half-light versus half-mass radius is further enhanced by the increased metal and dust column at smaller radii.  We note that outflows, if present, could reduce the steepness of the metallicity gradient.  Outflows are most efficiently launched from regions with high $\Sigma_{SFR}$ (see Lehnert \& Heckman 1996 for evidence in nearby galaxies, and Newman et al. 2012 at $z \sim 2$).  They may therefore expel metals from the central regions and redistribute them over larger areas when the outflow material rains back on to the galaxy (Oppenheimer \& Dav\'{e} 2008).  Color gradients of high-redshift SFGs with red cores have recently been presented by Szomoru et al. (2011) as well.  Gradients in the gas-phase metallicity, as traced by $[NII]/H\alpha$, and the extinction towards HII regions, as traced by $H\alpha/H\beta$, are also observed in $z \sim 1-2$ SFGs using near-infrared IFU spectroscopy (F\"{o}rster Schreiber et al. 2006; Genzel et al. 2008, 2011; Jones et al. 2010; Queyrel et al. 2012; Yuan et al. 2011, but see Wright et al. 2010 for an alternative interpretation of radial $[NII]/H\alpha$ variations).  Further evidence for inside-out disk formation, at $z \sim 1$, was presented by Hathi et al. (2009).  Superposing on top of this smooth disk model the short-lived star-forming clumps discussed in Section\ \ref{nature_fate.sec}, we obtain a set of toy galaxies that share the major observational trends described in this paper, and summarized again below.

\section {Summary}
\label{summary.sec}

We analyzed the resolved colors of a complete sample of 323 SFGs at $0.5 < z < 1.5$ and 326 SFGs at $1.5 < z < 2.5$ with masses above $10^{10}\ M_{\sun}$.  We inferred their resolved stellar populations using stellar population synthesis modeling.  We constructed co-added normalized light and mass profiles, and divided them in three regimes: center, outer disk, and (off-center) clumps.  We find the following results:

$\bullet$ The fraction of SFGs with (off-center) clumps is a decreasing function of wavelength, and is even lower in the stellar mass maps (see Table\ \ref{clump.tab}).  For a given selection waveband, the fractional contribution of clumps to the total amount of light emitted by all SFGs above $10^{10}\ M_{\sun}$ is also a decreasing function of wavelength.  Clumps selected at rest-frame 2800\AA\ (equivalent to the observed $z_{850}$ band at $z \sim 2$) account for 19\% of all star formation integrated over the entire $z \sim 2$ SFG population above $10^{10}\ M_{\sun}$, while they account only for 7\% of the total stellar mass.  Similar trends are seen at $z \sim 1$.  We note that the importance of clumps in terms of the stellar mass distribution within high-redshift disk galaxies may in reality be even smaller, when subtracting the background light from the host (F\"{o}rster Schreiber et al. 2011b), and if some of the SFGs that do show deviations from a smooth stellar mass map are not disks, but instead objects in the process of merging.  Kinematic measurements are required to robustly identify their nature.

$\bullet$ The rest-frame optical colors of SFGs are not spatially uniform.  Off-center regions with elevated surface brightness exhibit bluer colors compared to the underlying disk, and the reddest colors are found in the center.  In terms of stellar population properties, the 7-band ACS+WFC3 color information is broken down into a 'median galaxy model' with a red core due to an increased extinction and stellar age, and a green disk with young and blue, highly star-forming regions superposed.  Since no significant correlation between the stellar surface mass density and surface brightness at a given radius is found, we conclude that the stellar mass maps are smoother than the light profiles, implying that the star-forming "clumps" are necessarily short-lived ($\sim 100 - 200$ Myr).

$\bullet$ We measured sizes, concentrations, Gini and M20 coefficients on rest-frame $U_{rest}$, $V_{rest}$, and stellar mass maps of the same resolution, and find SFGs to be more concentrated, smaller, and smoother in their stellar mass distribution than what is inferred from the surface brightness distribution at a fixed wavelength.  We note that this has long been known for disk galaxies in the nearby universe, and agrees with model predictions for high-redshift disks by Dutton (2009) and Dutton et al. (2011).  High-resolution, multi-wavelength imaging is essential in reconstructing their physical structure, even in the era of JWST.

$\bullet$ While stellar masses estimated from integrated multi-wavelength photometry are remarkably consistent with those obtained from resolved SED modeling at $\sim 0.7$ kpc resolution, larger biases may occur in determining other stellar population properties when only using integrated photometry (e.g., resolved SED modeling gives older integrated ages of the bulk of the stars than inferred from the integrated light, particularly for SFGs with young stellar populations).

Our results are consistent with an inside-out growth scenario in combination with a "Christmas tree" model in which star-forming clumps emerge through gravitational instabilities in gas-rich disks, and are quickly disrupted through efficient feedback from star formation.  However, the presence of blue and young star-forming clumps superposed on a redder underlying disk is also consistent with a scenario of inward clump migration, provided this takes place rapidly on timescales of 1 - 1.5 orbital times, as predicted by theory (e.g., Dekel et al. 2009b) and seen in cosmological simulations (Ceverino et al. 2010, 2012) that do not incorporate clump-disrupting outflows.  Analyzing the combined constraints from individual clump properties, global stellar profiles, and integrated gas mass fractions for observed and mock-observed simulated SFGs in concert, will be required to discriminate between the two scenarios and models with different feedback efficiencies.  Furthermore, an interesting question remains to what degree torques induced by (even short-lived) star-forming clumps may channel gas inward to the center (Bournaud et al. 2011), and contribute to bulge formation.  Direct kinematic observations of inward gas motions would be required to empirically address this question.

$ $\\

S. W. acknowledges Philip F. Hopkins, Daniel Ceverino, and Aaron A. Dutton for stimulating discussions.  We thank the referee for his very careful reading and insightful suggestions.  Support for Program number HST-GO-12060 was provided by NASA through a grant from the Space Telescope Science Institute, which is operated by the Association of Universities for Research in Astronomy, Incorporated, under NASA contract NAS5-26555.


\onecolumngrid

\vspace{0.2in}
\begin{center} APPENDIX A  TESTING RESOLVED SED MODELING\end{center}

\begin{figure}[h]
\centering
\epsscale{0.76}
\plotone{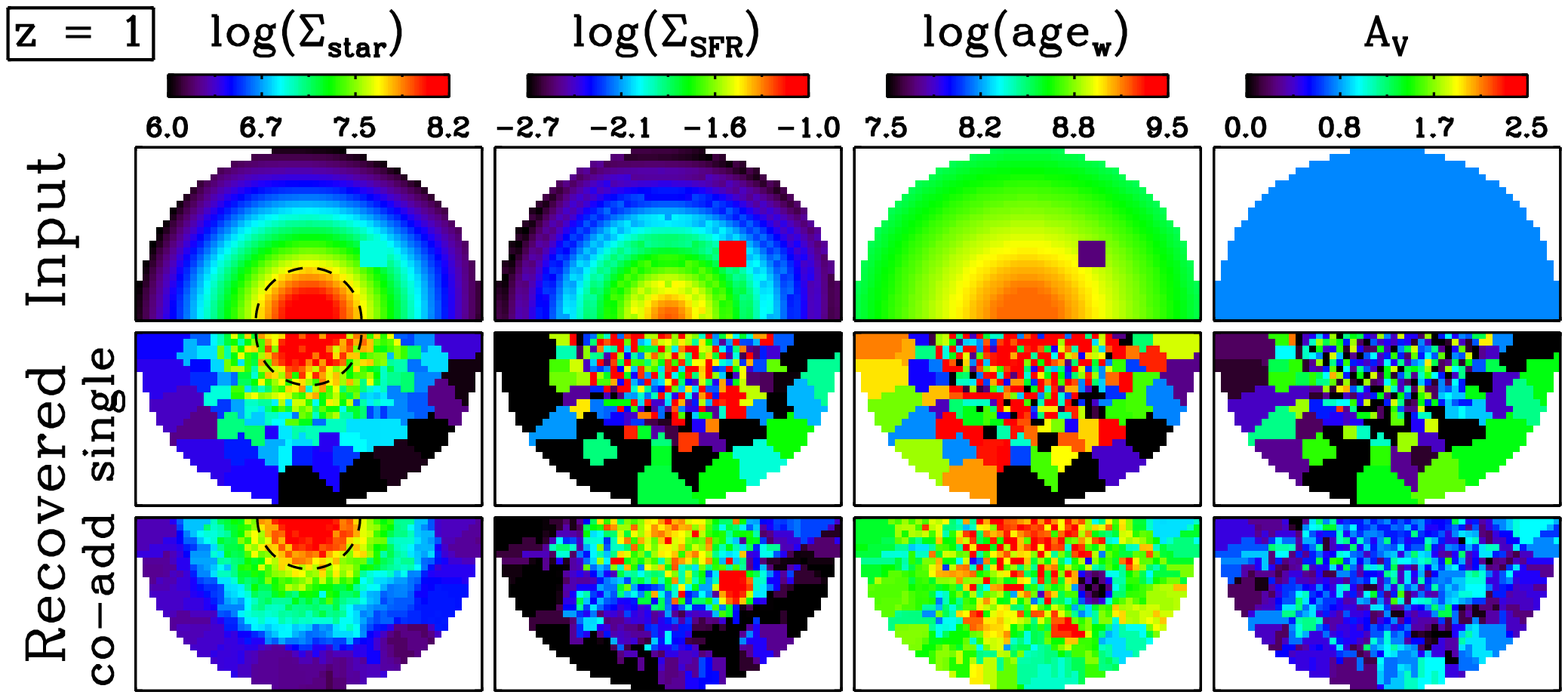} 
\vspace{0.2in}
\plotone{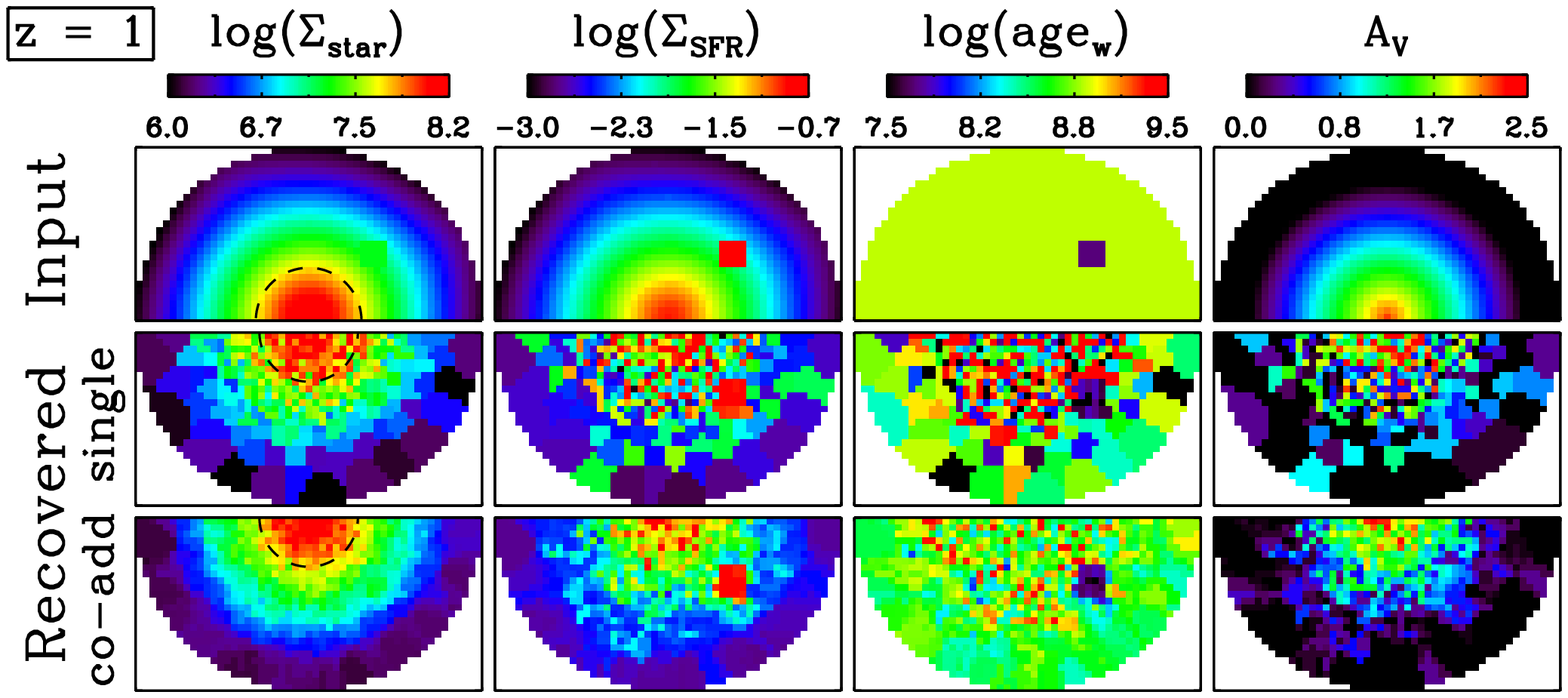} 
\vspace{0.2in}
\plotone{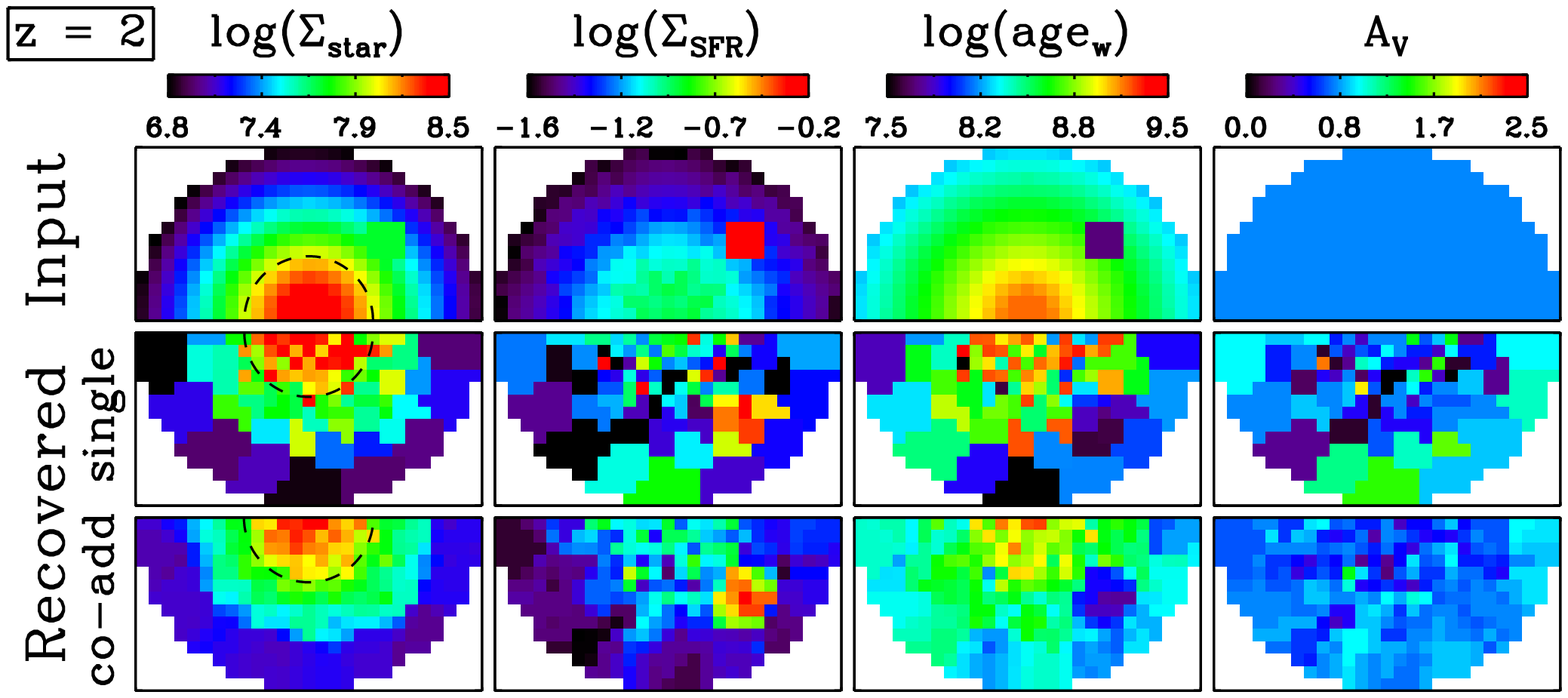} 
\vspace{0.2in}
\plotone{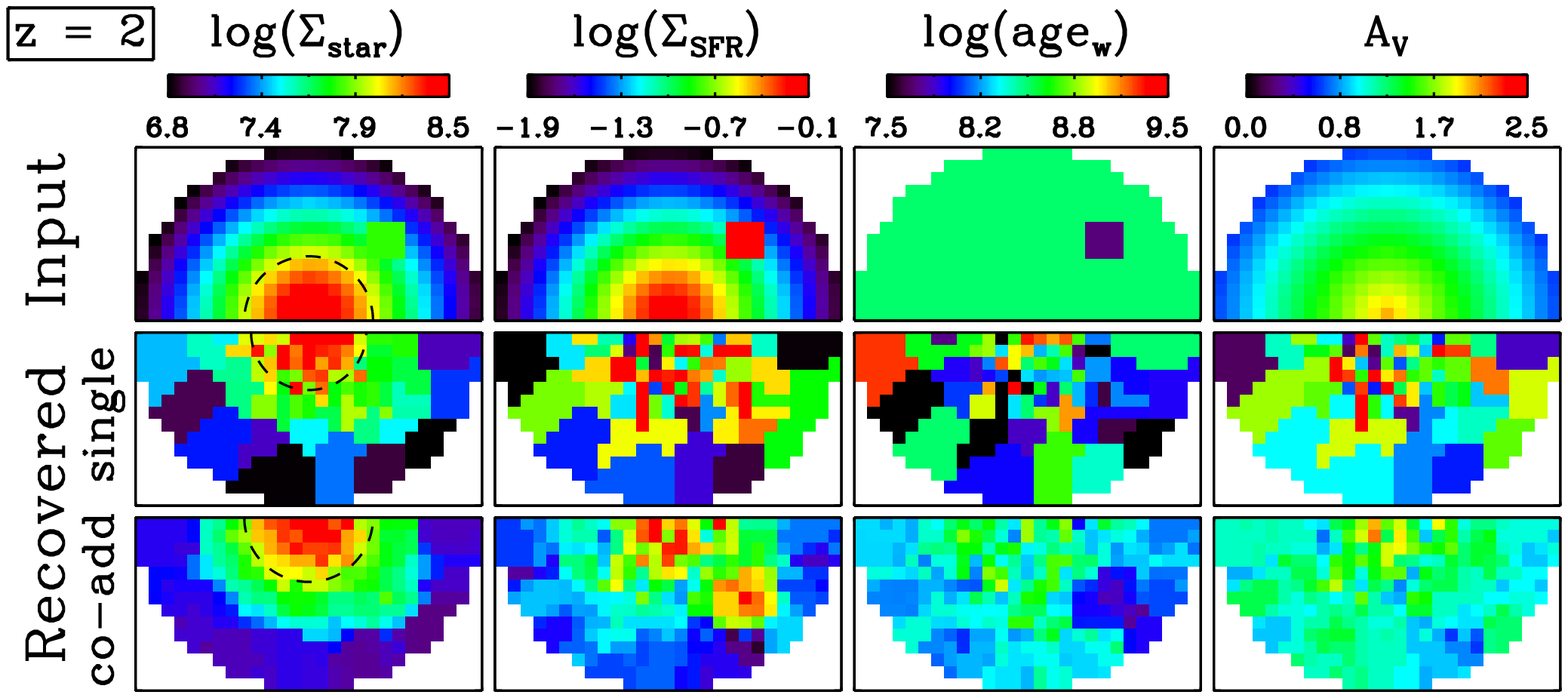} 
\caption{Comparison between input stellar population profiles of model SFGs, and the recovered profiles when applying our resolved SED modeling technique to mock observations of the model galaxies at $z = 1$ and $z = 2$.  The mock photometry was computed for the same observed wavebands as available for the real galaxies, and appropriate errors were applied.  The results for four mock galaxies are shown.  From top to bottom: a $z \sim 1$ SFG with a negative age gradient, a $z \sim 1$ SFG with a dust gradient, a $z \sim 2$ SFG with a negative age gradient, and a $z \sim 2$ SFG with a dust gradient.  In every case, we introduced a young clump with excess star formation surface density but only modest excess in stellar mass surface density.  The general characteristics of the stellar population profiles are well reproduced, particularly if the stellar population parameters derived from multiple realizations of the mock photometry are co-added (see text for more details).  The stellar mass maps are reconstructed most reliably, even on an individual object basis.  Black dashed half circles mark the input and inferred half-mass radii, which are in good agreement.
\label{mock.fig}}
\end{figure}

In order to test our ability to recover stellar population properties from the available multi-wavelength observations, we built model galaxies, and subjected their resolved photometry in $BVizYJH$ and integrated photometry in $U$, $K_s$, and IRAC to the same resolved SED modeling analysis applied to the real SFGs.  The  model galaxies have mass distributions following an exponential disk profile convolved with the WFC3 PSF.  They have an integrated stellar mass of $\log (M) = 10.5$, and SFR and size characteristic for the SFGs in our sample.  Multi-wavelength SEDs were computed in the observed frame for each pixel, and we perturbed the model fluxes randomly, with deviations drawn from a Gaussian with the photometric error in the respective band as standard deviation.  Next, we performed the Voronoi binning and computed the best-fit solution following the procedure outlined in Section\ \ref{binning.sec} and\ \ref{resolved.sec}, respectively.  In Figure\ \ref{mock.fig}, we present the results for 4 mock galaxies: a $z \sim 1$ SFG with an age gradient, a $z \sim 1$ SFG with a dust gradient, a $z \sim 2$ SFG with an age gradient, and a $z \sim 2$ SFG with a dust gradient.  In each case, we also introduced a young clump with an excess in star formation surface density for its radius.  For each toy galaxy, we show the input $\Sigma_{star}$, $\Sigma_{SFR}$, $age_w$, and $A_V$ maps in the top row.  Mirrored in the middle row are the reconstructed profiles.  We find that in all cases, the stellar mass maps are recovered most robustly.  This is consistent with previous work that indicated that also in integrated SED modeling, stellar mass is the most robust parameter (e.g., Papovich et al. 2001).  While noisier, the characteristic features of the $\Sigma_{SFR}$ map, its higher value in the center and at the clump location, are clearly reproduced.  The young clump is still present in the recovered age map, but, unsurprisingly, significant pixel-to-pixel variations in the age and dust maps are present, and correlated with each other due to the well-known age-dust degeneracy.  Additional constraints from resolved rest-frame optical emission line diagnostics (such as the $H\alpha$ equivalent width) will play a vital role here.  Age and dust are not entirely degenerate for the given photometric sampling of the SED.  This leads to a notable improvement when co-adding the SED modeling results of 10 mock observations of the same toy galaxy (bottom row for each model galaxy in Figure\ \ref{mock.fig}).  To obtain this result, we computed 10 realizations of the observed-frame photometry of the same model galaxy (i.e., with identical intrinsic properties, but independent noise).  Next, we carried out the resolved SED modeling for each realization.  Finally, we show the median recovered stellar population property at each location in the map.  The fact that co-adding just 10 mock observations leads to a significant improvement in the recovery of the intrinsic stellar population parameters boosts confidence in our use of co-added profiles to constrain the characteristic spatial variations in stellar populations of high-redshift SFGs.

\vspace{0.2in}
\begin{center} APPENDIX B  GALAXY-INTEGRATED STAR FORMATION HISTORIES\end{center}

\begin{figure}[h]
\centering
\epsscale{0.6}
\plotone{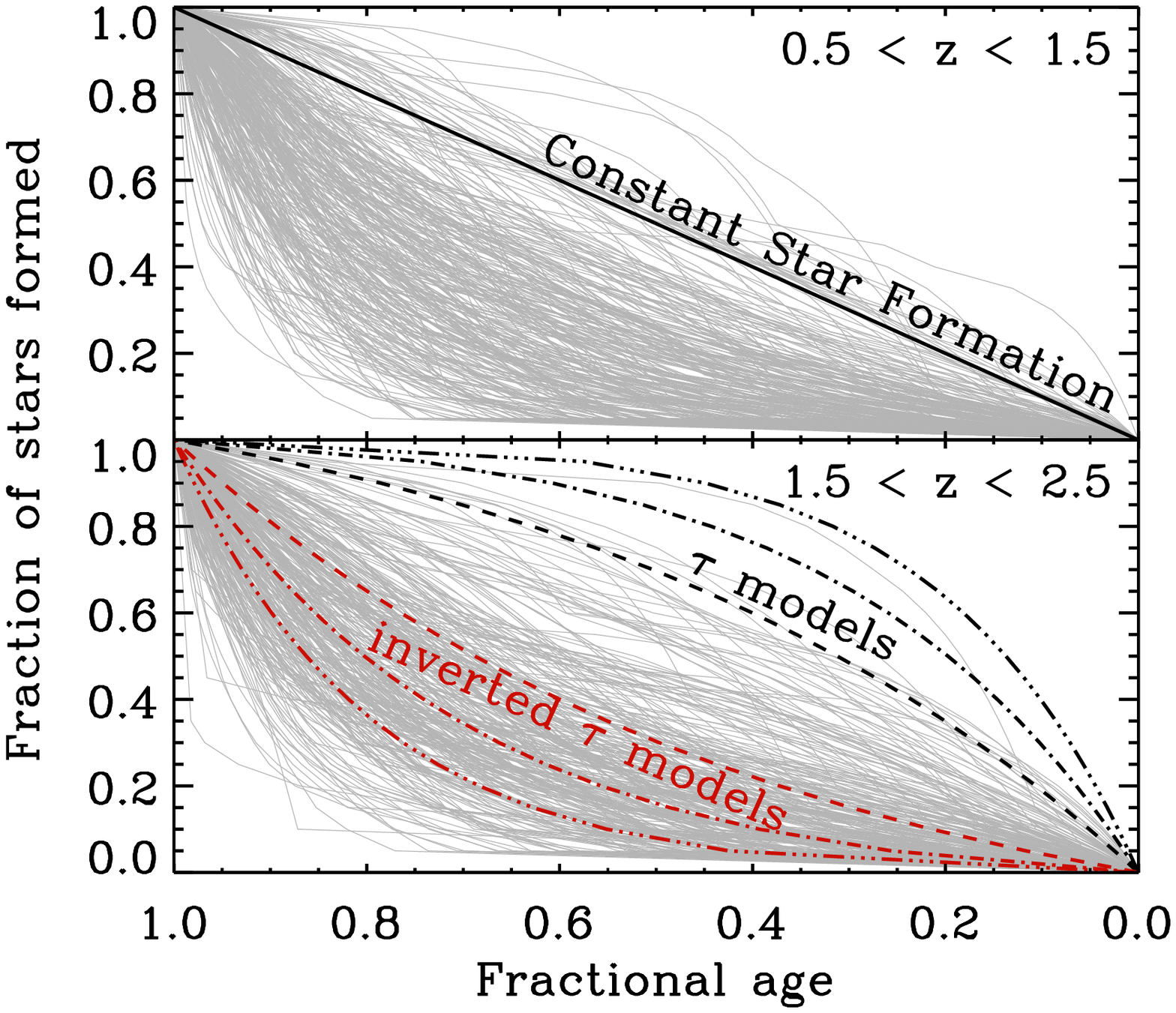}
\caption{
Effective galaxy-integrated SFHs inferred from resolved SED modeling of the SFGs in our sample (thin gray lines).  For comparison, we illustrate how the build-up of stellar mass proceeds for a constant star formation history (solid black line), and for a few exponentially declining (black) and increasing (red) star formation histories.  Dashed, dash-dotted, and dash-tripple-dotted lines mark galaxies with $tau = 300$ Myr, observed at 0.5 Gyr, 1 Gyr, and 1.5 Gyr after the onset of star formation.  Most tracks fall below the line of constant star formation history, implying that star formation initially proceeded at a slower pace, and continued at a higher rate at later times.
\label{galaxy_integrated_SFH.fig}}
\end{figure}

In modeling the broad-band SEDs of each spatial bin, we allowed a range of SFHs that are exponentially decreasing with time, with e-folding times down to 300 Myr (see Section\ \ref{resolved.sec}).  Since each spatial bin in a galaxy is assigned its own age and $\tau$, the effective SFH of the galaxy as a whole, obtained by summing the best-fit $\tau$ models of all spatial bins, can in principle be arbitrarily complex.  In order to investigate the characteristic SFH of galaxies residing on the so-called main sequence of star formation, we contrast the galaxy-integrated SFHs inferred from our resolved SED modeling analysis (thin gray lines) to simple functional forms in Figure\ \ref{galaxy_integrated_SFH.fig}.  We consider a history in which the SFR is constant over time (solid black line), exponentially decreasing ($exp{-t/\tau}$, so-called $\tau$ models), and exponentially increasing ($exp{t/\tau}$, so-called inverted $\tau$ models).  The latter two functional forms can span a range of tracks in Figure\ \ref{galaxy_integrated_SFH.fig} depending on the value of $\tau$ and the epoch of observation.  We illustrate their characteristic shape  for $\tau = 300$ Myr, as observed 0.5 Gyr, 1 Gyr, and 1.5 Gyr after the onset of star formation.  In the vast majority of cases, the tracks of observed SFGs lie below the line of constant star formation.  We conclude that, similar to inverted $\tau$ models, the galaxy-integrated star formation histories of the SFGs in our sample exhibit in most cases a slow onset of star formation.  As the SFGs grow more massive, they build up stars at a faster rate.  Eventually, at least after star formation in the last spatial bin turns on, the SFR will by construction decline over time (as no rising histories were allowed for individual pixels).  However, the point of our exercise is that this happens relatively late in the galaxy's history from the onset of star formation till the epoch of observation.

\vspace{0.2in}
\begin{center} APPENDIX C  DISTURBED MORPHOLOGIES AT ACS AND WFC3 RESOLUTION\end{center}

\begin{figure}[h]
\centering
\epsscale{0.6}
\plotone{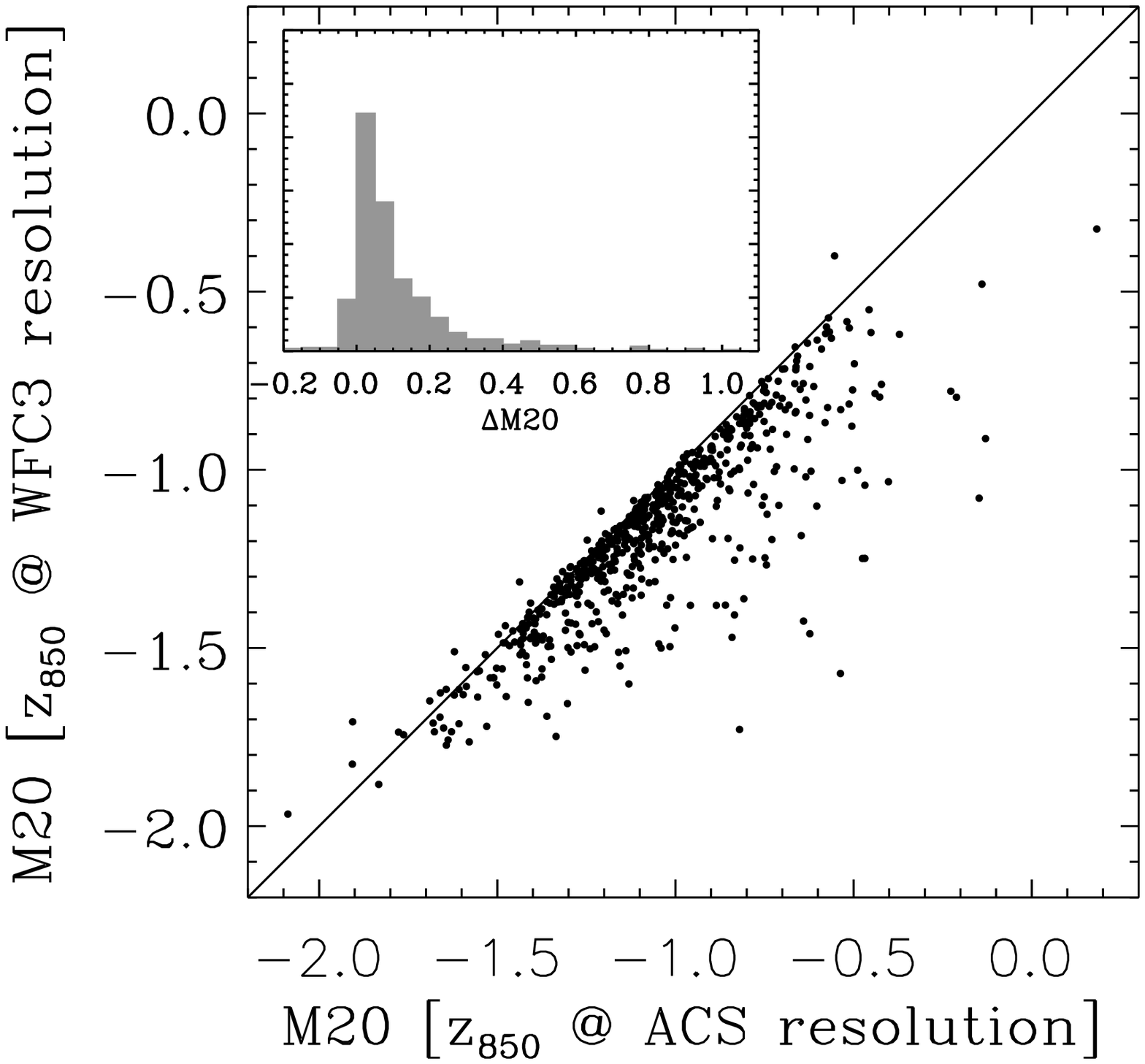}
\caption{
Comparison of the M20 coefficient of SFGs at $0.5 < z < 2.5$ measured on the observed $z_{850}$ image at the native ACS resolution and on the same waveband image after it was convolved to the WFC3 $H_{160}$ resolution.  While the majority of sources shows only a minor shift towards lower M20 after convolution, smoothing and blending of clumps by beam smearing can lead to a significantly more regular appearance for a subset of the SFGs.
\label{M20_compare.fig}}
\end{figure}
Throughout this paper, we study resolved properties from multi-wavelength imaging, PSF-matched to the WFC3 $H_{160}$-band resolution.  Implicitly, a consequence of this approach is that any substructure present below this resolution limit will not be picked up in our analysis.  In this respect, our approach differs from that by Guo et al. (2011b), who first select clumps in the $z_{850}$ band at the native ACS resolution, and subsequently study their multi-wavelength properties on PSF-matched images (i.e., at WFC3 resolution).  While our comparison between structural properties quantified on mass profiles and on light profiles at different wavebands is internally consistent, it is still useful to consider the impact of WFC3 versus ACS resolution on the presence of irregular morphologies.  We illustrate this in Figure\ \ref{M20_compare.fig}, where we compare the non-parametric M20 coefficient measured on the observed $z_{850}$ map, at the native ACS and the WFC3 resolution respectively.  We present the results for all SFGs in the $0.5 < z < 2.5$ redshift interval, and find that for most of them the difference in the M20 measurements is modest.  However, the distribution shows a tail towards large $\Delta \rm M20$, consisting of objects that intrinsically are more irregular (i.e., have a higher M20 as seen at the ACS resolution) than would be inferred from the resolution adopted in this paper.

$ $\\
\begin{references}
{\footnotesize

\reference{} Abraham, R. G., van den Bergh, S.,\& Nair, P. 2003, ApJ, 588, 218
\reference{} Abraham, R. G., Tanvir, N. R., Santiago, B. X., Ellis, R. S., Glazebrook, K.,\& van den Bergh, S. 1996, MNRAS, 279, 47
\reference{} Bell, E. F., van der Wel, A., Papovich, C., et al. 2011, submitted to ApJ, (arXiv1110.3786)
\reference{} Bertin, E.,\& Arnouts, S. 1996, A\&AS, 117, 393
\reference{} Bouch\'{e}, N., Dekel, A., Genzel, R., et al. 2010, ApJ, 718, 1001
\reference{} Bournaud, F., Dekel, A., Teyssier, R., Cacciato, M., Daddi, E., Juneau, S.,\& Shankar, F. 2011, ApJ, 741, 33
\reference{} Bournaud, F., Daddi, E., Elmegreen, B. G., et al. 2008, A\&A, 486, 741
\reference{} Bournaud, F., Elmegreen, B. G.,\& Elmegreen, D. M. 2007, ApJ, 670, 237
\reference{} Brammer, G. B., van Dokkum, P. G.,\& Coppi, P. 2008, ApJ, 686, 1503
\reference{} Bruzual, G,\& Charlot, S. 2003, MNRAS, 344, 1000
\reference{} Bullock, J. S., Dekel, A., Kolatt, T. S., Kravtsov, A. V., Klypin, A. A., Porciani, C.,\& Primack, J. R. 2001, ApJ, 555, 240
\reference{} Bundy, K., Fukugita, M., Ellis, R. S., et al. 2009, ApJ, 697, 1369
\reference{} Cacciato, M., Dekel, A.,\& Genel, S. 2012, MNRAS, 421, 818
\reference{} Calzetti, D., Armus, L., Bohlin, R. C., Kinney, A. L., Koornneef, J.\& Storchi-Bergmann, T. 2000, ApJ, 533, 682
\reference{} Cappellari, M.,\& Copin, Y. 2003, MNRAS, 342, 345
\reference{} Ceverino, D., Dekel, A., Mandelker, N., Bournaud, F., Burkert, A., Genzel, R.,\& Primack, J. 2012, MNRAS, 420, 3490
\reference{} Ceverino, D., Dekel, A.,\& Bournaud, F. 2010, MNRAS, 404, 2151
\reference{} Chabrier, G. 2003, PASP, 115, 763
\reference{} Conselice, C. J., Yang, C.,\& Bluck, A. F. L. 2009, MNRAS, 394, 1956
\reference{} Daddi, E., Dickinson, M., Morrison, G., et al. 2007, ApJ, 670 156
\reference{} Dav\'{e}, R., Finlator, K.,\& Oppenheimer, B. D. 2012, MNRAS, 421, 98
\reference{} Dekel, A., Birnboim, Y., Engel, G., et al. 2009a, Nature, 457, 451
\reference{} Dekel, A., Sari, R.,\& Ceverino, D. 2009b, ApJ, 703, 785
\reference{} Dekel, A.,\& Birnboim, Y. 2006, MNRAS, 368, 2
\reference{} de Ravel, L., Le F\`{e}vre, O., Tresse, L., et al. 2009, A\&A, 498, 379
\reference{} Dickinson, M., Hanley, C., Elston, R., et al. 2000, ApJ, 531, 624
\reference{} Dutton, A. A. 2009, MNRAS, 396, 121
\reference{} Dutton, A. A., van den Bosch, F. C.,\& Dekel, A. 2010, MNRAS, 405, 1690
\reference{} Dutton, A. A., van den Bosch, F. C., Faber, S. M., et al. 2011, MNRAS, 410, 16600
\reference{} Elbaz, D., Daddi, E., LeBorgne, D., et al. 2007, A\&A, 468, 33
\reference{} Elmegreen, B. G., Elmegreen, D. M., Fernandez, M. X.,\& Lemonias, J. J. 2009, ApJ, 692, 12
\reference{} Elmegreen, M. E., Elmegreen, B. G., Ravindranath, S.,\& Coe, D. A. 2007, ApJ, 658, 763
\reference{} Elmegreen, B. G.,\& Elmegreen, D. M. 2005, ApJ, 627, 632
\reference{} Epinat, B., Tasca, L., Amram, P., et al. 2012, A\&A, 539, 92
\reference{} F\"{o}rster Schreiber, N. M., Shapley, A. E., Erb, D. K., et al. 2011a, ApJ, 731, 65
\reference{} F\"{o}rster Schreiber, N. M., Shapley, A. E., Genzel, R., et al. 2011b, ApJ, 739, 45
\reference{} F\"{o}rster Schreiber, N. M., Genzel, R., Bouch\'{e}, N., et al. 2009, ApJ, 706, 1364
\reference{} F\"{o}rster Schreiber, N. M., Genzel, R., Lehnert, M. D., et al. 2006, ApJ, 645, 1062
\reference{} F\"{o}rster Schreiber, N. M., Genzel, R., Lutz, D.,\& Sternberg, A. 2003, ApJ, 599, 193
\reference{} Franx, M., van Dokkum, P. G., F\"{o}rster Schreiber, N. M., Wuyts, S., Labb\'{e}, I.,\& Toft, S. 2008, ApJ, 688, 770
\reference{} Fu, J., Guo, Q., Kauffmann, G.,\& Krumholz, M. R. 2010, MNRAS, 409, 515
\reference{} Genel, S., Naab, T., Genzel, R., et al. 2012, ApJ, 745, 11
\reference{} Genel, S., Genzel, R., Bouch\'{e}, N., et al. 2008, ApJ, 688, 789
\reference{} Genzel, R., Newman, S., Jones, T., et al. 2011, ApJ, 733, 101
\reference{} Genzel, R., Tacconi, L. J., Gracia-Carpio, J., et al. 2010, MNRAS, 407, 2091
\reference{} Genzel, R., Burkert, A., Bouch\'{e}, N., et al. 2008, ApJ, 687, 59
\reference{} Genzel, R., Tacconi, L. J., Eisenhauer, F., et al. 2006, Nature, 442, 786
\reference{} Giavalisco, M., Ferguson, H. C., Koekemoer, A. M., et al. 2004, ApJ, 600, 93
\reference{} Griffiths, R. E., Casertano, S., Ratnatunga, K. U., et al. 1994, ApJ, 435, 19
\reference{} Grogin, N. A., Kocevski, D. D., Faber, S. M., et al. 2011, ApJS, 197, 35
\reference{} Guo, Y., Giavalisco, M., Cassata, P., et al. 2011a, ApJ, 735, 18
\reference{} Guo, Y., Giavalisco, M., Ferguson, H. C., et al. 2011b, submitted to ApJ (arXiv1110.3800)
\reference{} Hathi, N. P., Ferreras, I, Pasquali, A., Malhotra, S., Rhoads, J. E., Pirzkal, N., Windhorst, R. A.,\& Xu, C. 2009, ApJ, 690, 1866
\reference{} Hopkins, P. F., Keres, D., Murray, N., Quataert, E.,\& Hernquist, L. 2011, submitted to MNRAS (arXiv1111.6591)
\reference{} Immeli, A., Samland, M., Gerhard, O.,\& Westera, P. 2004a, A\&A, 413, 547
\reference{} Immeli, A., Samland, M., Westera, P.,\& Gerhard, O. 2004b, ApJ, 611, 20
\reference{} Jones, T., Ellis, R., Jullo, E.,\& Richard, J. 2010, ApJ, 725, 176
\reference{} Jonsson, P., Groves, B. A.,\& Cox, T. J. 2010, MNRAS, 403, 17
\reference{} Jonsson, P. 2006, MNRAS, 372, 2
\reference{} Kartaltepe, J. S., Dickinson, M. Alexander, D. M., et al. 2011, submitted to ApJ (arXiv1110.4057)
\reference{} Kennicutt, R. C., Calzetti, D., Walter, F., et al. 2007, ApJ, 671, 333
\reference{} Keres, D., Katz, N., Fardal, M., Dav\'{e}, R.,\& Weinberg, D. H. 2009, MNRAS, 395, 160
\reference{} Keres, D. Katz, N., Weinberg, D. H.,\& Dav\'{e}, R. 2005, MNRAS, 363, 2
\reference{} Kocevski, D. D., Faber, S. M., Mozena, M., et al. 2012, ApJ, 744, 148
\reference{} Koekemoer, A. M., Faber, S. M., Ferguson, H. C., 2011, ApJS, 197, 36
\reference{} Kriek, M., van Dokkum, P. G., Labb\'{e}, I., Franx, M., Illingworth, G. D., Marchesini, D.,\& Quadri, R. F. 2009, ApJ, 700, 221
\reference{} Krumholz, M.,\& Dekel, A. 2011, submitted to ApJ (arXiv1106.0301)
\reference{} Krumholz, M.,\& Burkert, A. 2010, ApJ, 724, 895
\reference{} Lee, K.-S., Dey, A., Reddy, N., et al. 2011, ApJ, 733, 99
\reference{} Lehnert, M. D., \& Heckman, T. M. 1996, ApJ, 462, 651
\reference{} L\'{o}pez-Sanjuan, C., Le F\`{e}vre, O., de Ravel, L., et al. 2009, A\&A, 501, 505
\reference{} Lotz, J. M., Papovich, C., Faber, S. M., et al. 2011b, submitted to ApJ (arXiv1110.3821)
\reference{} Lotz, J. M., Jonsson, P., Cox, T. J., Croton, D., Primack, J. R., Somerville, R. S.,\& Stewart, K. 2011a, ApJ, 742, 103
\reference{} Lotz, J. M., Davis, M., Faber, S. M., et al. 2008, ApJ, 672, 177
\reference{} Lotz, J. M., Primack, J.,\& Madau, P. 2004, AJ, 128, 163
\reference{} Lowenthal, J. D., Koo, D. C., Guzman, R., et al. 1997, ApJ, 481, 673
\reference{} Maraston, C., Pforr, J., Renzini, A., Daddi, E., Dickinson, M., Cimatti, A.,\& Tonini, C. 2010, MNRAS, 407, 830
\reference{} Marchesini, D., van Dokkum, P. G., Quadri, et al. 2007, ApJ, 656, 42
\reference{} Mo, H. J., Mao, S.,\& White, S. D. M. 1998, MNRAS, 295, 319
\reference{} Murray, N. 2011, ApJ, 729, 133
\reference{} Neistein, E., van den Bosch, F. C.,\& Dekel, A. 2006, MNRAS, 372, 933
\reference{} Noeske, K. G., Weiner, B. J., Faber, S. M., et al. 2007, ApJ, 660, 43
\reference{} Noguchi, M. 1999, ApJ, 514, 77
\reference{} Oppenheimer, B. D.,\& Dav\'{e}, R. 2008, MNRAS, 387, 577
\reference{} Papovich, C., Bassett, R., Lotz, J. M., et al. 2012, ApJ, 750, 93
\reference{} Papovich, C., Finkelstein, S. L., Ferguson, H. C., Lotz, J. M.,\& Giavalisco, M. 2011, MNRAS, 412, 1123
\reference{} Papovich, C., Dickinson, M., Giavalisco, M., Conselice, C. J.,\& Ferguson, H. C. 2005, ApJ, 631, 101
\reference{} Papovich, C., Dickinson, M.,\& Ferguson, H. C. 2001, ApJ, 559, 620
\reference{} Queyrel, J., Contini, T., Kissler-Patig, M., et al. 2012, A\&A, 539, 93
\reference{} Renzini, A. 2009, MNRAS, 398, 58
\reference{} Robaina, A. R., Bell, E. F., Skelton, R. E., et al. 2009, ApJ, 704, 324
\reference{} Robertson, B., Bullock, J. S., Cox, T. J., Di Matteo, T., Hernquist, L., Springel, V.,\& Yoshida, N. 2006, ApJ, 645, 986
\reference{} Rodighiero, G., Daddi, E., Baronchelli, I., et al. 2011, ApJ, 739, 40
\reference{} Shapiro, K. L., Genzel, R., F\"{o}rster Schreiber, N. M., et al. 2008, ApJ, 682, 231
\reference{} Shapley, A. E., Steidel, C. C., Erb, D. K., Reddy, N. A., Adelberger, K. L., Pettini, M., Barmby, P.,\& Huang, J. 2005, ApJ, 626, 698
\reference{} Springel, V., White, S. D. M., Jenkins, A., et al. 2005, Nature, 435, 629
\reference{} Szomoru, D., Franx, M., Bouwens, R. J., van Dokkum, P. G., Labb\'{e}, I., Illingworth, G. D.,\& Trenti, M. 2011, ApJ, 735, 22
\reference{} Tacconi, L. J., Genzel, R., Neri, R., et al. 2010, Nature, 463, 781
\reference{} Thompson, R. I., Storrie-Lombardi, L. J., Weymann, R. J., et al. 1999, AJ, 117, 17
\reference{} Welikala, N., Connolly, A. J., Hopkins, A. M., Scranton, R.,\& Conti, A. 2008, ApJ, 677, 970
\reference{} Welikala, N., Hopkins, A. M., Robertson, B. E., et al. 2011, submitted to ApJ (arXiv1112.2657)
\reference{} Williams, R. J., Quadri, R. F.,\& Franx, M. 2011, ApJ, 738, L25
\reference{} Williams, R. J., Quadri, R. F., Franx, M., et al. 2009, ApJ, 691, 1879
\reference{} Windhorst, R. A., Cohen, S. H., Hathi, N. P., et al. 2011, ApJS, 193, 27
\reference{} Windhorst, R. A., Fomalont, E. B., Kellermann, K. I., Partridge, R. B., Richards, E., Franklin, B. E., Pascarelle, S. M.,\& Griffiths, R. E. 1995, Nature, 375, 471
\reference{} Wright, S. A., Larkin, J. E., Graham, J. R.,\& Ma, C.-P. 2010, ApJ, 711, 1291
\reference{} Wuyts, S., F\"{o}rster Schreiber, N. M., Lutz, D., et al. 2011a, ApJ, 738, 106
\reference{} Wuyts, S., F\"{o}rster Schreiber, N. M., van der Wel, A., et al. 2011b, ApJ, 742, 96
\reference{} Wuyts, S., Cox, T. J., Hayward, C. C., et al. 2010, ApJ, 722, 1666
\reference{} Wuyts, S., Franx, M., Cox, T. J., et al. 2009, ApJ, 700, 799
\reference{} Wuyts, S., et al. 2007, ApJ, 655, 51
\reference{} Yuan, T.-T., Kewley, L. J., Swinbank, A. M., Richard, J.,\& Livermore, R. C. 2011, ApJ, 732, 14
\reference{} Zibetti, S., Charlot, S.,\& Rix, H.-W. 2009, MNRAS, 400, 1181

}
\end {references}



\end {document}